\documentclass[a4paper,11pt]{article}
\pdfoutput=1
\usepackage[T1]{fontenc}
\usepackage{jcappub} 
\usepackage{natbib}
\setcitestyle{square,comma,numbers,sort&compress}
\usepackage{enumerate}
\usepackage{tabularx}
\usepackage[left = 1.0in, top=1.0in,right=1.0in,bottom=1.0in,a4paper]{geometry}
\usepackage[dvipsnames]{xcolor}
\usepackage{comment}
\usepackage[caption=false]{subfig}
\usepackage{cancel}
\usepackage{bbold}

\usepackage[section]{placeins}
\usepackage{listings}
\usepackage{amsbsy}
\usepackage{bm}
\usepackage{tikz}
\usepackage{tikz-feynman}
\usepackage{mathptmx}
\usepackage{pifont}
\usepackage{booktabs}
\usepackage{adjustbox}
\usepackage{standalone}
\usepackage{bm}
\usepackage[table]{xcolor}
\usepackage{tikz-cd}
\usepackage{amsthm}
\newtheorem{theorem}{Theorem}

\hypersetup{
  colorlinks=true,
  linkcolor=black,
  citecolor=black,
  urlcolor=black
}

\tikzfeynmanset{compat=1.0.0}
\usetikzlibrary{shapes.misc}
\tikzset{cross/.style={cross out, draw=black, minimum size=2*(#1-\pgflinewidth), inner sep=0pt, outer sep=0pt},cross/.default={3pt}}
\usetikzlibrary{decorations.pathreplacing, arrows.meta} 
\usetikzlibrary{arrows.meta, positioning}
\usepackage{caption}
\usepackage{amsmath}
\newcolumntype{Y}{>{\centering\arraybackslash}X}

\definecolor{myRED}{rgb}{0.8, 0.25, 0.33}
\usepackage[normalem]{ulem}
\usepackage{dsfont}
\usepackage{pbox}
\usepackage{graphicx}
\usepackage{multirow}
\usepackage{tikz}
\usetikzlibrary{arrows}
\usepackage{enumitem} 
\usepackage{ulem}
\usepackage{soul}
\usepackage{lipsum}

\newcommand{\KET}[1]{|#1 \rangle}

\def\LW{0.8cm}
\newcommand{\LISTQ}[5]{\makebox[\LW][r]{$#1$}&\makebox[\LW][r]{$#2$}&\makebox[\LW][r]{$#3$}&\makebox[\LW][r]{$#4$}&\makebox[\LW][r]{$#5$}}

\title{\boldmath\huge  
Reality-constrained Minimal Yukawa Structure in SO(10) GUT
}

\def\SU{\mathrm{SU}}
\def\SO{\mathrm{SO}}

\newcommand{\SMIRREP}[3]{(\mathbf{#1},\mathbf{#2},#3)}
\newcommand{\PSIRREP}[3]{(\mathbf{#1},\mathbf{#2},\mathbf{#3})}
\newcommand{\GIRREP}[1]{\mathbf{#1}}
\newcommand{\SUUIRREP}[2]{(\mathbf{#1},#2)}

\newcommand{\MAT}[1]{\mathbf{#1}}
\newcommand{\MATY}[1]{\mathbf{Y}_{#1}}

\newcommand{\VEV}[1]{\textcolor{red}{#1}}
\newcommand{\VEVEW}[3]{\textcolor{blue}{v_{#1}^{#2 #3}}}
\newcommand{\VEVEWT}[3]{\textcolor{blue}{\tilde{v}_{#1}^{#2 #3}}}
\newcommand{\VEVEWPS}[2]{\textcolor{blue}{v_{#1}^{#2}}}
\newcommand{\MATS}[1]{\left(\begin{smallmatrix} #1 \end{smallmatrix}\right)}

\newcommand{\INT}[2]{(#1,#2)}
\newcommand{\BR}[1]{\mathrm{Br}(#1)}

\def\DIMLEFTMARGIN{0.5cm} 

\author[a]{Shaikh Saad}
\author[b]{and Vasja Susič}

\emailAdd{shaikh.saad@ijs.si}
\emailAdd{vasja.susic@lnf.infn.it}

\affiliation[a]{Jožef Stefan Institute,\\ Jamova 39, P.O.~Box 3000, SI-1001 Ljubljana, Slovenia}

\affiliation[b]{Istituto Nazionale di Fisica Nucleare, Laboratori Nazionali di Frascati, \\
C.P.~13, Via Enrico Fermi 54, 00044 Frascati, Italy}

\abstract{
We investigate the minimal Yukawa sector of grand unified theories based on $\mathrm{SO}(10)$ symmetry, consisting of a Higgs structure with representations $\mathbf{10}_{\mathbb{R}}\oplus \mathbf{120}_{\mathbb{R}}\oplus\mathbf{126}$. Taking $\mathbf{10}_\mathbb{R}$ and $\mathbf{120}_{\mathbb{R}}$ to be real scalars, we derive the associated $\mathrm{SO}(10)$ reality conditions for their weak-doublet constituents and revisit previously reported fermion mass relations. Our analysis revises earlier results by 
introducing a relative sign difference between the reality constraints on the two weak doublets in $\mathbf{120}_{\mathbb{R}}$, yielding a new parameter (a magnitude) in the fermion mass relations. 
We establish this result through four complementary derivations: an explicit $\SO(10)$ tensorial calculation of invariants, an analytic reformulation via the Pati-Salam embedding, an algorithm for transporting reality structures from parent to daughter irreps, and an $\SU(5)$-based calculation using the $\SO(10)$ oscillator method, all giving consistent results.
The methods for determining reality conditions can be applied to any parent–daughter representation pair of $\mathrm{SO}(10)$ and its Pati–Salam subgroup, while the transport algorithm generalizes to arbitrary groups $H\subset G$.
Incorporating the correct mass relations, we perform an extensive numerical scan and find that the model successfully reproduces SM fermion masses and mixings, including recent precision measurements of solar oscillation parameters by \texttt{JUNO}. It accommodates both octants of $\theta_{23}$ while mildly disfavoring $\delta_{\mathrm{PMNS}}\sim\INT{140^\circ}{220^\circ}$. The model predicts a strongly hierarchical right-handed neutrino spectrum $(10^{5},10^{12},10^{15})$ GeV and a neutrinoless double beta decay parameter $m_{\beta\beta}\sim 3$–$4$ meV, just below future experimental sensitivity. Proton decay is dominated by $p\to\pi^+\overline{\nu}$ and $p\to\pi^0 e^+$, making these channels testable in upcoming experiments. 
}

\makeatletter
\gdef\@fpheader{}
\makeatother
\begin{document}
\maketitle
\flushbottom

\newpage
\section{Introduction \label{sec:introduction}}
The group $\SO(10)$ has long been identified as an attractive possibility for the gauge symmetry of a Grand Unified Theory (GUT)~\cite{Pati:1974yy, Georgi:1974sy, Georgi:1974yf, Georgi:1974my, Fritzsch:1974nn}. The feature, which makes it perhaps the most attractive GUT possibility, is that alongside gauge coupling unification it also provides for matter unification: each generation of Standard Model (SM) fermions, together with an extra right-handed neutrino, can be embedded into a single spinorial representation $\mathbf{16}$ of $\SO(10)$.

This remarkable economy of correctly reproducing the chiral SM fermions comes with very specific structural constraints for any realistic\footnote{
A substantial body of work has been devoted to achieving a realistic description of the fermion mass spectrum within renormalizable $\SO(10)$ GUTs.
Refs.~\cite{Babu:1992ia,Joshipura:2011nn,Dueck:2013gca,Babu:2016cri,Babu:2020tnf} have investigated both non-supersymmetric and supersymmetric $\SO(10)$ realizations. Studies focusing exclusively on the non-supersymmetric framework include Refs.~\cite{Altarelli:2013aqa,Babu:2016bmy,Ohlsson:2018qpt,Ohlsson:2019sja,Mummidi:2021anm,Saad:2022mzu,Haba:2023dvo,Kaladharan:2023zbr,Babu:2024ahk,Babu:2025wop}. Purely supersymmetric constructions, on the other hand, have been analyzed in Refs.~\cite{Babu:1995fp,Bajc:2001fe,Bajc:2002iw,Fukuyama:2002ch,Goh:2003sy,Goh:2003hf,Dutta:2004hp,Bertolini:2004eq,Bertolini:2005qb,Babu:2005ia,Dutta:2005ni,Bertolini:2006pe,Aulakh:2006hs,Grimus:2006rk,Bajc:2008dc,Fukuyama:2015kra,Babu:2018tfi,Babu:2018qca,Mohapatra:2018biy,Mimura:2019yfi}.
} Yukawa sector in an $\SO(10)$ GUT. Since the product of two spinorial representations decomposes into the irreducible representations (irreps)
\begin{align}
    \GIRREP{16}\otimes\GIRREP{16} &= \GIRREP{10}_{s} \oplus \GIRREP{120}_{a} \oplus \GIRREP{126}_{s}, \label{eq:16-times-16}
\end{align}
where $s$ and $a$ denote whether an irrep is part of the symmetric or antisymmetric combination, respectively. A renormalizable $\SO(10)$ Yukawa sector can thus utilize the following scalar irreps:
$\GIRREP{10}$, $\GIRREP{120}$, and $\GIRREP{126}$.
In supersymmetric frameworks, the fields are inherently complex; in this work, however, we focus exclusively on a non-supersymmetric and renormalizable theory.
While the $\GIRREP{126}$ is necessarily a complex representation, the $\GIRREP{10}$ and $\GIRREP{120}$ can be taken real. 

Assuming the simplest fermion sector that consists of $3$ chiral families of $\mathbf{16}$ (in particular, we assume no additional vector-like pairs $\GIRREP{16}\oplus\overline{\GIRREP{16}}$), it is well known~\cite{Babu:2016bmy} that minimal\footnote{
    ``Minimality'' here refers to a setup that employs the fewest scalar degrees of freedom in the Yukawa sector and introduces fewest parameters required to describe the Standard Model fermion masses.
    Two models may be minimal simultaneously, e.g.~when each excels over the other in one of the criteria.
}  
choices for a realistic renormalizable Yukawa sector are the following: 
(i) $\GIRREP{10}_{\mathbb{C}}\oplus\GIRREP{126}_{\mathbb{C}}$,
and (ii) $\GIRREP{10}_{\mathbb{R}}\oplus\GIRREP{120}_{\mathbb{R}}\oplus\GIRREP{126}_{\mathbb{C}}$. The former has a smaller number of degrees of freedom, while the latter involves a smaller number of parameters for the fermion fit. Each irrep has a subscript $\mathbb{R}/\mathbb{C}$ unambiguously denoting whether it is taken as real or complex. 

We are concerned in this paper with option (ii), and in particular the consequences of taking the representations $\GIRREP{10}$ and $\GIRREP{120}$ real. Some related subtleties have hitherto remained un(der)appreciated in the existing phenomenological GUT literature, and it is the goal of this paper to clarify, derive, and rectify this issue. 

For such a Yukawa sector, an analysis at the Pati-Salam (PS) $\SU(4)_C\times\SU(2)_L\times\SU(2)_{R}\equiv 4_C\,2_L\,2_R$ level\footnote{
    The subscript labels $C$, $L$ and $R$ in Pati-Salam factors refer to ``color'', ``left'' and ``right'', as usual.
    }  
already implies the mass matrices for the various SM sectors (U-up, D-down, E-charged lepton, $\nu_D$-Dirac neutrino, $\nu_{R}$-Majorana neutrino) to come out as 
\begingroup
\allowdisplaybreaks
\begin{align}
    \MAT{M}_U&=
        \MATY{10}\,\VEVEWT{10}{u}{1} + 
        \MATY{120}\,(\VEVEWT{120}{u}{1}+\VEVEWT{120}{u}{15}) + 
        \MATY{126}\, \VEVEWT{126}{u}{15}, \label{eq:MU-naive}\\
    \MAT{M}_D &=
        \MATY{10}\,\VEVEWT{10}{d}{1} +
        \MATY{120}\,(\VEVEWT{120}{d}{1}+\VEVEWT{120}{d}{15}) +
        \MATY{126}\, \VEVEWT{126}{d}{15}, \\
    \MAT{M}_E &=
        \MATY{10}\,\VEVEWT{10}{d}{1} +
        \MATY{120}\,(\VEVEWT{120}{d}{1}-3\,\VEVEWT{120}{d}{15}) 
        -3\,\MATY{126}\, \VEVEWT{126}{d}{15}, \\ 
    \MAT{M}_{\nu_D} &=
        \MATY{10}\,\VEVEWT{10}{u}{1} + 
        \MATY{120}\,(\VEVEWT{120}{u}{1}-3\,\VEVEWT{120}{u}{15}) 
        -3\,\MATY{126}\, \VEVEWT{126}{u}{15},  \label{eq:MND-naive} \\
    \MAT{M}_{\nu_{R}}&= \MAT{Y}_{126}\,\VEV{\tilde{\sigma}}. \label{eq:MNR-naive}
\end{align}
\endgroup
In the above expressions, family indices are suppressed and we instead use a bold font for $3\times 3$ matrices, while the vacuum expectation values (VEVs) are colored in \textcolor{blue}{blue} or \textcolor{red}{red} depending on whether they are switched on at the electroweak (EW) scale or seesaw scale, respectively.
Subscripts indicate the $\SO(10)$ origin of the quantity, a superscript $u$ or $d$ indicates the EW VEV is from a SM (i.e.~$\SU(3)_C\times\SU(2)_L\times\mathrm{U}(1)_Y\equiv 3_C\,2_L\,1_Y$) irrep $\SMIRREP{1}{2}{+\tfrac{1}{2}}$ or $\SMIRREP{1}{2}{-\tfrac{1}{2}}$, while a superscript $1$ or $15$ indicates the EW VEV comes from the Pati-Salam irrep $\PSIRREP{1}{2}{2}$ or
$\PSIRREP{15}{2}{2}$.
All VEVs carry a tilde sign to indicate that they are not canonically normalized,\footnote{
    We provide expressions with canonically normalized VEVs, along with a discussion of its importance, in Section~\ref{subsec:result-SO10}.
    } 
albeit associated pairs of $u$ and $d$ EW VEVs have the same normalization. The relations in Eqs.~\eqref{eq:MU-naive}--\eqref{eq:MND-naive} also correctly reproduce the Clebsch coefficients, in particular the factor $-3$ by Georgi-Jarlskog~\cite{Georgi:1979df} for the $\PSIRREP{15}{2}{2}$-VEVs when transitioning between the quark and lepton sectors ($U\to\nu_D$ or $D\to E$).

Having a real irrep $\GIRREP{10}$ and $\GIRREP{120}$ implies that the $u$- and $d$-type EW VEVs are not independent, but are related via
\begin{align}
    \VEVEWT{10}{d}{1}&=s_{3}\,\VEVEWT{10}{u}{1}^{*}, &
    \VEVEWT{120}{d}{1}&=s_{1}\,\VEVEWT{120}{u}{1}^{*}, &
    \VEVEWT{120}{d}{15}&=s_{2}\,\VEVEWT{120}{u}{15}^{*}, \label{eq:sign-definitions}
\end{align}
where the signs $s_{i}$ can have values $\pm 1$, depending on the choice of real structure. Crucially, the signs are not intrinsic to Pati-Salam, which admits either option, but are instead set by the parent reality condition from the $\SO(10)$ irrep. The correct signs $s_{i}$ can thus be derived only by a calculation using $\SO(10)$ information.

The existing literature~\cite{Babu:2016bmy} derives Eqs.~\eqref{eq:MU-naive}--\eqref{eq:MNR-naive} from Pati-Salam considerations and tacitly assumes $s_{1,2,3}=+1$. We show in this paper, however, that a full $\SO(10)$ theory calculation yields instead $s_{3}=s_{1}=1$ and $s_{2}=-1$. While overall signs associated to each $\SO(10)$ irrep are conventional, i.e.~physics does not change under $s_{3}\mapsto -s_{3}$ or $s_{1,2}\mapsto - s_{1,2}$, the relative sign $s_{1}s_{2}=-1$ is physical, as we demonstrate in our analysis. This observation has important implications: a careful reexamination of the Yukawa sector using our formalism shows that, once the independent physical inputs are properly identified, the revised reality conditions introduce one additional parameter (magnitude) relative to previous analyses.

\textbf{Summary of numerical results}---
Incorporating these corrections, we perform a comprehensive numerical analysis demonstrating that the model can successfully reproduce the observed charged fermion and neutrino mass spectra. Special attention is paid to the neutrino oscillation parameters. Our numerical studies show that the resulting neutrino oscillation parameters are fully consistent with recent high-precision measurements of solar oscillation parameters reported by \texttt{JUNO}~\cite{JUNO:2025gmd}. We further explore the implications of the model for neutrino observables that remain experimentally uncertain, in particular the atmospheric mixing angle $\theta_{23}$, which is subject to the octant ambiguity, and the leptonic CP-violating phase $\delta_{\mathrm{PMNS}}$, which is yet to be measured. We find that the model can incorporate values of $\theta_{23}$ in both octants, while slightly disfavoring values of $\delta_{\mathrm{PMNS}}$ in the range $\sim \INT{140^\circ}{220^\circ}$. While current \texttt{T2K}~\cite{T2K:2024status} and \texttt{NO$\nu$A}~\cite{2817608} data already disfavors CP conservation, \texttt{DUNE}, \texttt{T2HK}, and \texttt{ESSnuSB} will provide precise measurements to rigorously test our model and further constrain the parameter space of the theory. In addition, the model predicts suppressed neutrinoless double beta decay rates, with the effective Majorana mass parameter $m_{\beta\beta}$ lying in the range $3$--$4\,\mathrm{meV}$, right below the sensitivity of forthcoming experimental searches.    The framework also leads to a strongly hierarchical right-handed neutrino mass spectrum, $(M_1, M_2, M_3) \sim 10^{(5,12,15)}\,\mathrm{GeV}$, a direct consequence of reproducing the observed fermion masses and mixings. Finally, it yields proton decay signatures with relatively large branching ratios in the $p \rightarrow \pi^+ \overline{\nu}$ and $p \rightarrow \pi^0 e^+$ channels, making these modes promising targets for future experimental searches~\cite{Dev:2022jbf} such as \texttt{DUNE}, \texttt{THEIA}, and \texttt{Hyper-Kamiokande}.  

We organize the paper as follows: we derive/compute the signs $s_{i}$ from the reality conditions in two different ways in Section~\ref{sec:reality-conditions}, investigate the impact of these conditions on the analysis and predictions from the minimal $\SO(10)$ Yukawa sector in Section~\ref{sec:yukawa-sector}, and conclude in Section~\ref{sec:conclusions}. We provide in addition four technical appendices: a general overview of spinorial representations and their construction is given in Appendix~\ref{app:spinors}, our basis conventions in Appendix~\ref{app:basis}, two further complementary derivations of the minus sign
in Appendix~\ref{app:alternatives}, while general expressions relevant for proton decay are provided in Appendix~\ref{app:PD}.

Note: the breakdown into two main sections allows a reader interested exclusively in the technicalities of the real structure to focus on Section~\ref{sec:reality-conditions}, while a reader interested in the physical implications of the 
revised mass relations may proceed directly to Section~\ref{sec:yukawa-sector}.

\section{Deriving the reality conditions \label{sec:reality-conditions}}

\subsection{Group theory: preliminaries and conventions \label{subsec:group-theory}}

This section introduces the terminology and necessary tools for our group theory computation. Although none of these topics are novel, a dedicated introduction ensures the paper is self-contained with consistent notation and conventions; we do not shy away from technical details where we find them to facilitate conceptual clarity or prevent computational pitfalls.

The standard reference for information on Lie groups and their irreps is e.g.~Slansky~\cite{Slansky:1981yr}, while references for more specialized topics are given when these topics are encountered.
We provide a structured overview of the relevant topics below:
\begin{enumerate}[leftmargin=*, itemsep=0cm,label=(\arabic*),ref=\arabic*]
    \item \label{item:group-theory-PS-in-SO10}
        \textit{Description of Pati-Salam inside $\SO(10)$:}\\[3pt]
        The maximal subgroup of $\SO(10)$ crucial for our analysis is $\SO(6)\times\SO(4)$. We embed it so that $\SO(6)$ rotates the first 6 components of a real $10$-dimensional vector (the fundamental representation of $\SO(10)$), while $\SO(4)$ rotates the last $4$ components. We recognize $\SO(6)\times\SO(4)$ as the Pati-Salam group by accounting for the following isomorphisms between low-rank Lie algebras: $\SO(6)\cong \SU(4)$ and $\SO(4)\cong \SU(2)\times\SU(2)$. These connections are made explicit later upon the introduction of gamma matrices. A broader overview of such treatment and the associated formalism can be found in Ref.~\cite{Aulakh:2002zr}.

        Note: notationally we do not distinguish between Lie groups and algebras; also, since we are considering spinor representations, one should indeed be taking the Lie groups $\mathrm{Spin}(2n)$ rather than $\SO(2n)$, for which spinor matrices are only projective representations. This includes taking, strictly speaking, $\mathrm{Spin}(10)$ for the GUT group.
    \item \textit{Index notation for tensors:}\\[3pt]
        Due to a proliferation of different types of indices, we need to set sensible notational conventions for our purposes. We distinguish different index types by labeling them with different letters, as summarized in Table~\ref{tab:indices}. We elaborate on some subtleties below:
        \begin{table}[htb]
            \centering
            \caption{The conventions for labeling indices of all the relevant bases, representations and groups in Section~\ref{sec:reality-conditions} and Appendix~\ref{subapp:result-alternative-su5xu1}. We only use lower indices for real bases ($\mathbb{R}$), while both upper and lower indices are utilized for complex bases ($\mathbb{C}$). 
            \label{tab:indices}}
            \begin{tabular}{lllll}
                \toprule
                group & index labels & range & $\mathbb{R}/\mathbb{C}$ & irrep with index \\
                \midrule
                $\SO(2n)$ & $I,J,K,L$ & $1\ldots 2n$ & 
                    $\mathbb{R}$ & $\GIRREP{R}_{I}$ \\
                $\SO(10)$ & $p,q,r$ & $1\ldots 10$ & $\mathbb{R}$ & $\GIRREP{10}_{p}$\\
                & $i,j,k,l,m,n$ & $1\ldots 10$ & $\mathbb{C}$ & $\GIRREP{10}^{i}, \GIRREP{10}^{*}{}_{i}$\\
                & $X,Y,Z$ & $1\ldots 32$ & $\mathbb{C}$ & $(\GIRREP{16}\oplus\GIRREP{\overline{16}})^{X}, (\GIRREP{\overline{16}}\oplus\GIRREP{16})_{X}$\\
                $\SO(6)$ & $a,b$ & $1\ldots 6$ & $\mathbb{R}$ & $\GIRREP{6}_{a}$\\
                $\SO(4)$ & $\mu,\nu,\lambda,\kappa$ & $1\ldots 4$ & $\mathbb{R}$ & $\GIRREP{4}_{\mu}$\\
                $\SU(4)_C$ & $A,B,C,D,E,F$ & $1\ldots 4$ & $\mathbb{C}$ & $\GIRREP{4}^{A}$, $\GIRREP{\bar{4}}_{A}$\\
                $\SU(2)_L$ & $\alpha,\beta$ & $1\ldots 2$ & $\mathbb{C}$ & $\GIRREP{2}^{\alpha}$, $\GIRREP{\bar{2}}_{\alpha}$\\
                $\SU(2)_R$ & $\dot{\alpha},\dot{\beta}$ & $1\ldots 2$ & $\mathbb{C}$ & $\GIRREP{2}^{\dot{\alpha}}$, $\GIRREP{\bar{2}}_{\dot{\alpha}}$\\[6pt]
                $\SU(5)$ & $x,y,z$ & $1\ldots 5$ & $\mathbb{C}$ & $\GIRREP{5}^{x}$, $\GIRREP{\bar{5}}_{x}$\\[6pt]
                \bottomrule
            \end{tabular}
        \end{table}
        \begin{enumerate}[leftmargin=\DIMLEFTMARGIN,label=(\theenumi.\arabic*),ref=\theenumi.\arabic*]
            \item \textit{Indices of real bases in orthogonal groups:}\\[3pt]
            The components of the representation $\GIRREP{10}$ of $\SO(10)$ (the fundamental) are labeled in the real basis by indices starting with $p$. This real basis is the basis in which the generators of $\SO(10)$ are real matrices, and for $\GIRREP{10}_{\mathbb{R}}$ the components are real. Note that complexifying into $\GIRREP{10}_{\mathbb{C}}$ makes the components complex, but we still refer to the basis as the real basis. Similar real bases are defined for $\GIRREP{6}$ of $\SO(6)$ with index labels starting with $a$, and $\GIRREP{4}$ of $\SO(4)$ with greek indices starting with $\mu$. All real bases are self-conjugate, and thus distinguishing upper/lower indices is not necessary, simplifying the notation.
            \item \label{item:complex-basis} 
            \textit{Indices of complex bases in orthogonal groups:}\\[3pt]
                The representation $\GIRREP{10}$ of $\SO(10)$ can also be written in a complex basis, which is designed to split $\GIRREP{10}=\GIRREP{5}\oplus\GIRREP{\bar{5}}$ under the branching rule of $\SO(10)\to\SU(5)$. The components in this basis are labeled by an upper index (and of the conjugate with a lower index), starting with the letter $i$. The components $\GIRREP{10}^{i}$ will always be complex valued, whereby for $i=1\ldots 5$ they transform as a $\GIRREP{5}$ of $\SU(5)$, and for $i=6\ldots 10$ they transform as a $\GIRREP{\bar{5}}$. Taking a real irrep $\GIRREP{10}_{\mathbb{R}}$, the two sets are related via 
                $\GIRREP{\bar{5}}=\GIRREP{5}^*$. The advantage of the complex basis is that each entry has well-defined quantum numbers, i.e.~the Cartan generators are diagonal in this basis. Although analogous complex bases exist in the case of $\SO(6)$ and $\SO(4)$, we shall have no need of them in this work and thus refrain from specifying their conventions.\par
                Explicitly, the real and complex basis in $\SO(10)$ are related by a transformation matrix $P^{i}{}_{p}$, which transforms components from a real to a complex basis, see Appendix~A in \cite{Antusch:2019avd} for more details. 
                The upper and lower indices of the complex basis are related by an exchange $\GIRREP{5}\leftrightarrow \GIRREP{5}^*$, therefore the indices can be raised/lowered by the use of a matrix $P_{ij}$ (again using $P$ as a generic label for basis transformations of $\GIRREP{10}$ of $\SO(10)$; we use index labels for specifying which transformation we mean). Clearly $P^{ij}P_{jk}=\delta^{i}{}_{k}$ holds, since $P^{ij}$ and $P_{ij}$ by definition transform between the complex and anticomplex basis, and are hence inverses of each other. Explicitly we have
                \begin{align}
                    P^{ij}&=P_{ij}=
                        \begin{pmatrix}
                            0 & \mathbb{1}_{5}\\
                            \mathbb{1}_{5} & 0\\
                        \end{pmatrix}. \label{eq:Pmatrix}
                \end{align}      
            \item \textit{Indices of bases in unitary groups:}\\[3pt]
                The fundamental representations of unitary groups are complex, so we always have to distinguish between the fundamental and anti-fundamental version; we label their components by upper and lower indices, respectively. We label the indices of $\SU(4)_C$ with capital letters starting with $A$, while we label the $\SU(2)_L$ with greek letters and $SU(2)_R$ with dotted greek letters (both starting at $\alpha$). Upon conjugation, their upper/lower position is changed, but they retain their original dottedness property.\footnote{
                    The notation of dotted and undotted indices for $\SU(2)_L\times\SU(2)_R$ is reminiscent of the notation for left- and right-chiral Weyl indices in the Lorentz algebra $\SO(1,3)$, see e.g.~\cite{Martin:1997ns} for an introduction. Unlike $\SO(4)\cong \SU(2)_L\times\SU(2)_R$, where $L$ and $R$ are independent real Lie groups, i.e.~the undotted and dotted indices are unrelated, the left- and right-chiral objects of the Lorentz group are related by complex conjugation, i.e.~an undotted Weyl index is replaced with a dotted Weyl one upon conjugation, and vice versa. This is a crucial difference between $\SO(4)$ and $\SO(1,3)$.  
                    \label{foot:Euclidean-vs-Minkowski}
                    }
            \item \textit{Invariant tensors:}\\[3pt]
                A group $\SU(n)$ or $\SO(n)$ has the completely antisymmetric tensor $\epsilon$ with $n$ indices as an invariant tensor. For our purposes, the relevant invariant tensors for group theory constructions will be those of Pati-Salam factors:
                \begin{align}
                    \epsilon_{ABCD},\quad
                    \epsilon^{ABCD},\quad
                    \epsilon_{\alpha\beta},\quad
                    \epsilon^{\alpha\beta},\quad
                    \epsilon_{\dot{\alpha}\dot{\beta}},\quad
                    \epsilon^{\dot{\alpha}\dot{\beta}}. 
                    \label{eq:epsilon-tensors}
                \end{align}
                We use the conventions $\epsilon_{1234}=\epsilon^{1234}=1$ in accordance with the Euclidean signature (for the gauge group), such that $\epsilon^{ACDE}\epsilon_{BCDE} = 3!\,\delta^{A}{}_{B}$, while $\epsilon_{12}=-\epsilon^{12}=1$, so that $\epsilon^{\alpha\gamma} \epsilon_{\gamma\beta}=\delta^{\alpha}{}_{\beta}$ (an analogous relation holds also for dotted indices). The conventions for $\epsilon_{2}$ are modified relative to the Euclidean convention for $\epsilon_n$ with $n\geq 3$, so that they are more convenient for raising/lowering indices --- a unique feature of $\SU(2)$.      
        \end{enumerate}
    \item \textit{Reality conditions:} \label{item:reality-conditions}
        \begin{enumerate}[leftmargin=\DIMLEFTMARGIN,label=(\theenumi.\arabic*),ref=\theenumi.\arabic*]
        \item \label{item:real-structure} 
            \textit{What is a real structure?}\\[3pt]
            We find that conceptual clarity regarding reality conditions is improved if this topic is treated more abstractly than strictly required for deriving the necessary relations. The treatment can be found in standard textbooks, see e.g.~\cite{fulton1991representation}.  
            \par
            In formal mathematical language, a \textit{real structure} on a complex vector space $\mathcal{V}$ is an antilinear map $\rho: \mathcal{V}\to \mathcal{V}$, such that it is an involution, i.e.~$\rho^{2}=\mathrm{Id}_{\mathcal{V}}$. It allows one to split the complex vector space into two real vector spaces: $\mathcal{V} = \mathcal{V}_{\mathbb{R}}\oplus i \mathcal{V}_{\mathbb{R}} \cong \mathcal{V}_{\mathbb{R}}\otimes \mathbb{C}$, where $\mathcal{V}_{\mathbb{R}}=\{v\in \mathcal{V}; \rho(v)=v\}$ is the set of all fixed points by the real structure $\rho$. The equation $\rho(v)=v$ is referred to as the \textit{reality condition}, and it is typically expressed as a condition on the components $v^{i}$ of the vector $v$ in some given basis $e_{i}$, where $v=v^{i}\,e_{i}$. Intuitively, the real structure is a generalization of complex conjugation by possibly composing it with a suitable linear map; it is the underlying structure of any imposed reality condition, and it can be expressed in any basis of $\mathcal{V}$.
            \par
            If $\rho:\mathcal{V}\to\mathcal{V}$ is a real structure, so is its phase-transform $\rho_{\theta}:= e^{i\theta}\rho$, since $(e^{i\theta}\rho)^{2}=e^{i\theta}\rho\circ e^{i\theta}\rho=e^{i\theta}e^{-i\theta}\rho\circ\rho = \mathrm{Id}_{\mathcal{V}}$, where we used the antilinearity and involution property of $\rho$. 
            Thus a given real structure $\rho$ generates a $\mathrm{U}(1)$ family of phase-related real structures.
            The fixed-point subspace associated with $\rho_{\theta}$ is $\mathcal V_{\mathbb R}^{(\theta)}:=e^{i\theta/2}\mathcal V_{\mathbb R}$, so a generic phase rotates the real form inside $\mathcal V$. 
            If we require the decomposition $\mathcal V=\mathcal V_{\mathbb R}\oplus i\mathcal V_{\mathbb R}$ to remain the same as an unordered pair of summands, the phase freedom reduces to $\theta\in\{0,\pi\}$, i.e.~to a choice of sign $\pm 1$. The minus sign exchanges the two summands: if $\rho(v)=v$, then by antilinearity $\rho(iv)=-iv$, and hence $-\rho(iv)=iv$. Thus $i\mathcal V_{\mathbb R}=\mathcal{V}_{\mathbb{R}}^{(\pi)}$ is the real subspace associated with $-\rho$. More generally, all phase-related choices are equivalent in the sense that the real forms $\mathcal V_{\mathbb R}$ and $\mathcal V_{\mathbb R}^{(\theta)}$ are naturally isomorphic as real vector spaces.        
            \par 
            A similar concept is that of a \textit{pseudoreal} (also called a quaternionic or symplectic) structure, which is a map $\rho:\mathcal{V}\to\mathcal{V}$ that is antilinear and an anti-involution, i.e.~$\rho^{2}=-\mathrm{Id}_{\mathcal{V}}$.
        \item \label{item:real-structure-irreps}
            \textit{Real structures and representation theory:}\\[3pt]
            In the context of representation theory of Lie groups (or Lie algebras), an irreducible representation $\mathbf{R}$ of a real Lie group $G$ is \textit{real} or \textit{complex}, depending on whether the underlying vector space $\mathcal{V}$ being acted upon is a real or complex vector space. Complex irreps are either of \textit{complex type} or are \textit{self-conjugate}, with the latter further subdividing into those of \textit{real type} and \textit{pseudoreal type}. Irreps of real type admit a $G$-equivariant real structure (a real structure that commutes with the group action), while those of pseudoreal type admit a $G$-equivariant quaternionic structure. 
            It turns out the admitted $G$-equivariant structure is always unique up to a phase.   
            \par
            Intuitively, a $G$-equivariant real structure $\rho$
            allows one to select a ``real subspace'' of a complex irrep of real type, thereby obtaining a real representation. This is implemented explicitly by imposing a reality condition. 
            If we restrict the phase freedom to real coefficients, there are two possible choices differing by a sign, namely $\pm\rho$. The corresponding reality conditions pick out, respectively, what may be viewed as the ``real'' or ``imaginary'' components with respect to $\rho$. In this context, we shall refer to these as two versions of a real irrep inside the underlying complex irrep.
            \par
            In quantum field theory, one can switch between the two choices independently for each irrep by redefining its fields as $\phi\mapsto i\phi$ and absorbing any factors of $i$ appearing in invariant operators involving that irrep into the associated numerical coefficients.  
            The choice is thus merely conventional, albeit relevant for proper bookkeeping. For example, upon decomposing an irrep of $G$ under a subgroup $H$, the choice of real structure for the parent $G$-irrep induces a correlated set of reality conditions on its $H$-irreducible constituents; once the parent convention is fixed, the relative signs between these induced reality conditions carry physical information.
        \item \textit{Real structures in $\SO(10)$:}\\[3pt]
            Let us now consider the machinery of real structures on a few examples directly relevant for this paper. 
            \par 
            Consider the complex irrep $\GIRREP{10}$ of $\SO(10)$; it is of real type,             and hence admits an $\SO(10)$-equivariant real structure. Its underlying vector space is $\mathcal{V}=\mathbb{C}^{10}$. Since by definition the $\SO(10)$ acts on its fundamental representation $\GIRREP{10}$ by real rotation matrices of size $10\times 10$, and complex conjugation commutes with the action by these real matrices, the real structure can be defined as $\rho(v):= v^*$ for $v\in\mathbb{C}^{10}$. We make the sign choice of $+\rho$ rather than $-\rho$ for the real structure, thus picking $\mathbb{R}^{10}$ as our real vector space.
            \par
            The reality condition $\rho(v)=v$ can be imposed in any basis.
            Using a real basis (so $*$ does not conjugate it), the real vector space $V_{\mathbb{R}}=\mathbb{R}^{10}$ implies the components in this basis are real, hence explicitly
            \begin{align}
                \GIRREP{10}_{p}&= \GIRREP{10}_{p}^*, \label{eq:reality-condition-10-r}
            \end{align}
            which is expressed in the complex basis as 
            \begin{align}
                \GIRREP{10}^i &= P^{ij}\,(\GIRREP{10}^*)_{j}.  \label{eq:reality-condition-10-c}
            \end{align}
            Notice that the latter condition is more complicated than mere complex conjugation, since in the complex basis the basis vectors transformed non-trivially (with the matrix $P^{ij}$ from Eq.~\eqref{eq:Pmatrix}). What is preserved is the fixed-point condition of the real structure $\rho$.   
            \par 
            The next irrep we consider is $\GIRREP{120}\subset\GIRREP{10}^{\otimes 3}$. More precisely, $\GIRREP{120}=\Lambda^{3}\GIRREP{10}$ in the notation of exterior forms,  
            i.e.~an antisymmetric product of $3$ irreps $\GIRREP{10}$. The real structure $\rho$ on the $\GIRREP{10}$ can thus be extended by linearity to $\GIRREP{120}$; the reality conditions 
            expressed in the real and complex basis are then explicitly
            \begin{align}
                \GIRREP{120}_{pqr} &= \GIRREP{120}^{*}_{pqr}, &
                \GIRREP{120}^{ijk} &= P^{il}\,P^{jm}\,P^{kn}\;\GIRREP{120}^{*}{}_{lmn}.
                \label{eq:reality-condition-120-rc}
            \end{align}
            Examples of real structure on Pati-Salam irreps can be found in Section~\ref{subsec:result-PS}.
        \end{enumerate}
    \item \textit{Gamma matrices and spinor representations:}\\[3pt]
        We define our notation and list the relevant properties for the cases of $\SO(4)$, $\SO(6)$ and $\SO(10)$ below. The setup is based on the conclusions of the general theory in $\SO(2n)$, which is summarized in  Appendix~\ref{app:spinors}. Some of the technical details here extend the treatment from Ref.~\cite{Aulakh:2002zr}. 
        \begin{enumerate}[leftmargin=\DIMLEFTMARGIN,label=(\theenumi.\arabic*)] 
            \item \textit{Spinors in $\SO(4)$:}\\[3pt]
                In $\SO(4)$, there are $4$ gamma matrices $4\times 4$ that we label by $\bm{\Gamma}^{(4)}_{\mu}$. In the chiral basis with a canonical form for the charge operator, see Appendix~\ref{app:spinors-gamma} for an explicit construction, the gamma matrices take the block form
                \begin{align}
                    \bm{\Gamma}^{(4)}_{\mu} &= 
                        \begin{pmatrix}
                            0 & \bm{\sigma}_{\mu}\\
                            \overline{\bm{\sigma}}_{\mu} & 0\\
                        \end{pmatrix}
                        =
                        \begin{pmatrix}
                            0 & \sigma_{\mu}{}^{\alpha\dot{\alpha}} \\
                            \overline{\sigma}_{\mu\dot{\alpha}\alpha} & 0\\
                        \end{pmatrix}, 
                        \label{eq:gamma-SO4}
                \end{align}
                see Table~\ref{tab:indices} for index-labeling conventions. 
                In this construction, the basis has been adapted such that the rows describe the space $2_L\oplus \bar{2}_R$ in $\SU(2)_L\times\SU(2)_R$ (where an implicit choice of what is $\GIRREP{2}$ and what is $\overline{\GIRREP{2}}$ was made for each factor) and a conjugate basis to that in the columns. The off-diagonal $2\times 2$ blocks $\bm{\sigma}_{\mu}$ and $\overline{\bm{\sigma}}_{\mu}$ thus carry an appropriate index structure for this basis. These objects thus intertwine the description of an object as a $\GIRREP{4}$ of $\SO(4)$ (index $\mu$) and its description in terms of being a $(\GIRREP{2},\GIRREP{2})$ of $\SU(2)_L\times\SU(2)_R$, concretely implementing the Lie algebra isomorphism.  Regarding the index formalism, we reiterate that undotted and dotted indices here are independent, see footnote~\ref{foot:Euclidean-vs-Minkowski}.
                \par
                The gamma matrices $\bm{\Gamma}_{\mu}^{(4)}$ are Hermitian, and thus $\bm{\overline{\sigma}}=\bm{\sigma}^\dagger$. Furthermore, 
                there is an equivalence of irreps $\GIRREP{2}\sim\GIRREP{\bar{2}}$ in $\SU(2)$. This implies the objects behave under conjugation as 
                \begin{align}
                    (\sigma^{*})_{\mu\alpha\dot{\alpha}} &= 
                        \epsilon_{\alpha\beta}\,
                        \epsilon_{\dot{\alpha}\dot{\beta}} \,
                        \sigma_{\mu}{}^{\beta\dot{\beta}}
                        =
                        \overline{\sigma}_{\mu\dot{\alpha}\alpha}
                        , &
                    (\overline{\sigma}^{*})_{\mu}{}^{\dot{\alpha}\alpha} &= 
                        \epsilon^{\dot{\alpha}\dot{\beta}}\, 
                        \epsilon^{\alpha\beta}\,
                        \overline{\sigma}_{\mu\dot{\beta}\beta}
                        =
                        \sigma_{\mu}{}^{\alpha\dot{\alpha}}, 
                    \label{eq:offblock-so4-conjugation}
                \end{align}
                where the $2$-index $\epsilon$-tensor is used to raise and lower the fundamental indices of $\SU(2)$ factors. The underlying structures also imply the objects must satisfy the Clifford algebra relations
                \begin{align}
                    \sigma_{\mu}^{\alpha\dot{\alpha}} \,\overline{\sigma}_{\nu\dot{\alpha}\beta} 
                    + \sigma_{\nu}^{\alpha\dot{\alpha}} \,\overline{\sigma}_{\mu\dot{\alpha}\beta}  &
                        = 2\,\delta_{\mu\nu}\,\delta^{\alpha}{}_{\beta},\\
                    \overline{\sigma}_{\mu\dot{\alpha}\alpha}\,\sigma_{\nu}^{\alpha\dot{\beta}} +
                    \overline{\sigma}_{\nu\dot{\alpha}\alpha}\,\sigma_{\mu}^{\alpha\dot{\beta}}
                      &
                        = 2\,\delta_{\mu\nu}\,\delta_{\dot{\alpha}}{}^{\dot{\beta}},
                        \label{eq:offblock-so4-Clifford}
                \end{align}
                the orthogonality relation
                \begin{align}
                    \sigma_{\mu}{}^{\alpha\dot{\alpha}}\, \overline{\sigma}_{\nu\dot{\alpha}\alpha} &
                        = 2\, \delta_{\mu\nu},
                        \label{eq:offblock-so4-orthogonality}
                \end{align}
                and the completeness relation
                \begin{align}
                    \sigma_{\mu}{}^{\alpha\dot{\alpha}} \,\overline{\sigma}_{\mu\dot{\beta}\beta}&
                        = 2 \,\delta^{\alpha}{}_{\beta}\,\delta^{\dot{\alpha}}{}_{\dot{\beta}}.
                        \label{eq:offblock-so4-completeness}
                \end{align}
            \item \textit{Spinors in $\SO(6)$:}\\[3pt]
                In $\SO(6)$, there are $6$ gamma matrices $8\times 8$ that we label by $\bm{\Gamma}^{(6)}_{a}$. In the chiral basis constructed in Appendix~\ref{app:spinors-gamma}, the basis for rows corresponds to the split
                $\GIRREP{4}\oplus \GIRREP{\bar{4}}$ of the spinorial representation, while the columns are in the conjugate basis $\GIRREP{\bar{4}}\oplus\GIRREP{4}$. The  gamma matrices take the block form
                \begin{align}
                    \bm{\Gamma}^{(6)}_{a} &= 
                        \begin{pmatrix}
                            0 & \bm{\Sigma}_{a}\\
                            \overline{\bm{\Sigma}}_{a} & 0\\
                        \end{pmatrix}
                        =
                        \begin{pmatrix}
                            0 & \Sigma_{a}{}^{AB} \\
                            \overline{\Sigma}_{aAB} & 0\\
                        \end{pmatrix}, 
                        \label{eq:gamma-SO6}
                \end{align}
                in accordance with index-labeling conventions of Table~\ref{tab:indices}. The $\Sigma$-blocks are $4\times 4$ antisymmetric matrices 
                and the indices they carry are identified as $\SU(4)_C$ indices in the usual convention. The existence of this connection is due to the isomorphism of Lie algebras of $\SU(4)$ and $\SO(6)$, which also implies that $\Sigma$-blocks are intertwining objects between the two alternative  descriptions $\Lambda^{2} \GIRREP{4}=\GIRREP{6}$. Note that this remarkable connection is manifest in the usual conventions only in a proper $\SU(4)$-adapted basis in spinor space, in which the charge conjugation matrix $\MAT{C}$ takes canonical form, cf.~Appendix~\ref{app:spinors-gamma}.  
                \par
                Since $\bm{\Gamma}^{(6)}_{a}$ are Hermitian in this basis, the off-diagonal blocks are connected via $\overline{\bm{\Sigma}}=\bm{\Sigma}^\dagger$, which together with their antisymmetry in $\SU(4)$ indices leads to the set of relations
                \begin{align}
                    \Sigma_{a}{}^{AB} &= - \Sigma_{a}{}^{BA}, &
                    \overline{\Sigma}_{aAB} &= - \overline{\Sigma}_{aBA}, 
                    \label{eq:offblock-so6-antisymmetry}\\
                    \Sigma^*{}_{aAB} &= -\overline{\Sigma}_{aAB}, &
                    \overline{\Sigma}^*{}_{a}{}^{AB} &= - \Sigma_{a}{}^{AB}.
                    \label{eq:offblock-so6-conjugation}
                \end{align}
                Furthermore, they must also satisfy the $\SU(4)$ duality relations
                \begin{align}
                    \overline{\Sigma}_{aAB}&
                        =\tfrac{1}{2}\epsilon_{ABCD}\,\Sigma_{a}{}^{CD}, &
                    \Sigma_{a}{}^{AB}&=\tfrac{1}{2}\epsilon^{ABCD}\,\overline{\Sigma}_{aCD},
                    \label{eq:relation-spinors-so6-duality}
                \end{align}
                the Clifford algebra relation
                \begin{align}
                    \Sigma_{a}{}^{AB}\,\overline{\Sigma}_{bBC} + \Sigma_{b}{}^{AB}\,\overline{\Sigma}_{aBC} & = 2\,\delta_{ab}\,\delta^{A}{}_{C},
                    \label{eq:offblock-so6-Clifford}
                \end{align}
                the orthogonality relations
                \begin{align}
                    \tfrac{1}{2} \, \Sigma_{a}^{AB}\,\Sigma^{*}{}_{bAB}
                    \;=\; -\tfrac{1}{2} \, \Sigma_{a}^{AB}\,\overline{\Sigma}_{bAB} 
                    \;=\; -\tfrac{1}{4} \,\epsilon_{ABCD} \,\Sigma_{a}{}^{AB} \,\Sigma_{b}{}^{CD}
                    \;=\; -\tfrac{1}{4} \,\epsilon^{ABCD} \,\overline{\Sigma}_{aAB} \,\overline{\Sigma}_{bCD}
                    & \;=\; 2\,\delta_{ab},
                    \label{eq:offblock-so6-orthogonality}
                \end{align}
                and the completeness relations
                \begin{align}
                    \Sigma_{a}{}^{AB}\,\Sigma_{a}{}^{CD} 
                        & = -2\,\epsilon^{ABCD}, &
                    \overline{\Sigma}_{aAB}\,\overline{\Sigma}_{aCD}
                        &= -2\,\epsilon_{ABCD},\\
                    \Sigma_{a}{}^{AB} \,\Sigma^{*}{}_{aCD}    
                        &= 2\, \delta^{AB}_{CD},&
                    \overline{\Sigma}_{aCD} \,\overline{\Sigma}^{*}{}_{a}{}^{AB} 
                        &= 2\, \delta^{AB}_{CD},
                        \label{eq:offblock-so6-completness}
                \end{align}
                where $\delta^{AB}_{CD}:=\delta^{A}{}_{C}\,\delta^{B}{}_{D}-\delta^{A}{}_{D}\,\delta^{B}{}_{C}$ is the generalized Kronecker symbol.
                The numeric coefficients in all these relations reflect that any summation over a pair of common indices in two objects produces pairs of same-value terms, e.g.~a factor $\tfrac{1}{2}$ in Eq.~\eqref{eq:relation-spinors-so6-duality} appears due to summing over the pair of antisymmetric indices $CD$. 
            \item \textit{Spinors in $\SO(10)$:}\\[3pt]
                In $\SO(10)$, there are $10$ gamma matrices $32\times 32$ that we label by $\bm{\Gamma}_{p}\equiv \bm{\Gamma}^{(10)}_{p}$. The chiral 
                basis gives them in the $\GIRREP{16}\oplus\overline{\GIRREP{16}}$ split of spinorial representations for rows, and the conjugate of that for columns.
                Unlike $\SO(4)$ and $\SO(6)$, we do not need to consider any particular relations for such objects or their block off-diagonal structure. In the explicit $\SO(10)$ computation of Section~\ref{subsec:result-SO10}, we make use only of their tensor components $\Gamma_{p}{}^{X}{}_{Y}$ and the charge conjugation matrix $C_{XY}$  as constructed in Appendix~\ref{app:spinors-gamma}. 
        \end{enumerate}
\end{enumerate} 

\subsection{The reality-condition ambiguity in Pati-Salam language \label{subsec:result-PS}}

We now show how far a pure Pati-Salam computation can take us, and how the sign ambiguity discussed in the Introduction arises. The material here is straightforward, but we are purposefully presenting it here very explicitly and pedagogically, so that the discussion can be concrete.

Consider a description of the Yukawa sector in Pati-Salam $4_C\,2_L\,2_R$. As usual, the SM fermions of one generation (together with the right-handed neutrino $\nu^{c}$) are found in PS irreps
\begin{align}
    Q&\sim \PSIRREP{4}{2}{1}, &
    Q^{c}&\sim \PSIRREP{\bar{4}}{1}{2},
\end{align}
which can be written in terms of our index notation, cf.~Table~\ref{tab:indices}, and individual components as
\begin{align}
    Q^{A \alpha} &= 
    \begin{pmatrix}
        u_1 & u_2 & u_3 & \nu \\
        d_1 & d_2 & d_3 & e \\
    \end{pmatrix}^{\top}, &
    (Q^{c})_{A \dot{\alpha}} &=
    \begin{pmatrix}
        u_1^c & u_2^c & u_3^c & \nu^c \\
        d_1^c & d_2^c & d_3^c & e^c \\
    \end{pmatrix}^{\top}. \label{eq:irreps-PS-fermions}
\end{align}
Each of these fields carries also a family and a Weyl index (transforming as an L-chiral $(1/2,0)$ under the Lorentz group), suppressed in the above notation.

Let us consider two scalar PS irreps as a source of the SM Higgs:
\begin{align}
    \phi&\sim \PSIRREP{1}{2}{2}_{\mathbb{C}}, &
    \varphi&\sim \PSIRREP{15}{2}{2}_{\mathbb{C}}, \label{eq:irreps-PS-scalars}
\end{align}
with the initial assumption that they are complex. Each then contains an independent $u$-type and $d$-type doublet, i.e.~independent SM irreps $\SMIRREP{1}{2}{+\tfrac{1}{2}}$ and 
$\SMIRREP{1}{2}{-\tfrac{1}{2}}$, respectively. We label the EW VEVs by specifying their type in the superscript and the irrep (the $\SU(4)_C$ part) in the subscript. The EW VEVs in terms of components then become
\begin{align}
    \langle \phi^{\alpha\dot{\alpha}}\rangle &= 
        \begin{pmatrix}
            0 & -\VEVEWPS{1}{d} \\
            \VEVEWPS{1}{u} & 0 \\
        \end{pmatrix}, &
    \langle \varphi^{A}{}_{B}{}^{\alpha\dot{\alpha}} \rangle &= 
        \tfrac{1}{2\sqrt{3}}\,\mathrm{diag}(1,1,1,-3)\otimes 
        \begin{pmatrix}
            0 & -\VEVEWPS{15}{d} \\
           \VEVEWPS{15}{u} & 0 \\
        \end{pmatrix}, \label{eq:ansatz-PS-scalars}
\end{align}
where the $\PSIRREP{15}{2}{2}$ VEVs are in the $\MAT{T}_{15}$ direction of $\SU(4)_C$ (not breaking $\SU(3)_C$). The VEVs are normalized canonically in the sense that the quadratic invariants come out as
\begin{align}
    \langle (\phi)^{\alpha\dot{\alpha}} (\phi^*)_{\alpha\dot{\alpha}} \rangle &= 
        |\VEVEWPS{1}{u}|^{2} + |\VEVEWPS{1}{d}|^{2}, 
        \label{eq:normalization-PS-1}\\
    \langle (\varphi)^{A}{}_{B}{}^{\alpha\dot{\alpha}} (\varphi^*)_{A}{}^{B}{}_{\alpha\dot{\alpha}} \rangle &= 
        |\VEVEWPS{15}{u}|^{2} + |\VEVEWPS{15}{d}|^{2},
        \label{eq:normalization-PS-15}
\end{align}
while their sign/phase is conventional and set by Eq.~\eqref{eq:ansatz-PS-scalars}.

The Yukawa part of the Lagrangian in such a PS setup is written with explicit index contractions as 
\begin{align}
    \mathcal{L}_{Y} &= \phantom{+}
    \MATY{1}\; (Q)^{A\alpha}\,(Q^{c})_{A \dot{\alpha}}\,(\phi)^{\beta\dot{\alpha}} \;\epsilon_{\alpha\beta} 
    +2\sqrt{3}\; \MATY{15}\; (Q)^{A\alpha}\, (Q^{c})_{B\dot{\alpha}}\, (\varphi)^{B}{}_{A}{}^{\beta\dot{\alpha}} \;\epsilon_{\alpha\beta} \nonumber \\
    & \quad
    + \MAT{Y}'_{1}\; (Q)^{A\alpha}\,(Q^{c})_{A \dot{\alpha}}\,(\phi^*)_{\alpha\dot{\beta}} \;\epsilon^{\dot{\alpha}\dot{\beta}}
    +2\sqrt{3}\; \MAT{Y}'_{15}\; (Q)^{A\alpha}\, (Q^{c})_{B\dot{\alpha}}\, (\varphi^*)_{A}{}_{B}{}_{\alpha\dot{\beta}} \; \epsilon^{\dot{\alpha}\dot{\beta}} + h.c., \label{eq:Yukawa-PS}
\end{align}
where the family indices and Weyl indices carried by fermions have been suppressed; the $h.c.$ part adds conjugate terms, in which the pair of left-chiral fermions $Q$ and $Q^{c}$ is converted into a pair of right-chiral fermions $\overline{Q}$ and $\overline{Q}^c$. The Yukawa couplings $\MATY{1}$, $\MATY{15}$, $\MAT{Y}'_{1}$ and $\MAT{Y}'_{15}$ are independent and complex matrices, and numerical prefactors have been chosen such that coefficients of $1$ emerge later in the fermion mass matrix $\MAT{M}_U$. 

The reason for the appearance of the primed-Yukawa terms in Eq.~\eqref{eq:Yukawa-PS} is that representations $\phi$ and $\varphi$ of Eq.~\eqref{eq:irreps-PS-scalars} are self-conjugate, so an invariant can be formed either with their original or conjugate versions, analogous to how there are two terms $\GIRREP{16}_F\cdot \GIRREP{16}_{F}\cdot \GIRREP{10}$ and $\GIRREP{16}_F\cdot \GIRREP{16}_{F}\cdot \GIRREP{10}^*$ in the $\SO(10)$ Yukawa sector for a complex $\GIRREP{10}_\mathbb{C}$~\cite{Babu:1992ia}. Imposing a PQ symmetry by hand in $\SO(10)$ removes e.g.~the $\GIRREP{10}^*$ term~\cite{Babu:1992ia}, and analogously imposing PQ in the PS setup removes e.g.~the primed-Yukawa terms in Eq.~\eqref{eq:Yukawa-PS}~\cite{Saad:2017pqj}.

Computing explicitly by contracting the tensors in Eq.~\eqref{eq:Yukawa-PS}, we derive the well-known form of the fermion mass matrices~\cite{Mohapatra:1980qe,Pati:1983zp}:
\begin{align}
    \MAT{M}_U &= 
        \MAT{Y}_{1}\,\VEVEWPS{1}{u} + \MAT{Y}_{15} \,\VEVEWPS{15}{u}
        +\MAT{Y}'_{1} \,\VEVEWPS{1}{d}{}^* + \MAT{Y}'_{15} \,\VEVEWPS{15}{d}{}^*,
        \label{eq:result-PS-U}
        \\
    \MAT{M}_D &= 
        \MAT{Y}_{1} \,\VEVEWPS{1}{d} + \MAT{Y}_{15} \,\VEVEWPS{15}{d}
        +\MAT{Y}'_{1} \,\VEVEWPS{1}{u}{}^* + \MAT{Y}'_{15} \,\VEVEWPS{15}{u}{}^*,
        \label{eq:result-PS-D}
        \\
    \MAT{M}_E &= 
        \MAT{Y}_{1} \,\VEVEWPS{1}{d} -3\, \MAT{Y}_{15} \,\VEVEWPS{15}{d}
        +\MAT{Y}'_{1} \,\VEVEWPS{1}{u}{}^* -3\, \MAT{Y}'_{15} \,\VEVEWPS{15}{u}{}^*,
        \label{eq:result-PS-E}
        \\
    \MAT{M}_{\nu_{D}} &= 
        \MAT{Y}_{1} \,\VEVEWPS{1}{u} -3 \MAT{Y}_{15} \,\VEVEWPS{15}{u}
        +\MAT{Y}'_{1} \,\VEVEWPS{1}{d}{}^* -3 \MAT{Y}'_{15} \,\VEVEWPS{15}{d}{}^*.
        \label{eq:result-PS-ND}
\end{align}

Suppose now we take the representations $\phi$ and $\varphi$ to be real (in the sense of having a PS-covariant real structure, see item~\ref{item:reality-conditions} in Section~\ref{subsec:group-theory}), i.e.~we impose on the complex irreps of Eq.~\eqref{eq:irreps-PS-scalars} the reality conditions of the form
\begin{align}
    (\phi^{*})_{\alpha\dot{\alpha}} &= s'_{1} \;\epsilon_{\alpha\beta} \,\epsilon_{\dot{\alpha}\dot{\beta}}\, (\phi)^{\beta\dot{\beta}}, \label{eq:reality-phi}\\
    (\varphi^{*})_{A}{}^{B}{}_{\alpha\dot{\alpha}} &= s'_{15} \;\epsilon_{\alpha\beta} \,\epsilon_{\dot{\alpha}\dot{\beta}} \,(\varphi)^{B}{}_{A}{}^{\beta\dot{\beta}}, \label{eq:reality-varphi}
\end{align}
where $s'_{1}$ and $s'_{15}$ are signs of $\pm 1$; the choice for each is independent and reflects the real structure chosen for each complex irrep. Note that these reality conditions involve only invariant tensors, and hence the real structure commutes with the group action (it is PS-equivariant, as desired).
Inserting the ansatz of Eq.~\eqref{eq:ansatz-PS-scalars} into the reality conditions of Eqs.~\eqref{eq:reality-phi} and \eqref{eq:reality-varphi}, we relate the $u$- and $d$-type EW VEVs via
\begin{align}
    \VEVEWPS{1}{d} &= s'_{1}\;\VEVEWPS{1}{u}^{*}, &
    \VEVEWPS{15}{d} &= s'_{15}\;\VEVEWPS{15}{u}^{*}, \label{eq:reality-constraint-options-Pati-Salam}
\end{align}
where $\VEVEWPS{1}{u}$ and $\VEVEWPS{15}{u}$ are complex VEVs. This leads to fermion mass matrices
\begin{align}
    \MAT{M}_U &= 
        \MAT{Y}''_{1}\,\VEVEWPS{1}{u} + \MAT{Y}''_{15} \,\VEVEWPS{15}{u}, \label{eq:Yukawa-PS-real-U}\\
    \MAT{M}_D &= 
        s'_{1}\;\MAT{Y}''_{1} \,\VEVEWPS{1}{u}^* + s'_{15}\; \MAT{Y}''_{15} \,\VEVEWPS{15}{u}^*, \\
    \MAT{M}_E &= 
        s'_{1}\;\MAT{Y}''_{1} \,\VEVEWPS{1}{u}^* -3\,s'_{15} \; \MAT{Y}''_{15} \,\VEVEWPS{15}{u}^*,\\
    \MAT{M}_{\nu_{D}} &= 
        \MAT{Y}''_{1} \,\VEVEWPS{1}{u} -3\, \MAT{Y}''_{15} \,\VEVEWPS{15}{u}, \label{eq:Yukawa-PS-real-N}
\end{align}
with $\MAT{Y}''_{1}:=\MAT{Y}_{1}+s'_{1} \MAT{Y}'_{1}$ and $\MAT{Y}''_{15}:=\MAT{Y}_{1}+s'_{15} \MAT{Y}'_{15}$. The rearrangement into only two Yukawa matrices $\MAT{Y}''_{1}$ and $\MAT{Y}''_{15}$ shows that in the case of real irreps $\PSIRREP{1}{2}{2}$ and $\PSIRREP{15}{2}{2}$, the terms involving conjugate fields $\phi^*$ and $\varphi^*$ in Eq.~\eqref{eq:Yukawa-PS} can be skipped without loss of generality, as expected. 

Note that we first set the normalization in Eqs.~\eqref{eq:normalization-PS-1} and \eqref{eq:normalization-PS-15} for complex irreps; imposing the reality condition on them reveals the reality-constrained VEVs are no longer canonically normalized. Since there is only one EW VEV per scalar PS irrep, we can simply absorb the normalization factor into the Yukawa coefficients. This complication can be avoided if normalization is imposed after the reality condition is imposed, as we indeed do for the $\SO(10)$ computation in the next section.

The Pati-Salam scenario with reality-constrained Higgs representations is not realistic; the use of the same EW VEVs  in the up- and down-sectors is inconsistent with a large $t$-$b$ split in quark masses. Having derived the explicit result, however, allows us to interrogate its structural properties. We gather below the main inferred conceptual points:
\begin{itemize}[itemsep=0cm]
    \item 
        In Pati-Salam, we can freely choose the real structure of $\phi$ in Eq.~\eqref{eq:reality-phi} and $\varphi$ in Eqs.~\eqref{eq:reality-varphi},
        i.e.~we can independently choose the signs $s'_{1}$ and $s'_{15}$. 
    \item 
        Although the choice of $s'_{1,15}$ changes the mass-matrix expressions of Eq.~\eqref{eq:Yukawa-PS-real-U}--\eqref{eq:Yukawa-PS-real-N}, this has no physical impact, as discussed in item~\ref{item:real-structure-irreps} of Section~\ref{subsec:group-theory}. The transformation \hbox{$\phi \mapsto i\phi$} and \hbox{$\MAT{Y}''_{1}\mapsto -i\MAT{Y}''_{1}$} leaves the Yukawa terms unchanged; the fermion mass matrices also remain unchanged due to $\VEVEWPS{1}{u}\mapsto i\VEVEWPS{1}{u}$ in addition to the Yukawa matrix transformation. On the other hand, the transformation also induces the change of real structure $s'_{1}\mapsto -s'_{1}$ in Eq.~\eqref{eq:reality-phi} of the PS irrep $\PSIRREP{1}{2}{2}$. Analogous considerations hold for the transformation $\varphi \mapsto i\varphi$ and $\MAT{Y}''_{15}\mapsto -i\MAT{Y}''_{15}$, under which $s'_{15}\mapsto -s'_{15}$. In other words, the signs $s'_{1,15}$ can be absorbed by phase redefinitions of Yukawa couplings and EW VEVs. 
    \item
        In an $\SO(10)$ theory, the irreps from Eq.~\eqref{eq:16-times-16} relevant for the Yukawa sector decompose under $\SO(10)\to 4_C\,2_L\,2_R$ as
        \begin{align}
            \GIRREP{10} &= \PSIRREP{1}{2}{2} \oplus \PSIRREP{6}{1}{1}, \label{eq:irrep-decomposition-10}\\
            \GIRREP{120} &= \PSIRREP{1}{2}{2} \oplus 
                \PSIRREP{15}{2}{2} \oplus
                \PSIRREP{6}{3}{1} \oplus
                \PSIRREP{6}{1}{3} \oplus
                \PSIRREP{10}{1}{1} \oplus 
                \PSIRREP{\overline{10}}{1}{1}, \label{eq:irrep-decomposition-120}\\
            \GIRREP{126} &= \PSIRREP{15}{2}{2} \oplus
                \PSIRREP{6}{1}{1} \oplus 
                \PSIRREP{10}{3}{1} \oplus
                \PSIRREP{\overline{10}}{1}{3}. \label{eq:irrep-decomposition-126}
        \end{align}
        The $\GIRREP{126}$ is an irrep of complex type, 
        so its PS part $\PSIRREP{15}{2}{2}$ is always complex. If irreps $\GIRREP{10}$ and $\GIRREP{120}$ are taken real, they impose the reality condition on PS pieces $\PSIRREP{1}{2}{2}$ and $\PSIRREP{15}{2}{2}$ contained inside, i.e.~the sign $s'_{1}$ or $s'_{15}$ for each instance of a real PS irrep is determined by the parent $\SO(10)$ irrep in a non-trivial way that needs to be computed (we explicitly carry out such a computation in Section~\ref{subsec:result-SO10-PS}). Although there is a sign choice for every $\SO(10)$ irrep, that choice is inherited by all its PS parts, therefore the relative signs for PS pieces from the same parent are fixed by $\SO(10)$ symmetry. 
\end{itemize}

\subsection{Reality constraint through an explicit $\SO(10)$ computation \label{subsec:result-SO10}}

Let us now consider the minimal Yukawa sector of $\SO(10)$, i.e.~a $\SO(10)$ GUT with the fermion sector consisting of $3\times\mathbf{16}$ and scalars present in the Yukawa sector consisting of $\GIRREP{10}_{\mathbb{R}}\oplus\GIRREP{120}_{\mathbb{R}}\oplus\GIRREP{126}_{\mathbb{C}}$. We shall perform an explicit computation at the $\SO(10)$-level, including the imposition of reality conditions; this is the most direct way to obtain correct mass matrices, with no need to consider Pati-Salam language from Section~\ref{subsec:result-PS}.

We suppress the family and Weyl indices, and write the fermion irreps as
\begin{align}
    \Psi^{X}\sim \GIRREP{16},
\end{align}
where the upper index formally runs through $1$ to $32$, see Section~\ref{subsec:group-theory} and Table~\ref{tab:indices} for notation, but only the first $16$ entries in the basis $\GIRREP{16}\oplus\GIRREP{\overline{16}}$ are switched on (equivalent to using the projection operator $\MAT{P}_-$ defined in Eq.~\eqref{eq:projection-operator} from the Appendix).

The scalar representations are for simplicity simply labeled by their dimension, and they have an antisymmetric index structure; we express them in the complex basis as
\begin{align}
    \GIRREP{10}^{i}, && 
    \GIRREP{120}^{[ijk]}, && 
    \GIRREP{126}^{[ijklm]},
\end{align}
and the $\GIRREP{126}$ also satisfies the self-duality condition 
\begin{align}
    \GIRREP{126}^{i_{1}i_{2}i_{3}i_{4}i_{5}}
    &= 
    -\tfrac{i}{5!}\;
    P^{i_{1}j_{1}}\,
    P^{i_{2}j_{2}}\,
    P^{i_{3}j_{3}}\,
    P^{i_{4}j_{4}}\,
    P^{i_{5}j_{5}}\;
    \epsilon_{j_{1}j_{2}j_{3}j_{4}j_{5}k_{1}k_{2}k_{3}k_{4}k_{5}}\,
    \GIRREP{126}^{k_{1}k_{2}k_{3}k_{4}k_{5}}, \label{eq:126-selfduality}
\end{align}
which can be derived from the more familiar form of this condition expressed in the real basis as
\begin{align}
    \GIRREP{126}_{p_{1}p_{2}p_{3}p_{4}p_{5}}&=
        -\tfrac{i}{5!}\,
        \epsilon_{p_{1}p_{2}p_{3}p_{4}p_{5}q_{1}q_{2}q_{3}q_{4}q_{5}}
        \GIRREP{126}_{q_{1}q_{2}q_{3}q_{4}q_{5}}.
\end{align}
The components $\GIRREP{126}_{p_{1}p_{2}p_{3}p_{4}p_{5}}$ are taken complex despite being in the real basis, see e.g.~\cite{Antusch:2019avd}.
We make the choice of $+$ for the real structure on $\GIRREP{10}$ and $\GIRREP{120}$, leading to reality conditions in Eqs.~\eqref{eq:reality-condition-10-c} and \eqref{eq:reality-condition-120-rc}, respectively.

Given our convention for $\GIRREP{126}$ and $\GIRREP{\overline{126}}$, cf.~Eq.~\eqref{eq:16-times-16} that is consistent e.g.~with~\cite{Slansky:1981yr}, the Yukawa invariant is formed from $\GIRREP{16}\cdot\GIRREP{16}\cdot\GIRREP{\overline{126}}$. To that end, we prepare a form $\GIRREP{\overline{126}}$ with the conjugated entries of $\GIRREP{126}$ as follows:
\begin{align}
    \GIRREP{\overline{126}}^{ijklm} \equiv P^{ii'}\,P^{jj'}\,P^{kk'}\,P^{ll'}\,P^{mm'} \;(\GIRREP{126}^*)_{i'j'k'l'm'},
    \label{eq:126bar}
\end{align}
where both unprimed and primed indices refer to the complex basis.

The tensor entries of irreps can be related to fields with well-defined transformation properties by computing the quantum numbers under diagonal generators\footnote{
    The quantum numbers are well-defined for each $\SO(10)$ tensor entry if one uses fundamental indices in the complex basis and spinor indices in the chiral basis (as constructed in Appendix~\ref{app:spinors-gamma}).
    }
and their transformation properties under $\SO(10)$ generators. In this way, choosing an explicit SM embedding, we can identify the relevant field entries in $\SO(10)$ objects and carry out explicit computations. 

Given the decompositions of Eq.~\eqref{eq:irrep-decomposition-10}--\eqref{eq:irrep-decomposition-126} and the reality of $\GIRREP{10}$ and $\GIRREP{120}$, the list of PS irreps in the scalar sector containing EW doublets is as follows: $\PSIRREP{1}{2}{2}_{\GIRREP{10}_\mathbb{R}}$, $\PSIRREP{1}{2}{2}_{\GIRREP{120}_\mathbb{R}}$, $\PSIRREP{15}{2}{2}_{\GIRREP{120}_\mathbb{R}}$ and $\PSIRREP{15}{2}{2}_{\GIRREP{\overline{126}}_\mathbb{C}}$, where their subscript denotes the $\SO(10)$ origin. Their EW VEVs are then labeled as $\VEVEW{10}{u}{1}$, $\VEVEW{120}{u}{1}$, $\VEVEW{120}{u}{15}$, $\VEVEW{\overline{126}}{u}{15}$ and  $\VEVEW{\overline{126}}{d}{15}$, with $\SO(10)$ origin in the subscript, and the superscript denoting the $u$- or $d$-type and the $\SU(4)$ origin of PS. Notice that we consider the fields and VEVs in $\GIRREP{\overline{126}}$ of Eq.~\eqref{eq:126bar}  rather than $\GIRREP{126}$, with implications on the definition of $u$- and $d$-labels (so that $u$ appears in the up-sector masses, as usual), as well as on the SM-singlet VEV $\VEV{\sigma}$ (so that it appears unconjugated in the Majorana mass). The color coding for EW and seesaw VEVs introduced in Section~\ref{sec:introduction} will help with readability. 

We provide information on which tensor entries carry these VEVs in Appendix~\ref{app:basis}.
They are normalized such that 
\begin{align}
    \langle \GIRREP{10}^{i}\;\GIRREP{10}^{*}{}_{i} \rangle &
        = \big|\VEVEW{10}{u}{1}\big|^{2}, \label{eq:normalization-10}\\
    \frac{1}{3!}\,\langle \GIRREP{120}^{ijk} \GIRREP{120}^{*}{}_{ijk} \rangle&
        = \big|\VEVEW{120}{u}{1}\big|^{2} + \big|\VEVEW{120}{u}{15}\big|^{2}, \label{eq:normalization-120}
        \\
    \frac{1}{5!}\,\langle \GIRREP{\overline{126}}^{ijklm} \GIRREP{\overline{126}}^{*}{}_{ijklm}  \rangle&
        = \frac{1}{2} \big|\VEV{\sigma}\big|^2  
        + \big|\VEVEW{\overline{126}}{u}{15}\big|^{2} 
        + \big|\VEVEW{\overline{126}}{d}{15}\big|^{2}. \label{eq:normalization-126}
\end{align}
This normalization implies that taking the coefficient appearing on the left-hand side of the equation also as a prefactor for writing the kinetic terms, the EW VEVs are canonically normalized. 
As discussed later, setting such a normalization for the EW VEVs is important for a full theory with a fixed scalar sector, in which the doublet mass matrix is predicted. The normalization for $\VEV{\sigma}$ is not canonical and is chosen so that it is in line with a future work in preparation. 

We now perform the computation of the Yukawa-sector terms and fermion mass matrices in $\SO(10)$, where no PS ambiguities related to the real structure from Section~\ref{subsec:result-PS} arise. Suppressing family and Weyl indices, and using the machinery of Section~\ref{subsec:group-theory}, the Yukawa part of the Lagrangian is written as
\begin{align}
\mathcal{L}_Y&= 
    \phantom{+}\tfrac{1}{2}\;\MATY{10}\;\Psi^{X}\,C_{XY}\,(\bm{\Gamma}_{i})^{Y}{}_{Z}\,\Psi^{Z} \,\GIRREP{10}^{i} \nonumber \\
    &\quad 
    +\tfrac{1}{12}\;\MATY{120}\; \Psi^{X}\,C_{XY}\,(\bm{\Gamma}_{i} \bm{\Gamma}_{j} \bm{\Gamma}_{k})^{Y}{}_{Z}\,\Psi^{Z} \,\GIRREP{120}^{ijk} \nonumber\\
    &\quad 
    +\tfrac{1}{160\sqrt{3}} \; \MATY{126} \; \Psi^{X}\,C_{XY}\,(\bm{\Gamma}_{i}\bm{\Gamma}_{j}\bm{\Gamma}_{k}\bm{\Gamma}_{l}\bm{\Gamma}_{m})^{Y}{}_{Z}\,\Psi^{Z} \,\GIRREP{\overline{126}}^{ijklm} + h.c.. \label{eq:Yukawa-Lagrangian-SO10}
\end{align}
Since $\Psi$ is a spinorial representation, we use the $\SO(10)$ gamma matrices $\bm{\Gamma}_{p}$, written explicitly in components as $\Gamma_{p}{}^{X}{}_{Y}$, and the charge conjugation matrix $C_{XY}$, see Appendix~\ref{app:spinors} for their construction. In Eq.~\eqref{eq:Yukawa-Lagrangian-SO10}, we transformed the fundamental index of the gamma matrices from the real into the anti-complex basis via $\bm{\Gamma}_{i} := P_{ip}\,\bm{\Gamma}_{p}$, and a sequence of such objects in parentheses denotes their product as matrices in spinor space.

The Yukawa matrices in Eq.~\eqref{eq:Yukawa-Lagrangian-SO10} are $3\times 3$ complex matrices satisfying  
\begin{align}
    \MAT{Y}_{10}^\top & =\MAT{Y}_{10}, &
    \MAT{Y}_{120}^\top &= -\MAT{Y}_{120}, & 
    \MAT{Y}_{126}^\top &= \MAT{Y}_{126}.
\end{align}
Their (anti-)symmetric properties are imposed by the behavior of the associated $\SO(10)$ invariant under the exchange of the two spinorial fields (which carry family indices), cf.~Eq.~\eqref{eq:16-times-16}. We chose their numeric prefactors in the Lagrangian \textit{a posteriori}, such that the fermion mass matrices come out with 
coefficient $1$ in $\MAT{M}_{U}$ when possible (implying that all Yukawa couplings of canonically normalized fields also appear with $\mathcal{O}(1)$ coefficients). A physical advantage of such a normalizing convention for $\MAT{Y}$ is that the usual perturbativity limit of $|\MAT{Y}|\lesssim\mathcal{O}(1)$ applies.

Having explicitly implemented all objects in Eq.~\eqref{eq:Yukawa-Lagrangian-SO10} and contracting them in \textit{Mathematica}, we obtain the following fermion mass matrices:
\begin{align}
    \MAT{M}_U&=
        \MATY{10}\,\VEVEW{10}{u}{1} + 
        \MATY{120}\,(\VEVEW{120}{u}{1}+\tfrac{1}{\sqrt{3}}\VEVEW{120}{u}{15}) + 
        \MATY{126}\, \VEVEW{\overline{126}}{u}{15}, \label{eq:result-MU}\\
    \MAT{M}_D &=
        \MATY{10}\,\VEVEW{10}{u}{1}^* +
        \MATY{120}\,(\VEVEW{120}{u}{1}^*-\tfrac{1}{\sqrt{3}}\VEVEW{120}{u}{15}^*) +
        \MATY{126}\, \VEVEW{\overline{126}}{d}{15}, \label{eq:result-MD} \\
    \MAT{M}_E &=
        \MATY{10}\,\VEVEW{10}{u}{1}^* +
        \MATY{120}\,(\VEVEW{120}{u}{1}^* +\sqrt{3}\,\VEVEW{120}{u}{15}^*) 
        -3\,\MATY{126}\, \VEVEW{\overline{126}}{d}{15}, \label{eq:result-ME} \\ 
    \MAT{M}_{\nu_D} &=
        \MATY{10}\,\VEVEW{10}{u}{1} + 
        \MATY{120}\,(\VEVEW{120}{u}{1}-\sqrt{3}\,\VEVEW{120}{u}{15}) 
        -3\,\MATY{126}\, \VEVEW{\overline{126}}{u}{15}, \label{eq:result-MND}\\
    \MAT{M}_{\nu_{R}}&= 2\sqrt{3}\;\MATY{126}\,\VEV{\sigma}. \label{eq:result-MNR}
\end{align}

This is the complete $\SO(10)$ result given our conventions, which can be used as a starting point for a fermion fit, see Section~\ref{sec:yukawa-sector} for subsequent steps and results. We conclude here, however, by discussing its structural features: 
\begin{itemize}[itemsep=-0.1cm,leftmargin=\DIMLEFTMARGIN]
    \item There can always be an overall sign ambiguity for any real $\SO(10)$ irrep, and an overall phase ambiguity for any complex irrep. These are not physical, so we can choose the phase of $\GIRREP{126}$, e.g., such that $\VEV{\sigma}$ is real and positive. 
    \item All EW VEVs are in general complex; $\MAT{M}_{U,\nu_D}$ contain those of $u$-type, while $\MAT{M}_{D,E}$ contain those of $d$- or $u^*$-type, the latter being a feature of real $\SO(10)$ irreps. All EW VEVs have been normalized according to Eqs.~\eqref{eq:normalization-10}--\eqref{eq:normalization-126} --- this is their canonical normalization if kinetic terms are chosen with the same coefficients as those on the left-hand sides of the equations. This is an important improvement over the arbitrarily normalized tilde-labeled EW VEVs from Eqs.~\eqref{eq:MU-naive}--\eqref{eq:MNR-naive} in Section~\ref{sec:introduction}; although typically not crucial for the fermion fit itself, this feature becomes relevant in a full model with a fixed scalar sector. The EW doublet mass matrix there is predicted, and its (almost) massless mode determines the admixture of the SM Higgs inside the original doublets. Canonical normalization of both fields and VEVs ensures the latter to be proportional to the admixture weights, and they in addition satisfy the constraint (following~\cite{Antusch:2025fpm} for the precise numerical value of the SM Higgs VEV) 
    \begin{align}
        \sum_{k}|\VEVEW{k}{}{}|^2 &= (175.6\,\mathrm{GeV})^{2},
    \end{align}
    where the index $k$ is taken over all EW VEVs of the theory --- including those in scalar irreps not involved in the Yukawa sector. 
    \item The transition $U\to \nu_D$ and $D\to E$ in fermion sectors exhibits a Clebsch coefficient of $1$ for VEVs in $\PSIRREP{1}{2}{2}$ of PS, and a Clebsch $-3$
    for $\PSIRREP{15}{2}{2}$ of PS, as we know they should from the Pati-Salam result of Section~\ref{subsec:result-PS}.
    \item Since the $\GIRREP{120}$ of $\SO(10)$ contains two PS irreps, the relative factor $\tfrac{1}{\sqrt{3}}$ between $\VEVEW{120}{u}{15}$ and $\VEVEW{120}{u}{1}$ is the Clebsch coefficient under $\SO(10)\to 4_C\,2_L\,2_R$. This information cannot be extracted from a pure Pati-Salam computation; in other words, only an $\SO(10)$ calculation can relate $\MAT{Y}''_{1}$ and $\MAT{Y}''_{15}$ in Eqs.~\eqref{eq:Yukawa-PS-real-U}--\eqref{eq:Yukawa-PS-real-N} to a common origin from the coupling $\MAT{Y}_{120}$.
    \item Matching the results to the ansatz of non-normalized EW VEVs in Eqs.~\eqref{eq:MU-naive}--\eqref{eq:MND-naive} and conjugation relations in Eq.~\eqref{eq:sign-definitions}, we deduce the following: the reality condition on the $\GIRREP{10}$ ($+$ version) imposes $s_{3}=1$, while the reality condition on $\GIRREP{120}$ (again $+$ version) imposes $s_{1}=1$ and $s_{2}=-1$. 
    These results are partly just correct bookkeeping,
    since multiplication of an irrep by $i$ induces opposite signs for pieces from that irrep. The result $s_{1}s_{2}=-1$, however, is a structural constraint from $\SO(10)$, as discussed already at the end of Section~\ref{subsec:result-PS}; it is independent of the $\pm$ choice of real structure on $\GIRREP{120}$ and contains physical information. This is the main revision of previously reported results in Ref.~\cite{Babu:2016bmy} for the Yukawa setup we study. 
\end{itemize}

\subsection{Expressing the reality constraint in Pati-Salam language \label{subsec:result-SO10-PS}}

The transition from the $U$- to the $D$- sector in Eqs.~\eqref{eq:result-MU} and \eqref{eq:result-MD} of our $\SO(10)$ result can be compared to 
the Pati-Salam analogs in Eqs.~\eqref{eq:result-PS-U} and \eqref{eq:result-PS-D}. This reveals the implications of our $\SO(10)$ result in PS language: imposing a real $\GIRREP{10}$ and $\GIRREP{120}$ results in the $s_{1}'=+1$ real version of $\PSIRREP{1}{2}{2}$ in Eq.~\eqref{eq:reality-phi}, while the real structure of $\PSIRREP{15}{2}{2}$ imposed by a real $\GIRREP{120}$ is $s'_{15}=-1$ in Eq.~\eqref{eq:reality-varphi}.

Obtaining this result in Section~\ref{subsec:result-SO10} relied 
on an explicit computer-assisted $\SO(10)$ calculation, and was derived in an indirect way (via computing invariants). As such the underlying reasons for $s'_{15}=-1$ remain opaque. We dedicate this section to a transparent albeit technical derivation of the signs $s'_{1,15}$, showing analytically how the $\SO(10)$ reality conditions manifest in Pati-Salam language. 

To achieve this, we require a way of translating the PS irrep back to its $\SO(10)$ description, with the $\SO(6)\times\SO(4)$ description of Pati-Salam serving as a crucial intermediate step.

First, we look at how translation between the two languages of $\SO(6)$ and $\SU(4)$ (due to the relation $\SO(6)\cong \SU(4)$) works in the case of specific representations of interest, and analogously for the relationship $\SO(4)\cong \SU(2)\times\SU(2)$.
\begin{enumerate}[leftmargin=\DIMLEFTMARGIN ,itemsep=0.1cm, label=(\alph*)]
    \item \label{item:description-so6-6}
        The representation $\GIRREP{6}$ of $\SO(6)$ is the fundamental representation with index structure $\GIRREP{6}_a$, while in $\SU(4)$ it corresponds to the antisymmetric product $\Lambda^{2}\GIRREP{4}$, hence has index structure $\GIRREP{6}^{[AB]}$. Furthermore, it also corresponds to the antisymmetric product  $\Lambda^{2}\GIRREP{\bar{4}}$, and should hence also have a dual description with index structure $\GIRREP{6}_{[AB]}$. The different descriptions of the same representation are related using the representation-theoretic objects introduced in Section~\ref{subsec:group-theory}, i.e.~the intertwiners $\Sigma,\overline{\Sigma}$ introduced in Eq.~\eqref{eq:gamma-SO6}, and the invariant tensors: 
        \begin{align}
            \GIRREP{6}_{a} &
                = \tfrac{1}{2}\,\overline{\Sigma}_{aAB}\,\GIRREP{6}^{AB}
                =\tfrac{1}{2}\,\Sigma_{a}{}^{AB}\,\GIRREP{6}_{AB}, \label{eq:translation-su4-6-a}\\ 
            \GIRREP{6}^{AB} &
                = \Sigma_{a}{}^{AB}\,\GIRREP{6}_{a} 
                = \tfrac{1}{2} \,\epsilon^{ABCD} \,\GIRREP{6}_{CD},
                \label{eq:translation-su4-6-b}\\
            \GIRREP{6}_{AB} &
                = \overline{\Sigma}_{aAB}\,\GIRREP{6}_{a} 
                = \tfrac{1}{2} \,\epsilon_{ABCD} \,\GIRREP{6}^{CD}.
                \label{eq:translation-su4-6-c}
        \end{align}
        The above expressions allow passing between any two of the three possible forms. Note that $\GIRREP{6}$ can in principle be complex; upon complex conjugation, its $\SU(4)$ forms change index height, e.g.~$(\GIRREP{6}^{AB})^*=(\GIRREP{6}^*)_{AB}$, which should not be confused with $\GIRREP{6}_{AB}$.
    \item \label{item:description-so6-15}
        The representation $\GIRREP{15}$ is the adjoint of $\SO(6)\cong \SU(4)$ and also has many alternative descriptions in terms of indices. In $\SO(n)$ groups the adjoint corresponds to an antisymmetric product of the fundamental representation (the generators are labeled as $\MAT{T}_{IJ}$, see Appendix~\ref{app:spinors-Clifford}), hence we have $\GIRREP{15} = \Lambda^{2}\GIRREP{6}$ for $\SO(6)$ and an index structure $\GIRREP{15}_{[ab]}$ (and no additional conditions). Performing the conversion of each $\SO(6)$ index to two antisymmetric $\SU(4)$ indices, cf.~Eq.~\eqref{eq:translation-su4-6-a}, there is an alternative index structure $\GIRREP{15}^{[AB][CD]}$ with the additional antisymmetry $\GIRREP{15}^{ABCD}=-\GIRREP{15}^{CDAB}$. Furthermore, in $\SU(n)$ groups, the adjoint is part of a product of the fundamental and anti-fundamental irrep, hence $\GIRREP{15}\subset\GIRREP{4}\otimes\GIRREP{\bar{4}}$, implying a possible index structure $\GIRREP{15}^{A}{}_{B}$ together with an $\SU(4)$-equivariant tracelessness condition $\GIRREP{15}^{A}{}_{B}\,\delta^{B}{}_{A}=0$. These descriptions are connected via the relations
    \begin{align}
        \GIRREP{15}_{ab}&
            = 
            \tfrac{1}{4}\,\overline{\Sigma}_{aAB} \, 
            \overline{\Sigma}_{bCD} \,\GIRREP{15}^{ABCD}, &
        \GIRREP{15}^{ABCD} &
            = \tfrac{1}{4}\,\Sigma_{a}{}^{AB}\,\Sigma_{b}{}^{CD}\, \GIRREP{15}_{ab},
            \label{eq:translation-su4-15-a}\\
        \GIRREP{15}^{A}{}_{E}& 
            =  \tfrac{1}{2}\, \epsilon_{EBCD}\,\GIRREP{15}^{ABCD}, &
        \GIRREP{15}^{ABCD} &
            = \tfrac{1}{2}\,\epsilon^{EBCD}\,\GIRREP{15}^{A}{}_{E} 
                - \tfrac{1}{2}\,\epsilon^{EACD}\,\GIRREP{15}^{B}{}_{E},
            \label{eq:translation-su4-15-b}
    \end{align}
        together leading to the $\GIRREP{15}_{ab}\leftrightarrow\GIRREP{15}^{A}{}_{B}$ translations
    \begin{align}
            \GIRREP{15}_{ab}& 
                = \tfrac{1}{4}\,\overline{\Sigma}_{aAB}\,
                \left(
                    \GIRREP{15}^{A}{}_{E}\,\Sigma_{b}{}^{EB}
                    -\GIRREP{15}^{B}{}_{E}\,\Sigma_{b}{}^{EA}
                \right),
                \label{eq:translation-su4-15-c}\\
            \GIRREP{15}^{A}{}_{E} &= 
                \tfrac{1}{8} \,\epsilon_{EBCD}\,
                \Sigma_{a}{}^{AB}\,\Sigma_{b}{}^{CD}\,\GIRREP{15}_{ab},
                \label{eq:translation-su4-15-d}
    \end{align}
        where the duality relation of \eqref{eq:relation-spinors-so6-duality} was used to derive Eq.~\eqref{eq:translation-su4-15-c}.
        \par
        We reiterate that all this translation gymnastics is structural --- it arises from representation theory. The key to deriving the starting relations of Eqs.~\eqref{eq:translation-su4-15-a} and \eqref{eq:translation-su4-15-b} is simple: given the index-structure of group-theoretic objects (gamma matrices and invariant tensors), there is a clear way how to express the indices carried by the $\GIRREP{15}$ on the left- with the indices carried by the same object on the right-hand sides of equations. Particular attention has to be given, however, to two aspects elaborated on below: tensor properties and normalization. 
        \par
        First, the expressions must be consistent with the tensor properties (antisymmetry, tracelessness, etc.) of each manifestation of the $\GIRREP{15}$ that was listed earlier. In particular for every equation, assuming the tensor properties of the $\GIRREP{15}$ on the right-hand side, the expression should automatically exhibit the tensor properties of the $\GIRREP{15}$ on the left-hand side. This condition, for example, required an explicit antisymmetrization of indices $A$ and $B$ in the second equation of \eqref{eq:translation-su4-15-b} by writing two terms.  
        \par
        Second, the overall normalization and phase of each manifestation $\GIRREP{15}_{ab}$, $\GIRREP{15}^{ABCD}$ and $\GIRREP{15}^{A}{}_{B}$ is not fixed \textit{a priori}. In lines \eqref{eq:translation-su4-15-a} and \eqref{eq:translation-su4-15-b}, the first (left-side) equation sets the convention for the relative normalization, while the second (right-side) equation must be the inverse of the first one, thus has fixed numerical coefficients. 
        The normalization conditions are related as
    \begin{align}
        \tfrac{1}{2}\,\GIRREP{15}_{ab} \GIRREP{15}^*_{ab} 
        &= \tfrac{1}{2}\,\GIRREP{15}^{ABCD} \GIRREP{15}^*{}_{ABCD} 
        = \GIRREP{15}^{A}{}_{B}\, \GIRREP{15}^{*}{}_{A}{}^{B}.
    \end{align}
    \item \label{item:description-so4-4}
    The irrep $(\GIRREP{2},\GIRREP{2})$ of $\SU(2)_L\times\SU(2)_R$ is equivalent to a $\GIRREP{4}$ of $\SO(4)$, suggesting $\GIRREP{4}^{\alpha\dot{\alpha}}\sim \GIRREP{4}_{\mu}$. The intertwiners between the two descriptions are the off-diagonal blocks $\sigma_{\mu}{}^{\alpha\dot{\alpha}}$ and $\overline{\sigma}_{\mu \dot{\alpha}\alpha}$ of the $\SO(4)$ gamma matrices introduced in  Eq.~\eqref{eq:gamma-SO4}. 
    \par
    Each $\SU(2)$ index can be raised or lowered by an appropriate two-index epsilon tensor, and one could also exchange the order of the indices; among these, we examine further only the description $\GIRREP{4}_{\dot{\alpha}\alpha}$, so that our considerations are symmetric with respect to $\sigma$ and $\overline{\sigma}$.
    \par
    Explicitly, the connection are
    \begin{align}
        \GIRREP{4}^{\alpha\dot{\alpha}} &
            = \sigma_{\mu}{}^{\alpha \dot{\alpha}}\,\GIRREP{4}_{\mu}
            = \epsilon^{\alpha\beta}\,\epsilon^{\dot{\alpha}\dot{\beta}} \,\GIRREP{4}_{\dot{\beta}\beta}, 
            \label{eq:translation-so4-4-a}\\
        \GIRREP{4}_{\dot{\alpha}\alpha} &
            = \overline{\sigma}_{\mu\dot{\alpha}\alpha}\,\GIRREP{4}_{\mu}
            = \epsilon_{\alpha\beta}\,\epsilon_{\dot{\alpha}\dot{\beta}} \,\GIRREP{4}^{\beta\dot{\beta}},
            \label{eq:translation-so4-4-b}\\  
        \GIRREP{4}_{\mu}&
            = \overline{\sigma}_{\mu\dot{\alpha}\alpha}\, \GIRREP{4}^{\alpha\dot{\alpha}} 
            = \sigma_{\mu}{}^{\alpha\dot{\alpha}}\,\GIRREP{4}_{\dot{\alpha}\alpha}, 
            \label{eq:translation-so4-4-c}
    \end{align}
    with normalizations due to Eq.~\eqref{eq:offblock-so4-conjugation} and \eqref{eq:offblock-so4-completeness} related via
    \begin{align}
        \GIRREP{4}_{\mu}\,\GIRREP{4}^{*}_{\mu} &
            = 2 \cdot\GIRREP{4}^{\alpha\dot{\alpha}}\,\GIRREP{4}^*{}_{\alpha\dot{\alpha}}
            = 2 \cdot\GIRREP{4}_{\dot{\alpha}\alpha}\,\GIRREP{4}^{*\dot{\alpha}\alpha}.
    \end{align}
\end{enumerate}

\noindent
Having established the connections between the descriptions of relevant representations, we turn to deriving the reality conditions imposed by $\SO(10)$ in PS language for the irreps containing EW VEVs in $\GIRREP{10}_{\mathbb{R}}$ and $\GIRREP{120}_{\mathbb{R}}$ of $\SO(10)$. The strategy will always be as follows: we know how to impose the reality condition in a $\SO(6)\times\SO(4)\subset\SO(10)$ description, so we should link the PS irrep to that description as well. The relevant $4_C\,2_L\,2_R\subset\SO(10)$ cases are as follows:
\begin{enumerate}[leftmargin=\DIMLEFTMARGIN,itemsep=0.1cm,label=(\roman*)]
    \item \label{item:122-10}
        \textit{The $\PSIRREP{1}{2}{2}$ in $\GIRREP{10}_{\mathbb{R}}$:}\\[3pt]
        The irrep is a $\SU(4)_C$ singlet, so only the $\SU(2)_L\times\SU(2)_R$ part is non-trivial, and can be linked to a $\GIRREP{4}$ of $\SO(4)$ in accordance with earlier item~\ref{item:description-so4-4}. Suppose we label the components of the irrep $\GIRREP{10}_{\mathbb{R}}$ of $\SO(10)$ in the real basis by $\Phi_{p}$. The part $\GIRREP{4}$ of $\SO(4)$ are simply the components $\Phi_{7,8,9,10}$, which we denote as $\Phi_{\mu}$. Note: $\Phi_p$ and $\Phi_\mu$ have same field label, but different index label (going over a different range, see Table~\ref{tab:indices}); we also imagine $\mu=1\ldots 4$, i.e.~$\mu=p-6$. We then have the connection from item~\ref{item:description-so4-4} as
        \begin{align}
            \Phi^{\alpha\dot{\alpha}}&= \sigma_{\mu}{}^{\alpha\dot{\alpha}} \Phi_{\mu}. \label{eq:Phi-express-122-10}
        \end{align}
        \par
        The reality condition $\Phi_{p}=\Phi_{p}^{*}$ for all components of the $\GIRREP{10}_{\mathbb{R}}$ imposes $\Phi_{\mu}^*=\Phi_{\mu}$ (since $\mu$ goes over a subset of index values of $p$). In PS language, the reality condition can then be expressed as
        \begin{align}
            (\Phi^*)_{\alpha\dot{\alpha}} 
                &= \sigma^*{}_{\mu\alpha\dot{\alpha}} \Phi^{*}_{\mu} 
                \label{eq:Phi-derivation-122-10-step1}\\
                &= (\epsilon_{\alpha\beta} \,\epsilon_{\dot{\alpha}\dot{\beta}} \,\sigma_{\mu}{}^{\beta\dot{\beta}})\,\Phi_{\mu}
                \label{eq:Phi-derivation-122-10-step2}\\
                &= +\epsilon_{\alpha\beta} \,\epsilon_{\dot{\alpha}\dot{\beta}}\,\Phi^{\beta\dot{\beta}}.
                \label{eq:Phi-derivation-122-10-step3}
        \end{align}
        We used the relation \eqref{eq:Phi-express-122-10} in Eq.~\eqref{eq:Phi-derivation-122-10-step1}, the conjugation relation \eqref{eq:offblock-so4-conjugation} and the $\SO(10)$ reality condition $\Phi_{\mu}^*=\Phi_{\mu}$ in Eq.~\eqref{eq:Phi-derivation-122-10-step2}, and reassembled $\Phi$ back into the original PS language again using \eqref{eq:Phi-express-122-10} in Eq.~\eqref{eq:Phi-derivation-122-10-step3}. We derived the version of PS reality condition from Eq.~\eqref{eq:reality-phi} with $s'_{1}=+1$, i.e.~taking a real $10$ imposes $s_{3}=+1$ in Eq.~\eqref{eq:sign-definitions}. 
    \item \label{item:122-120} 
        \textit{The $\PSIRREP{1}{2}{2}$ in $\GIRREP{120}_{\mathbb{R}}$:}\\[3pt]
        Suppose we again label irrep $\GIRREP{120}_{\mathbb{R}}$ by $\Phi$.
        The irrep $\GIRREP{120}$ can be expressed as an antisymmetric product $\Lambda^{3} \GIRREP{10}$, hence the index structure $\Phi_{[pqr]}$ and the reality condition $\Phi^{*}_{pqr}=\Phi_{pqr}$. Decomposing the $\GIRREP{10}$ into PS irreps, cf.~\eqref{eq:irrep-decomposition-10}, and requiring a result transforming trivially under $\SU(4)_C$, one can infer that the $\PSIRREP{1}{2}{2}$ within $\Phi$ is the product of PS parts $\Lambda^{3}\PSIRREP{1}{2}{2}$ in $\Lambda^{3}\GIRREP{10}$, hence its components can be written as $\Phi_{[\mu\nu\lambda]}$ with the induced reality condition $\Phi^{*}_{\mu\nu\lambda}=\Phi_{\mu\nu\lambda}$. This can be considered an $\SO(4)$ object, and using the $\SO(4)$ Hodge dual, we can express it as
        \begin{align}
            \Phi_{\mu} &= \tfrac{1}{3!}\,\epsilon_{\mu\nu\lambda\kappa}\,\Phi_{\nu\lambda\kappa},
        \end{align}
        with the associated reality condition $\Phi^*_{\mu}=\Phi_{\mu}$. The description is thus exactly the same as the $\PSIRREP{1}{2}{2}$ in $\GIRREP{10}_{\mathbb{R}}$, see item~\ref{item:122-10}, leading to the same result $s'_{1}=+1$ in the reality condition of Eq.~\eqref{eq:reality-phi}. This implies a real $\GIRREP{120}$ imposes $s_{1}=+1$ in Eq.~\eqref{eq:sign-definitions}. 
    \item \textit{The $\PSIRREP{15}{2}{2}$ in $\GIRREP{120}_{\mathbb{R}}$:}\\[3pt]
        We again label the components of $\GIRREP{120}_{\mathbb{R}}$ by $\Phi_{[pqr]}$, and the $\SO(10)$ reality condition is $\Phi^*_{pqr}=\Phi_{pqr}$. Given the $\SO(10)\to 4_C\,2_L\,2_R$ irrep decomposition in Eq.~\eqref{eq:irrep-decomposition-10}, the $\PSIRREP{15}{2}{2}$ can be found in the $(\Lambda^{2} \PSIRREP{6}{1}{1}) \otimes \PSIRREP{1}{2}{2}$-part of $\Lambda^{3}\GIRREP{10}$, implying the relevant components to be exactly those of $\Phi_{[ab]\mu}$ (with no additional conditions imposed). 
        This is consistent with the description of the $\GIRREP{15}$ of $\SU(4)$ from item~\ref{item:description-so6-15} and the description of the $(\GIRREP{2},\GIRREP{2})$ of $\SU(2)_L\times\SU(2)_R$ from item~\ref{item:description-so4-4}. 
        \par
        Combining the translation in Eq.~\eqref{eq:translation-su4-15-d} of the $\SU(4)_C$ part, and the translation in Eq.~\eqref{eq:translation-so4-4-a} of the $\SU(2)_L\times\SU(2)_R$ part, we can write 
        \begin{align}
            \Phi^{A}{}_{E}{}^{\alpha\dot{\alpha}} &= 
            \tfrac{1}{8}\,\epsilon_{EBCD}\,\Sigma_{a}{}^{AB}\,
            \Sigma_{b}{}^{CD}\,\sigma_{\mu}{}^{\alpha\dot{\alpha}}\,\Phi_{ab\mu}.
            \label{eq:Phi-express-1522-120}
        \end{align}
        Applying complex conjugation, we get
        \begin{align}
            (\Phi^{*})_{A}{}^{E}{}_{\alpha\dot{\alpha}} &
                = \tfrac{1}{8}\,\epsilon^{EBCD}\,\Sigma^{*}{}_{aAB}\,\Sigma^*{}_{bCD}
                \,\sigma^{*}{}_{\mu\alpha\dot{\alpha}}\,\Phi^{*}{}_{ab\mu} 
                \label{eq:Phi-derivation-1522-120-step1}
                \\
                &= \tfrac{1}{8}\,\epsilon^{EBCD}\,\overline{\Sigma}_{aAB}\,\overline{\Sigma}_{bCD}\,\epsilon_{\alpha\beta}\,\epsilon_{\dot{\alpha}\dot{\beta}}\,\sigma_{\mu}{}^{\beta\dot{\beta}}\,\Phi_{ab\mu} 
                \label{eq:Phi-derivation-1522-120-step2} \\
                &= \tfrac{1}{8}\,\epsilon_{ABCD}\,\Sigma_{a}{}^{CD}\,\Sigma_{b}{}^{EB}\,
                \epsilon_{\alpha\beta}\,\epsilon_{\dot{\alpha}\dot{\beta}}\,\sigma_{\mu}{}^{\beta\dot{\beta}}\,\Phi_{ab\mu}
                \label{eq:Phi-derivation-1522-120-step3} \\
                &=  -\tfrac{1}{8}\,\epsilon_{ABCD}\,\Sigma_{b}{}^{EB}\,\Sigma_{a}{}^{CD}\,
                \epsilon_{\alpha\beta}\,\epsilon_{\dot{\alpha}\dot{\beta}}\,\sigma_{\mu}{}^{\beta\dot{\beta}}\,\Phi_{ba\mu}
                \label{eq:Phi-derivation-1522-120-step4} \\
                &=  -\epsilon_{\alpha\beta}\,\epsilon_{\dot{\alpha}\dot{\beta}}\,\Phi^{E}{}_{A}{}^{\beta\dot{\beta}}.
                \label{eq:Phi-derivation-1522-120-step5}
        \end{align}
        The justification of the performed steps is as follows. To obtain \eqref{eq:Phi-derivation-1522-120-step2}, we applied twice the conjugation relation for $\Sigma$ from Eq.~\eqref{eq:offblock-so6-conjugation}, the conjugation relation for $\sigma$ from Eq.~\eqref{eq:offblock-so4-conjugation}, and the reality condition $\Phi^*_{ab\mu}=\Phi_{ab\mu}$ (following trivially from the $\SO(10)$ reality condition $\Phi^*_{pqr}=\Phi_{pqr}$). In the next step \eqref{eq:Phi-derivation-1522-120-step3}, we apply the $\Sigma \leftrightarrow \overline{\Sigma}$ duality relation of Eq.~\eqref{eq:relation-spinors-so6-duality} in two different ways: we combine $\epsilon$ and $\overline{\Sigma}_b$ into $\Sigma_b$, while $\overline{\Sigma}_a$ is expressed with $\Sigma_a$ and $\epsilon$.
        We then apply the antisymmetry $\Phi_{ab\mu}=-\Phi_{ba\mu}$ in Eq.~\eqref{eq:Phi-derivation-1522-120-step4}; this enables us to reassemble $\Phi_{ba\mu}$ back into PS language via \eqref{eq:Phi-express-1522-120} to obtain the final result of Eq.~\eqref{eq:Phi-derivation-1522-120-step5}.
        \par
        Comparing Eq.~\eqref{eq:Phi-derivation-1522-120-step5} with the possible PS reality conditions in Eq.~\eqref{eq:reality-varphi}, we see that we obtained $s'_{15}=-1$, implying that taking $\GIRREP{120}$ real imposes $s_{2}=-1$ in Eq.~\eqref{eq:sign-definitions}. This result represents the main correction to the previous literature. The appearance of the minus sign in step~\eqref{eq:Phi-derivation-1522-120-step4} can be ultimately traced to the $a\leftrightarrow b$ asymmetric form taken by the translation in Eq.~\eqref{eq:Phi-express-1522-120}, in which the $\Sigma$-object carrying the free upper $\SU(4)$ index ($A$) connects to a particular $\SO(6)$ index (first index $a$) of $\Phi$.\footnote{
            Although the form of Eq.~\eqref{eq:Phi-express-1522-120} is conventional up to a sign, its application in both disassembling $\Phi^{A}{}_{E}{}^{\alpha\dot{\alpha}}$ in \eqref{eq:Phi-derivation-1522-120-step1} and reassembling in \eqref{eq:Phi-derivation-1522-120-step5} ensures the sign in the final result is convention-independent.
        }
\end{enumerate}
Using the formalism and methods demonstrated in this section, analogous considerations can be used to analytically determine the reality conditions for other PS parts of a real $\SO(10)$ irrep if such a need arises. Alternatively, one can also use a general and novel algorithm presented in Appendix~\ref{app:rtransport}, in which explicit use of the embedding map of PS-irreps into $\SO(10)$-irreps is avoided.

\section{Minimal Yukawa Sector of $\SO(10)$ \label{sec:yukawa-sector}}

\subsection{Parametrization of the Yukawa expressions \label{subsec:parametrizations}}

\subsubsection{Setup \label{subsec:setup}}

As discussed in the Introduction, the minimal Yukawa sector of $\SO(10)$ unification comprises Higgs fields in the real $\GIRREP{10}$ and $\GIRREP{120}$ representations, together with a $\GIRREP{126}$ representation. After imposing the appropriate reality conditions, the resulting fermion mass matrices are given in Eqs.~\eqref{eq:result-MU}–\eqref{eq:result-MNR}. For convenience, we now express these matrices in a more compact and transparent form by introducing the following definitions:
\begin{align}
    \VEVEW{10}{u}{1}&
        \equiv |\VEVEW{10}{}{}| e^{i\varphi}, &
    \VEVEW{\overline{126}}{u}{15}&
        \equiv |\VEVEW{126}{u}{}| e^{i\phi_u}, &
    \VEVEW{\overline{126}}{d}{15}&
        \equiv |\VEVEW{126}{d}{}| e^{i\phi_d}, &
    \VEVEW{120}{u}{1} &=: \VEVEW{120}{}{1}, & 
    \VEVEW{120}{u}{15} &=: \VEVEW{120}{}{15},
\end{align}
and
\begin{align}
    \mathcal{M}_1 &
        := |\VEVEW{10}{}{}|\MATY{10}, &
    \mathcal{M}_2 &
        := |\VEVEW{126}{u}{}|\MATY{126}, &
    \mathcal{M}_3 &
        := \left(\VEVEW{120}{}{1}+\tfrac{1}{\sqrt{3}}\VEVEW{120}{}{15}\right)\MATY{120}.
\end{align}
Using the above definitions, Eqs.~\eqref{eq:result-MU}--\eqref{eq:result-MNR} become 
\begin{align}
    \MAT{M}_{U} &
        = e^{i\varphi} \mathcal{M}_1 + e^{i\phi_u} \mathcal{M}_2 + \mathcal{M}_3, \label{eq:MUpre} \\
    \MAT{M}_{D} &
        = e^{-i\varphi} \mathcal{M}_1+p e^{i\phi_d} \mathcal{M}_2+ q \mathcal{M}_3 , \\
    \MAT{M}_{E} &
        = e^{-i\varphi} \mathcal{M}_1-3 p e^{i\phi_d} \mathcal{M}_2+ r \mathcal{M}_3 ,  \\
    \MAT{M}_{\nu_D} &
        = e^{i\varphi} \mathcal{M}_1-3  e^{i\phi_u} \mathcal{M}_2+ \frac{3q^*-r^*}{q^*+r^*} \mathcal{M}_3 \label{eq:MDnupre}, \\
    \MAT{M}_{\nu_{R}} &
        = s \mathcal{M}_2, \label{eq:MRpre}  
\end{align}
where we defined four dimensionless VEV ratios $(p,q,r,s)$ as
\begin{align}
    p&
        :=\frac{|\VEVEW{126}{d}{}|}{|\VEVEW{126}{u}{}|}, 
        & 
    s&
        :=\frac{2\sqrt{3} \VEV{\sigma}}{|\VEVEW{126}{u}{}|},
        &
    q&
        :=\frac{\VEVEW{120}{}{1}^*-\tfrac{1}{\sqrt{3}}\VEVEW{120}{}{15}^*}{\VEVEW{120}{}{1}+\tfrac{1}{\sqrt{3}}\VEVEW{120}{}{15}}, 
        & 
    r&
        := \frac{\VEVEW{120}{}{1}^*+\sqrt{3}\VEVEW{120}{}{15}^*}{\VEVEW{120}{}{1}+\tfrac{1}{\sqrt{3}}\VEVEW{120}{}{15}}. \label{eq:ratios}
\end{align}
\noindent
Moreover, assuming type-I seesaw~\cite{Minkowski:1977sc,Yanagida:1979as,Glashow:1979nm,Gell-Mann:1979vob,Mohapatra:1979ia,Schechter:1980gr,Schechter:1981cv} dominance, the mass matrix of the light neutrinos is given by 
\begin{align}
    \MAT{M}_N &= - \MAT{M}^\top_{\nu_D} \MAT{M}^{-1}_{\nu_R} \MAT{M}_{\nu_D}. \label{eq:mnu}
\end{align}
Note that Eqs.~\eqref{eq:MUpre}-\eqref{eq:MDnupre} and Eq.~\eqref{eq:mnu} are written in the right-left (RL) convention.

\subsubsection{Parameter counting \label{subsec:parameter-count}}

Let us now count the number of independent input parameters present in Eqs.~\eqref{eq:MUpre}-\eqref{eq:MRpre}. 

The matrices $\mathcal{M}_{1,2}$ are complex and symmetric matrices, while $\mathcal{M}_{3}$ is complex and antisymmetric. Therefore, $\mathcal{M}_{1,2}$ each contain 6 magnitudes and 6 phases, whereas $\mathcal{M}_{3}$ contains 3 magnitudes and 3 phases. We in addition have the following parameters of VEV ratios: $p$ is real and positive, whereas $s$, $q$, and $r$ are complex and independent. Moreover, the above set of equations contains 3 phases $\phi_{u,d}$ and $\varphi$. In total, there are 19 magnitudes and 21 phases. 

Considering only physical observables in the fermion mass fit, however, reveals that Eqs.~\eqref{eq:MUpre}--\eqref{eq:MRpre} contain redundancies. These should be eliminated for ease of numerical computation. The redundancies consist of the following:
\begin{itemize}[itemsep=0cm,leftmargin=\DIMLEFTMARGIN]
\item 
    There is a redundancy of $\mathrm{U}(3)$ rotations in family space, which impacts the matrices $\mathcal{M}_{i}$. This can be used to bring, for example, the symmetric complex matrix $e^{i\phi_{u}}\mathcal{M}_2$ into a real, positive, and diagonal form $\mathcal{M}_2^\mathrm{diag}$, eliminating $6$ phases and $3$ magnitudes. The matrix $\mathcal{M}_1$ remains a general symmetric complex matrix, while $\mathcal{M}_3$ remains complex and antisymmetric. The factor $e^{i\phi_{u}}$ was picked up by the family rotation, so it is eliminated from the $U$- and $\nu_R$ sectors, but reappears in the $\nu_{R}$ sector as an overall factor $e^{-i\phi_u}$, while the $D$- and $E$-sectors depend only on the difference of phases $\Delta=\phi_d-\phi_u$.
\item   
    Without loss of generality, one of the EW VEVs can be chosen to be real, since one can redefine the overall phase of the SM Higgs doublet; we use this freedom to choose 
    $\VEVEW{10}{u}{1}$ real and positive, i.e.~we can eliminate one phase by setting $\varphi= 0$.
\item 
    After the previous points, $\phi_{u}$ and $s$ explicitly appear only in the $\nu_R$ sector as a product $e^{-i\phi_{u}}s$. Furthermore, $U(1)_{B-L}$ gauge freedom can be utilized to redefine the phase of the $(B-L)$-breaking VEV $\VEV{\sigma}$, and thereby the phase of $s$, while leaving EW VEV ratios $q$ and $r$ unchanged.
    We can either use the $B-L$ transformation to make $\VEV{\sigma}$ and $s$ real and positive, while leaving behind an overall phase $e^{-i\phi_{u}}$ in $\MAT{M}_{\nu_{R}}$ with no physical impact (convenient for fits in complete models), or fully remove the redundancy in fermion mass matrices by making $e^{-i\phi_{u}}s$ positive and real (for fits of Yukawa sector only, i.e.~this paper, where $\VEV{\sigma}$ or $s$ do not appear elsewhere).
\end{itemize}

Therefore, in the chosen basis of redundancies, we get the final parametrization
\begin{align}
&\MAT{M}_{U}= \mathcal{M}_1 + \mathcal{M}_2^\mathrm{diag} + \mathcal{M}_3, \label{eq:MU} \\
&\MAT{M}_{D}= \mathcal{M}_1+\hat p  \mathcal{M}_2^\mathrm{diag}+ q \mathcal{M}_3 , \\
&\MAT{M}_{E}= \mathcal{M}_1-3 \hat p  \mathcal{M}_2^\mathrm{diag}+ r \mathcal{M}_3 ,  \\
&\MAT{M}_{\nu_D}= \mathcal{M}_1-3  \mathcal{M}_2^\mathrm{diag}+ \frac{3q^*-r^*}{q^*+r^*} \mathcal{M}_3 \label{eq:MDnu}, \\
&\MAT{M}_{\nu_{R}} = s \mathcal{M}_2^\mathrm{diag}, \label{eq:MR}  
\end{align}
where we defined $\hat p= pe^{i\Delta\phi}$.

The independent parameters impacting observables are thus the following:
\begin{align}
    \mathbb{R}^{+}:&\qquad 
        (\mathcal{M}_2^\mathrm{diag})_{ii}, \ s; \label{eq:independent-inputs-R}\\
    \mathbb{C}:&\qquad 
        (\mathcal{M}_1)_{(ij)},\ 
        (\mathcal{M}_3)_{[ij]},\ 
        \hat{p},\  
        q,\ 
        r. \label{eq:independent-inputs-C}
\end{align}
This yields a final tally of \textbf{16 magnitudes} and \textbf{12 phases}, which amounts to one extra magnitude compared to the previous analysis of this Yukawa sector in Ref.~\cite{Babu:2016bmy}. In particular, the relative minus sign with which $\VEVEW{120}{}{15}^*$ appears in $q$ promotes that parameter from a pure phase to also have a magnitude.

\vspace{5pt}
\textbf{Comparison with the previous study}-- Before performing a numerical fit, we explicitly clarify why a new fit is required. In Eq.~\eqref{eq:ratios}, the parameters are defined in terms of ratios of VEVs. Due to the relative minus sign we have derived above, the parameter $q$, involves a ratio of two independent combinations of $\VEVEW{120}{}{1}$ and $\VEVEW{120}{}{15}$. As a result, $q=|q|e^{i\phi_q}$ is, in general, a complex parameter with nontrivial magnitude. In contrast, if this relative minus sign is not considered~\cite{Babu:2016bmy}, the same combination of VEVs appears in both the numerator and denominator up to complex conjugation, reducing the ratio to a pure phase, $q\to e^{i\phi_q}$ with $|q|=1$.

We now directly compare the fermion mass matrix expressions of our result with those of Ref.~\cite{Babu:2016bmy} in Table~\ref{tab:compare}.
In the special limit of $q\to e^{i\phi_{q}}$,
all mass matrices match except for the Dirac-neutrino one. In particular, the $\mathcal{M}_{3}$ coefficient of $\MAT{M}_{\nu_{D}}$ does not match in the limit: 
\begin{align}
&\frac{3q^*-r^*}{q^*+r^*}\to \frac{3e^{-i\phi_{q}}-r^*}{e^{-i\phi_{q}}+r^*} \neq r^* e^{i\phi_{q}}. 
\end{align}
This upsets the fit to the neutrino observables and necessitates a new combined fit.

\begin{table}[ht]
\centering
\begin{tabular}{lll}
\toprule
matrix & our expression & expression from Ref.~\cite{Babu:2016bmy} \\ \midrule
$\mathbf{M}_{U}$ & $\mathcal{M}_1 + \mathcal{M}_2^\mathrm{diag} + \mathcal{M}_3$ & $\mathcal{M}_1 + \mathcal{M}_2^\mathrm{diag} + \mathcal{M}_3$ \\ [1ex]
$\mathbf{M}_{D}$ & $\mathcal{M}_1+\hat p  \mathcal{M}_2^\mathrm{diag}+ q \mathcal{M}_3$ & $\mathcal{M}_1+\hat p  \mathcal{M}_2^\mathrm{diag}+ e^{i\phi_{q}} \mathcal{M}_3$ \\ [1ex]
$\mathbf{M}_{E}$ & $\mathcal{M}_1-3 \hat p  \mathcal{M}_2^\mathrm{diag}+ r \mathcal{M}_3$ & $\mathcal{M}_1-3 \hat p \mathcal{M}_2^\mathrm{diag}+ r \mathcal{M}_3$ \\ [1ex]
$\mathbf{M}_{\nu_D}$ & $\mathcal{M}_1-3  \mathcal{M}_2^\mathrm{diag}+ \frac{3q^*-r^*}{q^*+r^*} \mathcal{M}_3$ & $\mathcal{M}_1-3  \mathcal{M}_2^\mathrm{diag}+ r^* e^{i\phi_{q}} \mathcal{M}_3$ \\ [1ex]
$\mathbf{M}_{\nu_{R}}$ & $s \mathcal{M}_2^\mathrm{diag}$ & $s \mathcal{M}_2^\mathrm{diag}$ \\ \bottomrule
\end{tabular}
\caption{Comparison of fermion mass matrices in our result with those in Ref.~\cite{Babu:2016bmy}.  \label{tab:compare}}
\end{table}

\subsection{Fit to the SM Fermion Sector \label{subsec:fit}}

\subsubsection{Numerical Procedure \label{subsec:numerical-procedure}}
The fitting procedure to the SM fermion masses and mixings is as follows:
\begin{itemize}[itemsep=0cm,leftmargin=\DIMLEFTMARGIN]
\item For the charged fermions and neutrino  masses, we start with the GUT scale mass relations given in Eqs.~\eqref{eq:MU}--\eqref{eq:MR} and Eq.~\eqref{eq:mnu}. This set of equations contains the 28 free parameters counted in Section~\ref{subsec:parameter-count}, namely those listed in Eqs.~\eqref{eq:independent-inputs-R} and \eqref{eq:independent-inputs-C}.

\item We perform the renormalization group evolution (RGE) of the Yukawa couplings defined in Eqs.~\eqref{eq:MU}--\eqref{eq:MDnu}, together with the right-handed neutrino mass matrix of Eq.~\eqref{eq:MR}, by solving the full set of SM+Type-I seesaw RGEs between $M_{\rm GUT}$ and the electroweak scale $M_Z=91.1876\,\mathrm{GeV}$. For the analysis of this paper, we set a fixed value $M_{\rm GUT}=10^{16}\,\mathrm{GeV}$ for the GUT scale. The heavy right-handed neutrinos are successively integrated out at their respective mass thresholds during the running. The RG evolution is implemented using the public package \texttt{REAP}~\cite{Antusch:2005gp}. Since the model predicts an intermediate symmetry-breaking scale $M_{\mathrm{int}} \sim 10^{15}\,\mathrm{GeV}$ (see below), which lies close to the GUT scale, it is sufficient for our analysis to employ only the SM+Type-I seesaw RGEs below $M_{\rm GUT}$, i.e.~we do not employ a model-dependent EFT between $M_{\mathrm{int}}$ and $M_{\rm GUT}$.

\item At scale $M_{Z}$ we compute the $\chi^2$ function for observables, defined as
\begin{align}
\chi^2 = \sum_k \left( \frac{T_k - O_k}{E_k} \right)^2 ,
\end{align}
where $T_k$, $O_k$, and $E_k$ denote the theoretical prediction, the experimentally measured central value, and the corresponding $1$-$\sigma$ experimental uncertainty of the $k$-th observable, respectively. The sum over $k$ includes the masses of three up-type quarks, three down-type quarks, and three charged leptons; three CKM mixing angles and the CKM Dirac CP phase; two neutrino mass-squared differences, and three PMNS mixing angles. Since the Dirac CP phase in the PMNS matrix has not yet been measured, it is not included in the fit. The low-energy experimental inputs in the charged- and neutral-fermion sectors are listed in Table~\ref{tab:fit}, and are taken from Refs.~\cite{Antusch:2025fpm} and~\cite{NUFIT,Esteban:2020cvm}, respectively. For numerical stability of the fitting procedure, and due to the theoretical uncertainties associated to a fixed loop-order calculation, we enlarge the relative error to $1\,\%$ for observables which have been measured more precisely than that.  For illustrative purposes, this work is restricted to the case of normal neutrino mass ordering. 

\item The above steps compute $\chi^{2}$ as a function of input parameters. We perform a fit to the fermion masses and mixing parameters by minimizing $\chi^{2}$ with a differential evolution (DE) algorithm. As specified earlier, the PMNS phase $\delta_\mathrm{PMNS}$ (denoted sometimes $\delta$ for brevity) is not considered in the $\chi^{2}$. By adding a soft penalty in the form of a quadratic term $\chi^2_{\delta}\propto (\delta-\delta_{0})^{2}$, one can minimize $\chi^{2}+\chi^{2}_{\delta}$ to force a best fit for any target value $\delta_{0}$.
We compute best fits for target PMNS phases in intervals of $10^\circ$ (sometimes reducing steps to $5^\circ$ where warranted by rapid changes in $\chi^2$). To speed up this systematic scan, we first perform a minimization of $\chi^{2}$, then use the obtained fit to find a minimum to $\chi^{2}+\chi^{2}_{\delta}$ for a nearby $\delta_{0}$, using that fit in turn to help with minimizing to a new neighboring target of $\delta_{0}$, and so on. 
\par
We choose two satisfactory fits as benchmark (BM) points for further analysis --- denoted by BM I and BM II, see Table~\ref{tab:fit}, corresponding to $\delta_0=0^\circ$ (no CP violation) and $\delta_0=270^\circ$ (maximal CP violation), respectively. For these two BM solutions, we perform a Markov Chain Monte Carlo (MCMC) analysis and explore the parameter space in their vicinity --- the sets of obtained points are referred to as MCMC I and MCMC II, respectively. Note that the MCMC is not restricted to a hypersurface of a fixed PMNS phase, i.e.~we run MCMC with no $\chi^{2}_{\delta}$ penalty term. For each MCMC, we compute with $10$ parallel chains for a total of $7\cdot 10^5$ points (after excluding the burn-in stage of each chain). 
\par
Since the parameter space is $28$-dimensional and the number of observables included in the $\chi^{2}$ is $18$, cf.~Table~\ref{tab:fit}, the system is underdetermined and one generically expects the space of minima to be a $10$-dimensional hypersurface. The MCMC then explores the $28$-dimensional vicinity of this hypersurface. Given the dimensionalities involved, the number of collected points is not sufficient to fully explore the entire space of good $\chi^{2}$ values, as indicated by the somewhat localized nature of each MCMC in the PMNS observable. We partially compensate for this limitation by running two MCMCs initiated at different starting points and afterwards comparing results, while acknowledging that a comprehensive global analysis is beyond the scope of this paper. 
\end{itemize}

\begin{figure}[b!]
\centering
\includegraphics[width=0.48\textwidth]{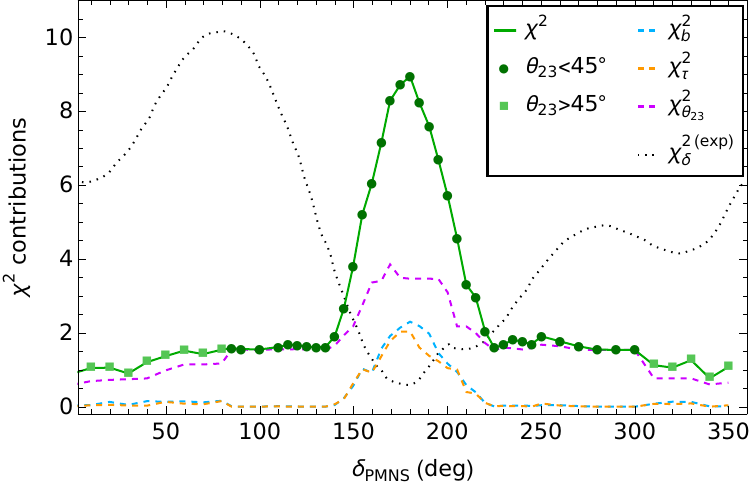}
\includegraphics[width=0.48\textwidth]{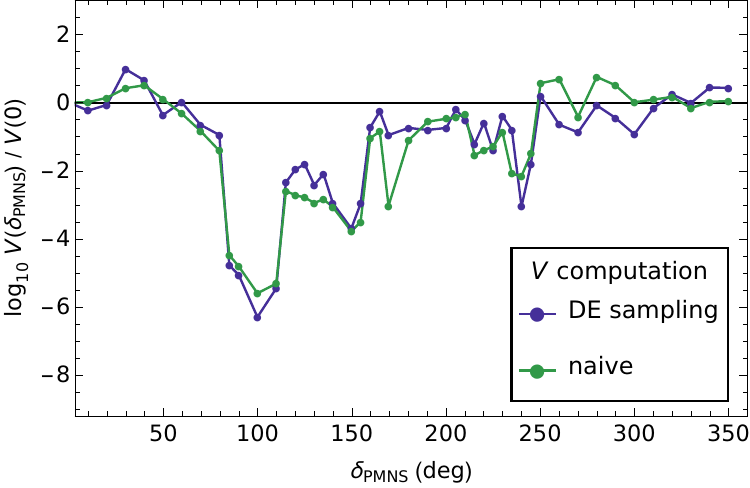}
\caption{Left: minimum $\chi^{2}$ for different target values of $\delta_\mathrm{PMNS}$ (solid line), and the dominant contributions to it from observables $y_{b}$, $y_{\tau}$ and $\sin^{2}\theta_{23}$ (dashed lines). The points on the solid curve denote the actual data, with circular/square markers denoting the $\theta_{23}$ octant in which the fit settled. For comparison we show also the inferred $\chi^{2}_{\delta}$ from the NuFIT 6.0~\cite{NUFIT} global fit of experimental results, which is not included in the total $\chi^{2}$ of fits. Right: the relative volume associated to the hypersurfaces on which the best-fit points lie; computed in two different ways, see main text.
\label{fig:chi2-scan-delta}
}
\end{figure}

\subsubsection{Fit Results}
Since the CP-violating phase has not yet been measured experimentally, in our numerical studies, we pay special attention to this observable and examine whether the model exhibits any preference for a particular range of values.
Following the procedure outlined above, see Section~\ref{subsec:numerical-procedure}, we obtain best $\chi^{2}$ values for fixed targets of the PMNS phase $\delta$.
The results are presented in the left panel of Figure~\ref{fig:chi2-scan-delta}. The results show that a good fit can be obtained for any target PMNS phase, consistent with prior investigations using the uncorrected Yukawa expressions~\cite{Saad:2022mzu,Babu:2024ahk}; most of the phase region admits $\chi^{2}<2$, while $\delta_\mathrm{PMNS}\in\INT{140^\circ}{220^\circ}$ still admit $\chi^{2}\lesssim 9$, with largest contributions to the $\chi^2$ coming from observables $y_b$, $y_\tau$ and $\sin^{2}\theta_{23}$. Note also that the NuFIT 6.0~\cite{NUFIT} result gives $\chi^{2}_{\sin^2\theta_{23}}$ as bimodal (the only such observable), and we denote for each best-fit data point which minimum it lands in by its marker-type (light square or dark circle).

Further information on how the PMNS phase predictions are distributed across the parameter space can be seen from the relative volumes shown in the right panel of Figure~\ref{fig:chi2-scan-delta}. The quantity plotted is $\log_{10}\frac{V(\delta)}{V(0)}$, where $V(\delta)$ is the volume associated to the hypersurface with fixed PMNS phase $\delta$ (we fix it by the soft penalty term $\chi^{2}_{\delta}$). This quantity is hard to compute reliably, and so we compare two different methods. The more proper and robust of the two methods looks for points with $\Delta(\chi^{2}+\chi^{2}_{\delta})<2$ using cycles of DE sampling (with settings to attempt to expand the volume and thus maximize exploration), with the volume then estimated from the covariance matrix $C$ of the points ($V\propto \sqrt{\det C}$), and observing how the relative volume of a moving window of points stabilizes as more samples are gathered. The other ``naive'' method assumes that the relative sensitivity in each input parameter is comparable for all points; the volume is thus modeled as a box around the found point, which simply scales with the input parameters $x_{i}$, i.e.~$V(\delta)\propto|\prod_{i}x_{i}|$. The relatively good agreement of both calculations suggests that a large part of the volume effect can be attributed simply to the magnitude of the inputs, rather than some systematic differences in the sensitivity to the inputs or a complicated shape of the space. The volume results indicate a lower-volume region around $\delta_\mathrm{PMNS}\approx 100^\circ$, implying that the MCMC sampling will disfavor this region (as it will somewhat disfavor the region with higher $\chi^{2}$ around $\delta_\mathrm{PMNS}\approx 180^\circ$).

\begin{figure}[htb]
\centering
\includegraphics[width=0.5\textwidth]{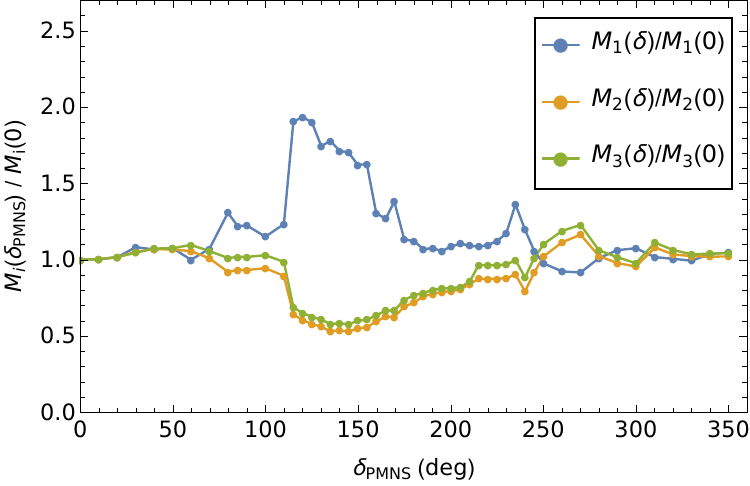}
\caption{An interesting correlation  $M_i (\delta_\mathrm{PMNS})$ between the heavy right-handed neutrino masses (normalized to $M_{i}(0)$) and the PMNS phase. 
\label{fig:correlation-M}
}
\end{figure}

The final result comparing best-fits at different $\delta_\mathrm{PMNS}$
is the set of correlations in heavy right-handed neutrino masses $M_{i}$ shown in Figure~\ref{fig:correlation-M}. The results show the predicted masses are surprisingly stable over different $\delta_\mathrm{PMNS}$, with a somewhat larger deviation of factor $\lesssim 2$ occurring in the region $\delta_\mathrm{PMNS}\in(120^\circ,160^\circ)$, whereby the upward deviation for $M_{1}$ and downward deviation for $M_{2,3}$ 
are mirrored. 

We now transition to presenting two benchmark points BM I and BM II of the fitted model, chosen at $\delta\approx 0^\circ$ and $\delta\approx 270^\circ$ as motivated in Section~\ref{subsec:numerical-procedure}. BM I has the following inputs:
\begingroup
\allowdisplaybreaks
\begin{align}
&
\left(\hat p, q, r, s\right) =
\left(
4.94824\times 10^{-3}\,e^{i\,1.19474},\;
1.71169\,e^{i\,1.78105},\;
1.69701\,e^{-i\,1.56906},\;
1.15577\times 10^{13}
\right),  \label{eq:fit01a} 
\\
&\mathcal{M}_2^\mathrm{diag}= \left(
\begin{array}{ccc}
4.25817\times 10^{-9} & 0 & 0 \\
0 & 0.295873 & 0 \\
0 & 0 & 85.9099 \\
\end{array}
\right) \;\rm{GeV},
\\
&
\mathcal{M}_1=
\begin{pmatrix}
0.000461675\, e^{i\,2.33318} 
&
0.00411085\, e^{i\,2.50199} 
&
0.0051718\, e^{i\,2.75482}
\\[6pt]
0.00411085\, e^{i\,2.50199} 
&
0.0592637\, e^{-i\,1.94638} 
&
0.544621\, e^{i\,2.76814}
\\[6pt]
0.0051718\, e^{i\,2.75482} 
&
0.544621\, e^{i\,2.76814} 
&
0.0672658\, e^{-i\,2.04878}
\end{pmatrix}
\;\rm{GeV},
\\
&
\mathcal{M}_3=\begin{pmatrix}
0
&
0.000195923\, e^{-i\,1.95839}
&
0.000309166\, e^{-i\,1.85675}
\\[6pt]
0.000195923\, e^{i\,1.1832}
&
0
&
0.320083\, e^{i\,0.963284}
\\[6pt]
0.000309166\, e^{i\,1.28484}
&
0.320083\, e^{-i\,2.17831}
&
0
\end{pmatrix} \;\rm{GeV}. \label{eq:fit01b}
\end{align}
\endgroup
\noindent
The other benchmark point BM II has inputs
\begingroup
\allowdisplaybreaks
\begin{align}
&
\left(\hat p, q, r, s\right) =
\left(
3.77898\times 10^{-3}\,e^{i\,1.38218},\;
1.73504\,e^{i\,1.72877},\;
1.84027\,e^{-i\,1.57242},\;
1.42482\times 10^{13}
\right),  \label{eq:fit02a}
\\
&\mathcal{M}_2^\mathrm{diag}= \left(
\begin{array}{ccc}
3.16730\times 10^{-9} & 0 & 0 \\
0 & 0.279523 & 0 \\
0 & 0 & 85.6150 \\
\end{array}
\right) \;\rm{GeV},
\\
&
\mathcal{M}_1=
\begin{pmatrix}
0.000474066\, e^{i\,2.56459} 
&
0.00418855\, e^{i\,2.93309} 
&
0.00558972\, e^{i\,2.96218}
\\[6pt]
0.00418855\, e^{i\,2.93309} 
&
0.0584201\, e^{-i\,1.68856} 
&
0.561849\, e^{i\,2.68735}
\\[6pt]
0.00558972\, e^{i\,2.96218} 
&
0.561849\, e^{i\,2.68735} 
&
0.292784\, e^{-i\,1.75857}
\end{pmatrix}
\;\rm{GeV},
\\
&
\mathcal{M}_3=\begin{pmatrix}
0
&
0.000209375\, e^{-i\,2.11125}
&
0.000271847\, e^{-i\,2.05453}
\\[6pt]
0.000209375\, e^{i\,1.03035}
&
0
&
0.336203\, e^{i\,0.943189}
\\[6pt]
0.000271847\, e^{i\,1.08706}
&
0.336203\, e^{-i\,2.1984}
&
0
\end{pmatrix} \;\rm{GeV}.\label{eq:fit02b}
\end{align}
\endgroup

The fitted values of the observables obtained from the parameters in Eqs.~\eqref{eq:fit01a}--\eqref{eq:fit01b} and 
\eqref{eq:fit02a}--\eqref{eq:fit02b} are listed in Table~\ref{tab:fit}. The 3rd and 5th column display the fitted values, while the 4th and 6th column show the corresponding pulls. The second column contains the experimental values of the observables at the $M_Z$ scale, together with their $1\sigma$ uncertainties. The charged-fermion observables are taken from Ref.~\cite{Antusch:2025fpm}, and the neutrino observables from Refs.~\cite{NUFIT,Esteban:2020cvm}. Note that the model does not allow a purely type-II seesaw solution~\cite{Babu:2016bmy}. Although both normal and inverted neutrino mass orderings are in principle allowed~\cite{Saad:2022mzu}, in this work we focus on the normal ordering scenario. 

\begin{table}[htb]
\centering
\footnotesize
\resizebox{1\textwidth}{!}{
\begin{tabular}{|c|c|cc|cc|}
\hline
\textbf{Observables} & \textbf{Exp.~values} & \textbf{Fitted values (BM I)} & \textbf{Pulls  (BM I)} & \textbf{Fitted values (BM II)} & \textbf{Pulls  (BM II)} \\
\hline\hline

\rowcolor{blue!20}$y_u/10^{-6}$        & $7.09^{+1.56}_{-0.88}$ & 6.71209 & $-0.429$  & 6.91 & $-0.205$ \\  
\rowcolor{blue!20}$y_c/10^{-3}$        & $3.55^{+0.10}_{-0.09}$ & 3.55042 & $+0.0042$   & 3.55126 & $+0.0126$ \\ 
\rowcolor{blue!20}$y_t$                & $0.968^{+0.004}_{-0.004}$ & 0.96813 & $+0.0134$   & 0.967687 & $-0.0324$ \\ \hline

\rowcolor{red!20}$y_d/10^{-5}$        & $1.55^{+0.14}_{-0.07}$ & 1.5336 & $-0.234$   & 1.55352 & $+0.0251$ \\ 
\rowcolor{red!20}$y_s/10^{-4}$        & $3.10^{+0.26}_{-0.14}$ & 3.11936 & $+0.0745$   & 3.13008 & $+0.116$ \\ 
\rowcolor{red!20}$y_b/10^{-2}$        & $1.63^{+0.02}_{-0.01}$ & 1.6336 & $+0.180$   & 1.63258 & $+0.129$ \\ \hline

\rowcolor{yellow!20}$y_e/10^{-6}$        & $2.77705^{+0.00033}_{-0.00039}$ & 2.77967 & $+0.0943$   & 2.77646 & $-0.0212$ \\ 
\rowcolor{yellow!20}$y_\mu/10^{-4}$      & $5.85026^{+0.00076}_{-0.00075}$ & 5.84719 & $-0.0525$   & 5.84995 & $-0.00528$ \\ 
\rowcolor{yellow!20}$y_\tau/10^{-2}$     & $0.99370^{+0.00015}_{-0.00014}$ & 0.993145 & $-0.0558$   & 0.992753 & $-0.0953$ \\ \hline

\rowcolor{cyan!20}$\theta_{12}^{CKM}$           & $0.2270\pm 0.0008$ & 0.227135 & $+0.0396$   & 0.227085 & $+0.0175$ \\ 
\rowcolor{cyan!20}$\theta_{23}^{CKM}/10^{-2}$   & $4.194\pm 0.041$ & 4.19563 & $+0.0334$   & 4.19474 & $+0.0121$ \\ 
\rowcolor{cyan!20}$\theta_{13}^{CKM}/10^{-3}$   & $3.70\pm 0.08$ & 3.70714 & $+0.0892$   & 3.70146 & $+0.0182$ \\ 
\rowcolor{cyan!20}$\delta_{CKM}$                & $1.139\pm 0.023$ & 1.1391 & $+0.00423$   & 1.13898 & $-0.00100$ \\
\hline\hline

\rowcolor{orange!20}$\Delta m^2_{21} (\mathrm{eV}^2)/10^{-5}$ & $7.49\pm 0.19$ & 7.49184 & $+0.00966$   & 7.49229 & $+0.0120$ \\ 
\rowcolor{orange!20}$\Delta m^2_{31} (\mathrm{eV}^2)/10^{-3}$ & $2.534^{+0.025}_{-0.023}$ & 2.53391 & $-0.00349$   & 2.53428 & $+0.0112$ \\ \hline

\rowcolor{green!20}$\sin^2 \theta_{12}$          & $0.307^{+0.012}_{-0.011}$ & 0.306717 & $-0.0257$   & 0.307865 & $+0.0721$ \\ 
\rowcolor{green!20}$\sin^2 \theta_{23}$          & $0.561^{+0.012}_{-0.015}$ & 0.559953 & $-0.769$   & 0.463325 & $-1.243$ \\ 
\rowcolor{green!20}$\sin^2 \theta_{13}$          & $0.02195^{+0.00054}_{-0.00058}$ & 0.021945 & $-0.00856$   & 0.0219554 & $+0.0101$ \\

\hline\hline

$\chi^2$ & -- & -- & 0.89 & --  & 1.63 \\
\hline
\end{tabular}

}
\caption{
Experimental values of the observables at the $M_Z$ scale, along with their $1\sigma$ uncertainties, are taken from Refs.~\cite{Antusch:2025fpm} for the charged-fermion sector and Refs.~\cite{NUFIT,Esteban:2020cvm} for the neutrino sector. While the central experimental value of $\sin^2\theta_{23}$ along with its $1\sigma$ uncertainty is quoted in the table, our numerical procedure incorporates the full range covering both scenarios, namely $\theta_{23} < 45^\circ$ and $\theta_{23} > 45^\circ$; see Ref.~\cite{NUFIT}.  The fitted values corresponding to the benchmark solution are shown in the third column, with their pulls listed in the fourth column.} 
\label{tab:fit}
\end{table}

\begin{table}[htb]
\centering
\footnotesize
\resizebox{1\textwidth}{!}{
\begin{tabular}{c| c c| c c}
\hline
\textbf{Observable} & \textbf{BM I} & \textbf{$2\sigma$ HPD intervals (MCMC I)} & \textbf{BM II} & \textbf{$2\sigma$ HPD intervals (MCMC II)} \\
\hline

\rowcolor{teal!20}$m_1\,(\mathrm{meV})$ & 0.0441 & $\INT{0.0339}{0.0642}$ & 0.0457 & $\INT{0.0339}{0.0613}$ \\ 
\rowcolor{teal!20}$m_2\,(\mathrm{meV})$ & 8.66 & $\INT{8.17}{9.09}$ & 8.66 & $\INT{8.43}{8.88}$ \\
\rowcolor{teal!20}$m_3\,(\mathrm{meV})$ & 50.34 & $\INT{49.82}{50.85}$ & 50.34 & $\INT{49.82}{50.85}$ \\  \hline

\rowcolor{yellow!20}$m_{\beta\beta}\,(\mathrm{meV})$ & 3.58 & $\INT{3.00}{3.89}$ & 3.71 & $\INT{3.25}{3.92}$ \\  \hline

\rowcolor{red!20}$\delta_\mathrm{PMNS}\,(\mathrm{deg})$ & 0  & $\INT{-43.37}{43.93}$  
& 270 & $\INT{263.71}{313.67}$ \\ \hline

\rowcolor{green!20}$M_1/10^{4}\,(\mathrm{GeV})$  &  4.92  & $\INT{4.09}{7.73}$  
& 4.51 & $\INT{3.51}{7.80}$ \\ 

\rowcolor{green!20}$M_2/10^{12}\,(\mathrm{GeV})$  &  3.42  & $\INT{2.92}{3.68}$  
& 3.98 & $\INT{3.42}{4.33}$ \\ 

\rowcolor{green!20}$M_3/10^{15}\,(\mathrm{GeV})$ &  0.993  & $\INT{0.85}{1.06}$  
& 1.22 & $\INT{1.07}{1.32}$ \\ \hline

\rowcolor{orange!20}$\BR{p\rightarrow \pi^0 e^+}/\%$  &  44.75 & $\INT{41.80}{47.32}$ & 42.69 & $\INT{39.36}{46.18}$ \\ 
\rowcolor{orange!20}$\BR{p\rightarrow \pi^0 \mu^+}/\%$  &  0.33 & $\INT{0.21}{0.56}$ & 0.33 & $\INT{0.17}{0.70}$ \\
\rowcolor{orange!20}$\BR{p\rightarrow K^0 e^+}/\%$  &   0.54 & $\INT{0.49}{0.63}$ & 0.53 & $\INT{0.48}{0.60}$ \\
\rowcolor{orange!20}$\BR{p\rightarrow K^0 \mu^+}/\%$  &  2.43 & $\INT{2.17}{3.02}$ & 3.24 & $\INT{2.63}{3.90}$ \\
\rowcolor{orange!20}$\BR{p\rightarrow \eta^0 e^+}/\%$  &  0.043 & $\INT{0.037}{0.045}$ & 0.041 & $\INT{0.037}{0.044}$ \\
\rowcolor{orange!20}$\BR{p\rightarrow \eta^0 \mu^+}/\%$  &  0.00031 & $\INT{0.00020}{0.00053}$ & 0.00031 & $\INT{0.00012}{0.00080}$ \\
\rowcolor{orange!20}$\BR{p\rightarrow \pi^+\overline{\nu}}/\%$  &   50.89 & $\INT{48.41}{53.32}$ & 52.07 & $\INT{49.10}{54.75}$ \\
\rowcolor{orange!20}$\BR{p\rightarrow K^+ \overline{\nu}}/\%$ & 1.02  & $\INT{0.92}{1.10}$ & 1.10 & $\INT{1.00}{1.16}$ \\ \hline
\end{tabular}
}
\caption{Benchmark fit predictions and $2\sigma$ HPD intervals from MCMC data for a set of interesting observables.}
\label{tab:prediction}
\end{table}

\begin{figure}[b!]
\centering
\includegraphics[width=0.47\textwidth]{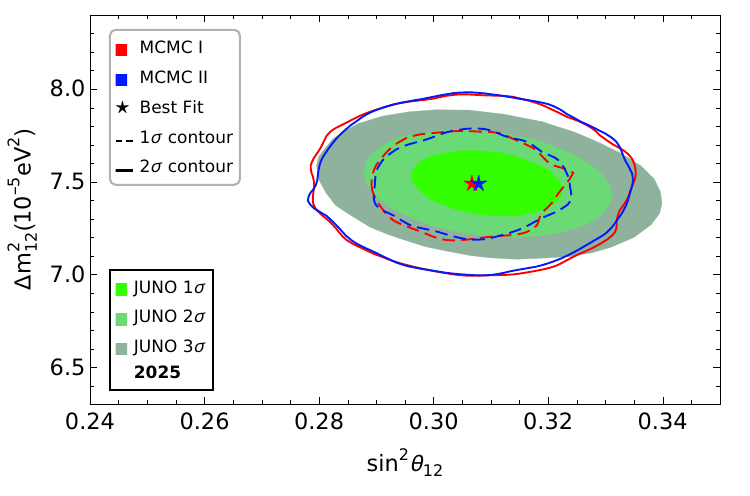}
\includegraphics[width=0.495\textwidth]{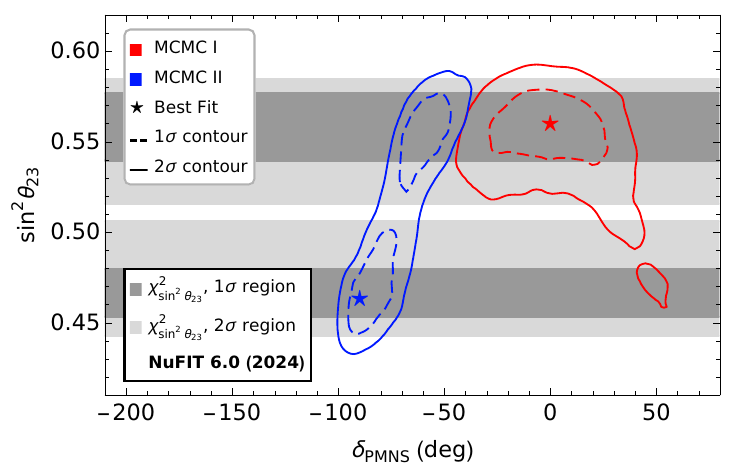}
\caption{ Left panel: MCMC results illustrating the correlation between the neutrino mixing angle $\theta_{12}$ and the mass-squared difference $\Delta m^2_{21}$. As can be seen from this plot, our results are fully consistent with the recent \texttt{JUNO} measurement~\cite{JUNO:2025gmd}. Right panel: the HPD contours for the correlation of 
the leptonic Dirac CP-violating phase $\delta_\mathrm{PMNS}$ and the atmospheric mixing $\sin^2\theta_{23}$, obtained from the MCMC analyses. It is to be pointed out that even though the MCMC did not explore the entire range, the model can accommodate any value $\delta_\mathrm{PMNS} \in [0,2\pi)$, see text for details. 
} \label{fig:MCMC:01}
\end{figure}

To explore the parameter space around the benchmark points, we perform an MCMC analysis as described in Section~\ref{subsec:numerical-procedure}, with the datasets referred to as MCMC I and MCMC II. The resulting predictions are compiled in Table~\ref{tab:prediction}, which specifies the predictions for a list of yet-to-be measured observables: the light neutrino masses $m_{i}$, the parameter $m_{\beta\beta}$ relevant for neutrinoless double beta decay, the PMNS phase $\delta_\mathrm{PMNS}$, the heavy right-handed neutrino masses $M_{i}$, and the branching ratios for various proton decay channels. The table consists of both the benchmark point predictions, as well as the $2$-$\sigma$ highest posterior density (HPD) intervals from the MCMC data.

\begin{figure}[t!]
\centering
\includegraphics[width=0.48\textwidth]{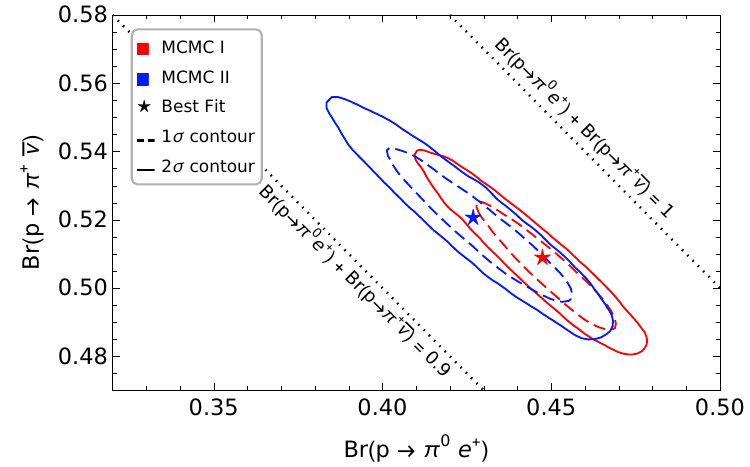}
\includegraphics[width=0.48\textwidth]{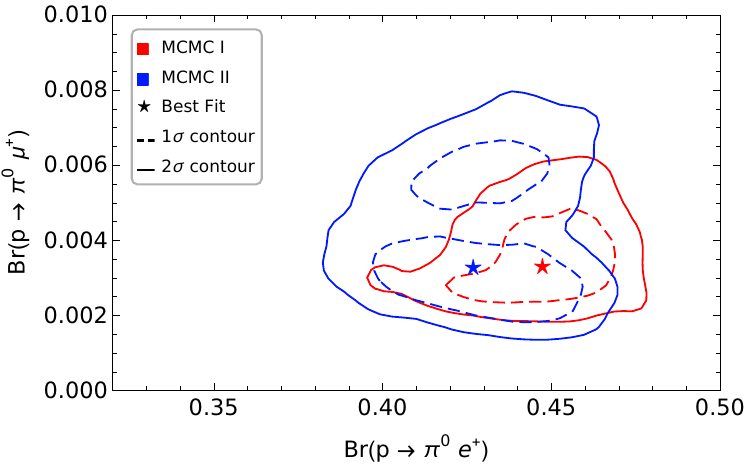}
\caption{Correlations in MCMC results in the branching ratio of $p\to \pi^{0} e^{+}$ with two other processes. Left panel: the correlation with the largest branching ratio $p\to \pi^{+} \overline{\nu}$. Right panel: the correlation with $p\to\pi^{0}\mu^{+}$, providing information about the flavor structure in the lepton sector. \label{fig:MCMC:03} 
}
\end{figure}

Finally, the detailed MCMC results are compiled in Figures~\ref{fig:MCMC:01}, \ref{fig:MCMC:02} and~\ref{fig:MCMC:03}, which show $1$- and $2$-$\sigma$ HPD regions of MCMC data as dashed and solid lines, respectively. The two datasets are shown in different colors --- \textcolor{red}{red} for MCMC I and \textcolor{blue}{blue} for MCMC II. We gather our commentary on these results below:
\begin{itemize}[itemsep=0cm,leftmargin=\DIMLEFTMARGIN]
    \item In the left panel of Fig.~\ref{fig:MCMC:01}, we illustrate the correlation between the neutrino mixing angle $\theta_{12}$ and the mass-squared difference $\Delta m^2_{21}$ with $1$- and $2$-$\sigma$ HPD contours. It is worth highlighting that the \texttt{JUNO} Collaboration has recently reported its first results on reactor neutrino oscillations based on $59.1$ days of data taking~\cite{JUNO:2025gmd}. Their measurements of solar oscillation parameters $\sin^2\theta_{12}$ and $\Delta m^2_{21}$ are consistent with previous experimental determinations, while achieving an approximately 1.6-fold improvement in precision compared to the combined results of earlier experiments (whose values we fit, cf.~Table~\ref{tab:fit}). We compare our MCMC predictions in the $\Delta m^2_{21}$–$\sin^2\theta_{12}$ plane with this initial \texttt{JUNO} dataset and find excellent agreement. 
\item
    The right panel in Fig.~\ref{fig:MCMC:01} displays the 
    HPD regions in the plane of $\delta_\mathrm{PMNS}$ (rephased to the interval $\INT{-180^\circ}{180^\circ}$) and $\sin^{2}\theta_{23}$. We reiterate that the former is not part of the $\chi^2$, while the latter is, having a bimodal distribution of good $\chi^{2}_{\sin^2\theta_{23}}$ values (shown by grey HPD bands). We clearly see from the plot that the initial best-fit points (denoted by star symbols) were at different minima of $\chi^{2}_{\sin^2\theta_{23}}$, but the MCMC then migrated across the barrier to explore both regions, especially in the case of MCMC II where the distribution is strongly bimodal. The fact that the regions are localized in $\delta_\mathrm{PMNS}$ confirms that MCMC explores well locally rather than globally, likely facing obstructions from different curvature in different regions of the good-$\chi^2$ hyperplane. 
\item 
    The left panel of Fig.~\ref{fig:MCMC:02} visually compares the MCMC results for the neutrinoless double beta decay parameter with the current \texttt{KamLAND-Zen} bound~\cite{KamLAND-Zen:2016pfg}  and  sensitivities for future experiments, namely \texttt{JUNO}~\cite{Zhao:2016brs},  \texttt{GERDA Phase II}~\cite{GERDA:2019cav}, as well as \texttt{nEXO}~\cite{nEXO:2021ujk}. 
    We see that the two MCMC results are mutually consistent and predict a value $m_{\beta\beta} \sim 3$--$4\mathrm{meV}$, i.e.~just below the projected sensitivities of upcoming experiments. 
\item 
    The right panel of Fig.~\ref{fig:MCMC:02} displays the distribution of right-handed neutrino masses, revealing a sharply-predicted and highly hierarchical~\cite{Babu:2016bmy} spectrum with $M_1 \approx 5 \times 10^4\,\mathrm{GeV}$, $M_2 \approx 3 \times 10^{12}\,\mathrm{GeV}$, and $M_3 \approx 10^{15}\,\mathrm{GeV}$. This characteristic hierarchy is consistent for both MCMCs, and as demonstrated also by the earlier Figure~\ref{fig:correlation-M}, it is a prediction of the model. We point out that the generic features of the results resemble those from the original fermion mass matrices (where the reality condition was not consistently taken into account) in Ref.~\cite{Babu:2016bmy}.

\begin{figure}[t!]
\centering
\includegraphics[width=0.5\textwidth]{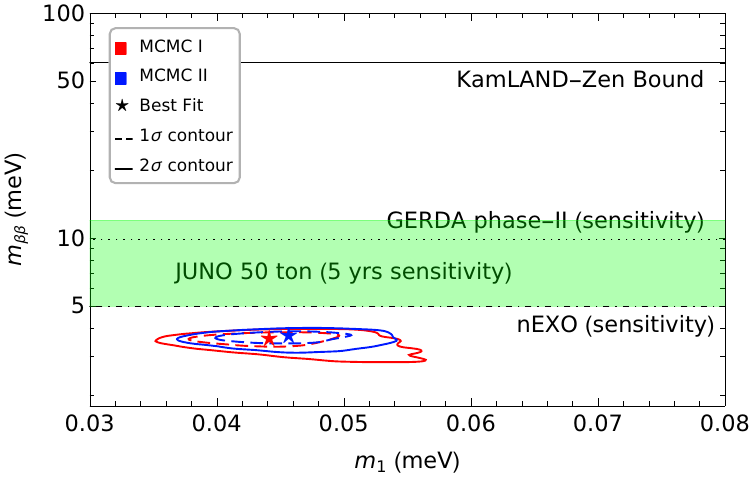}
\includegraphics[width=0.47\textwidth]{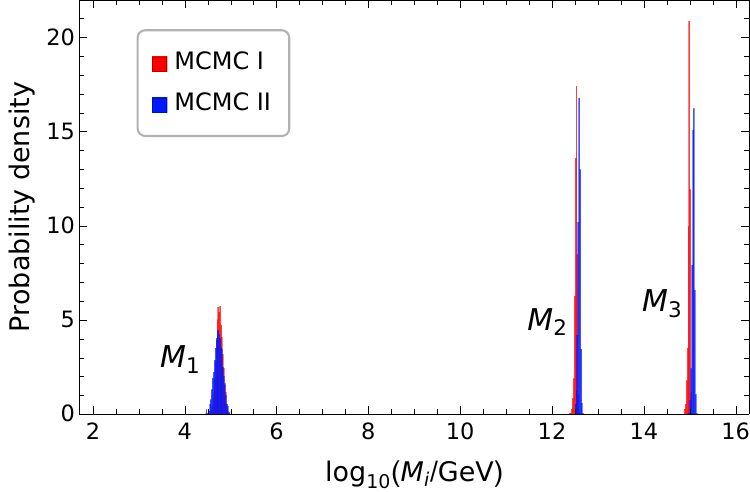}
\caption{ Left panel: The MCMC results indicate that the neutrinoless double beta decay parameter is expected to be small, in particular $m_{\beta\beta} \sim 3$--$4\,\mathrm{meV}$, which lies below the projected sensitivities of future experiments. Right panel: the distributions of the heavy sterile neutrino masses show the model predicts a sharp and extremely hierarchical mass spectrum, namely $M_1 \approx 5 \times 10^4\,\mathrm{GeV}$, $M_2 \approx 3 \times 10^{12}\,\mathrm{GeV}$, and $M_3 \approx 10^{15}\,\mathrm{GeV}$.   } \label{fig:MCMC:02}
\end{figure}

\item 
    While predicted ranges for branching ratios of proton decay have been provided already in Table~\ref{tab:prediction}, Figure~\ref{fig:MCMC:03} focuses on the most relevant model-dependent correlations between the channels, with all computational details for proton decay rates relegated to Appendix~\ref{app:PD}. 
    The left panel shows the anti-correlation between the two dominant modes $p \to \pi^+ \overline{\nu}$ and $p \to \pi^0 e^+$; while their branching ratios can vary somewhat relative to each other, they together amount to a robustly predicted $\approx 95\,\%$ of the total decay width. Observation of proton decay in any channel other than these two dominant modes would therefore disfavor the model. The right panel shows the correlation between 
    $p \to \pi^0 e^+$ and $p \to \pi^0 \mu^+$; although the latter process would be hard to measure due to its low branching ratio, the correlation in this pair of channels reveals the most information about the flavor structure in the leptonic sector,
    since replacing $e^+$ with $\mu^+$ in the final state directly probes the flavor coefficients $|\hat{c}(e_\beta,d^C)|^{2}$ and $|\hat{c}(e^C_\beta,d)|^{2}$  for $\beta=1,2$, cf.~Eq.~\eqref{CIA} and \eqref{CIA2} for definitions and Eq.~\eqref{eq:decay-pi-e} for the decay rate expression. 
    We also see that the MCMC II distribution is again bimodal, a feature that is however independent from previous bimodal properties of $\sin^2\theta_{23}$ or $\delta{\text{PMNS}}$; the only observable from Table~\ref{tab:prediction}, with which the bimodality of $\BR{p\to \pi^0 \mu^+}$ is somewhat coordinated, is the lowest right-handed neutrino mass $M_{1}$.
    \par
    We conclude the proton decay discussion by noting that its decay rate depends on the GUT scale $M_{\text{GUT}}$ and is therefore not a robust prediction on its own. Instead, we focus on branching ratios, which depend on flavor structures and thus constitute robust predictions of our numerical analysis. Note that for the analyses in this paper, we took a fixed GUT scale at $10^{16}\,\mathrm{GeV}$, for which we checked that all decay channels are predicted within experimental bounds, assuming e.g.~$g=\sqrt{4\pi/35}$. 
\item A comparison of the results across Table~\ref{tab:prediction} and Figures~\ref{fig:MCMC:01}, \ref{fig:MCMC:02} and \ref{fig:MCMC:03} for MCMC I and II shows that both datasets reproduce overlapping and consistent results, the main exception being the explored regions in $\delta_{\text{PMNS}}$, which was indeed the main discriminator for the two starting points.
\item 
    Finally, the observed baryon asymmetry of the universe can be reproduced via leptogenesis~\cite{Babu:2024ahk,Babu:2025wop}; however, we do not impose this constraint in the present analysis.
\end{itemize}

\section{Conclusions \label{sec:conclusions}}
In this work, we have studied the minimal Yukawa structure of $\SO(10)$ unification, where the Higgs sector contributing to fermion masses consists of real $\GIRREP{10}_\mathbb{R}$ and $\GIRREP{120}_\mathbb{R}$, along with a $\GIRREP{126}$ representation. 

We focused on the reality conditions imposed on $\SO(10)$ representations $\GIRREP{10}$ and $\GIRREP{120}$. We performed an explicit $\SO(10)$ computation in Section~\ref{subsec:result-SO10} and showed how the reality constraints manifest in fermion mass matrices. In particular, while the $\GIRREP{10}$ contributes equally to all fermion sectors (apart from conjugating the EW VEV in the down-quark and charged-lepton sectors), the reality constraints of $\GIRREP{120}$ introduce a relative minus sign in fermion mass relations when the conjugated EW VEV of the Pati-Salam $\PSIRREP{15}{2}{2}$ Higgs bi-doublet appears, but no such sign is introduced in terms with the conjugated EW VEV from the $\PSIRREP{1}{2}{2}$.
The relative minus sign is physical, and it revises previously reported fermion mass relations in the minimal Yukawa sector of $\SO(10)$ GUT.

The relative sign is imposed purely by $\SO(10)$ symmetry (and having an $\SO(10)$-equivariant reality condition on $\GIRREP{120}$). To ensure this conclusion is robust, we confirm this outcome by using four distinct and complementary methodologies:
\begin{enumerate}[nosep,label=(\arabic*)]
\item \label{item:method-tensor} A direct tensorial calculation of the relevant $\SO(10)$ invariants fixes the relative sign explicitly, which is presented in Section~\ref{subsec:result-SO10}.
\item \label{item:method-embedding} As discussed in Section~\ref{subsec:result-PS}, while a stand-alone Pati-Salam consideration by itself cannot determine this sign,  Section~\ref{subsec:result-SO10-PS} constructs the embedding of Pati-Salam into $\SO(10)$ and  derives the inherited reality condition from this map.
\item An independent $SU(5)$-based calculation using the oscillator construction of $\SO(10)$ reproduces the same result; see Section~\ref{subapp:result-alternative-su5xu1}.
\item \label{item:method-transport} Finally, we develop a novel systematic algorithm for transporting the real structure of irreducible representations from parent to daughter in Section~\ref{app:rtransport}, which further corroborates our result. 
\end{enumerate}
The agreement among these conceptually distinct derivations provides a nontrivial cross-check of the relative sign and its group-theoretic origin. The methods developed here extend well beyond the example they were applied on:
the formalism of method~\ref{item:method-embedding} can be used to analytically determine reality conditions for other parent-daughter pairs of $\SO(10)$ and PS representations, while method~\ref{item:method-transport} generalizes to real irreps of arbitrary group-subgroup pairs. Furthermore, our work demonstrates why reality conditions of daughter irreps need to be consistently linked to the real structure of their common parent irrep of the full gauge group.

The phenomenological implications of the revised fermion mass matrices were then studied in Section~\ref{sec:yukawa-sector}. After reexamining the  parametrization to identify the independent inputs, we find an additional parameter (a magnitude of a complex variable) appears with the revision. We performed a detailed numerical study to demonstrate that the model consistently reproduces the observed fermion masses, including both the charged fermion and neutrino sectors. The model predicts relatively small values of the neutrinoless double beta decay parameter, $m_{\beta\beta} \sim 3$--$4\,\mathrm{meV}$, which lie just below the projected sensitivities of upcoming experiments. At the same time, the predicted neutrino observables are fully consistent with the recent measurements of solar oscillation parameters $\sin^2\theta_{12}$ and $\Delta m^2_{21}$  reported by \texttt{JUNO}. We also investigate the model’s predictions for neutrino observables that are not yet precisely determined, in particular the atmospheric mixing angle $\theta_{23}$, which is subject to the octant ambiguity, and the leptonic CP-violating phase $\delta_{\mathrm{PMNS}}$. The model accommodates values of $\theta_{23}$ in both octants, while showing a mild preference against $\delta_{\mathrm{PMNS}}$ in the range $\sim \INT{140^\circ}{220^\circ}$. Future facilities such as \texttt{DUNE}, \texttt{T2HK}, and \texttt{ESSnuSB} are expected to provide high-precision measurements that will further scrutinize the model and significantly constrain its parameter space. Furthermore, the model predicts a highly hierarchical right-handed neutrino mass spectrum—a unique feature of the minimal scenario—given by $M_1 \approx 5 \times 10^4\,\mathrm{GeV}$, $M_2 \approx 3 \times 10^{12}\,\mathrm{GeV}$, and $M_3 \approx 10^{15}\,\mathrm{GeV}$, with the heaviest state surprisingly close to the GUT scale, suggesting immediate proximity of the intermediate and GUT scales. Finally, the model predicts the largest proton decay branching ratios for the modes $p \to \pi^+ \overline{\nu}$ and $p \to \pi^0 e^+$, cumulatively accounting for $\approx 95\,\%$ of the total decay width. These decay channels provide clear experimental signatures for testing the framework, in particular at \texttt{DUNE}, \texttt{THEIA}, and \texttt{Hyper-Kamiokande}.

\section*{Acknowledgments}
SS acknowledges the financial support
from the Slovenian Research Agency (research core funding No.~P1-0035 and N1-0321). SS gratefully acknowledges the warm hospitality of INFN, Frascati, where part of this work was conducted.
VS is supported by the European Union --- Next Generation EU and
by the Italian Ministry of University and Research (MUR) 
via the PRIN 2022 project n.~2022K4B58X --- AxionOrigins.

\appendix
\section{An overview of spinors in $\SO(2n)$} \label{app:spinors}

\subsection{Clifford algebra and the spinorial representation} \label{app:spinors-Clifford}

To elucidate the reality issue in the Yukawa sector of $\SO(10)$ GUT, it is necessary to consider spinor representations in $\SO(10)$ (fermions are present in a spinorial $\GIRREP{16}$ of $\SO(10)$), as well as in $\SO(6)$
and $\SO(4)$ (to fully enable explicit algebra isomorphisms $\SO(6)\cong \SU(4)_{C}$ and $\SO(4)\cong \SU(2)_L\times\SU(2)_R$). We thus start the discussion in the general setting of $\SO(2n)$, or strictly speaking $\mathrm{Spin}(2n)$, cf.~the note in item~\ref{item:group-theory-PS-in-SO10} of Section~\ref{subsec:group-theory}. This content is well known; here we partly follow~\cite{Wilczek:1981iz}, albeit with slightly different conventions for chirality (as elaborated at the end of App.~\ref{app:spinors-gamma}).
                          
A Clifford algebra in $(2n)$-dimensional Euclidean space is a set of $2n$ gamma matrices $\bm{\Gamma}^{(2n)}_{I}$, such that the following relation holds:
    \begin{align}
        \{\bm{\Gamma}^{(2n)}_{I},\bm{\Gamma}^{(2n)}_{J}\}&
            =2\, \delta_{IJ}\,\mathbb{1},  
            \label{eq:definition-clifford}   
    \end{align}
with the anti-commutator defined as $\{\mathbf{A},\mathbf{B}\}:=\mathbf{AB}+\mathbf{BA}$, and the index range is $I,J=1\ldots 2n$. The superscript $(2n)$ is merely a label specifying the dimension; we shall omit writing it for the general case from now on. Eq.~\eqref{eq:definition-clifford} states that all gamma matrices square to an identity and anti-commute for different indices $I$ and $J$. Any product of gamma matrices can thus be rearranged so that each gamma matrix (as labeled by the index) appears at most once and the index values are in ascending order;  such a rearrangement incurs a minus sign each time two gamma matrices switch order, and the set of such products constitute a basis of the Clifford algebra if viewed as a vector space. The Clifford algebra in $(2n)$ dimensions is thus a vector space of dimension $2^{2n}$. 

These gamma objects prove useful for Lie theory, since subsequently defining
    \begin{align}
        \MAT{T}_{IJ} &
            := \tfrac{i}{4}\,[\bm{\Gamma}_{I},\bm{\Gamma}_{J}] 
            \label{eq:definition-gen-spinor}
    \end{align}
reveals that the objects $\MAT{T}_{IJ}$ satisfy the commutation relations for generators of an orthogonal Lie algebra $\SO(2n)$:
    \begin{align}
        [\MAT{T}_{IJ},\MAT{T}_{KL}]&=
            -i\left(
            \delta_{JL}\,\mathbf{T}_{IK}
            +\delta_{IK}\,\mathbf{T}_{JL}
            -\delta_{JK}\,\mathbf{T}_{IL}
            -\delta_{IL}\,\mathbf{T}_{JK}
            \right).
    \end{align}
Since Eq.~\eqref{eq:definition-gen-spinor} implies $\mathbf{T}_{IJ}=-\mathbf{T}_{JI}$ and we have the range $I,J=1\ldots 2n$, there are $\binom{2n}{2}$ independent such objects. The subset of $n$ generators $\{ \MAT{T}_{12},\MAT{T}_{34},\ldots,\MAT{T}_{2n-1,2n}\}$, i.e.~those of the form $\MAT{T}_{I,I+1}$ for $I=1,3,5,\ldots,2n-1$, is mutually commuting, and it generates a Cartan subalgebra.

Two further structurally important objects are then constructed: the chiral element $\bm{\chi}\equiv\bm{\Gamma}_{\rm{FIVE}}$ and the charge conjugation element $\MAT{C}$:
\begin{enumerate}[leftmargin=0.5cm,itemsep=0pt,label=(\roman*)]
    \item
        The chiral element $\bm{\chi}$ is constructed by multiplying all gamma matrices as
        \begin{align}
            \bm{\chi} &
            := i^{n}\;\bm{\Gamma}_{1}\bm{\Gamma}_{2}\cdots \bm{\Gamma}_{2n}.
            \label{eq:chi-definition}
        \end{align}
        This convention differs from that of~\cite{Wilczek:1981iz} by a factor $(-1)^n$, see end of App.~\ref{app:spinors-gamma} for discussion.
        The following properties can be derived from Eq.~\eqref{eq:definition-clifford}:
        \begin{align}
            \{\bm{\chi},\bm{\Gamma}_{I}\}&=0, \label{eq:chi-property-1}\\
            \bm{\chi}^2 &= \mathbb{1}. \label{eq:chi-property-2}
        \end{align}
        Anti-commutation with all gamma matrices in Eq.~\eqref{eq:chi-property-1} implies that the basis elements of the Clifford algebra formed as products of gamma matrices either commute or anticommute with $\bm{\chi}$, depending on whether the products consist of an even or odd number of gamma matrices, respectively. Since there is an equal number of basis elements in each set, the two spaces have the same dimensionality, splitting the Clifford algebra into two pieces of equal dimension, referred to by convention as \textit{even} and \textit{odd} subspaces, which have respective eigenvalues $+1$ and $-1$ under the automorphism $\MAT{A}\mapsto \bm{\chi}\MAT{A}\bm{\chi}^{-1}$. 
        \par
        One can also define projection operators acting on the spinor representation space: 
        \begin{align}
            \mathbf{P}_{\pm} &:= \tfrac{1}{2}\left(\mathbb{1}\pm \bm{\chi}\right)
            \label{eq:projection-operator}
        \end{align}
        splits it into so-called chiral subspaces. The realtion $\mathbf{P}_\pm^{2}=\mathbf{P}_\pm$ holds due to Eq.~\eqref{eq:chi-property-2}. We make use of these concepts later in Appendix~\ref{app:spinors-gamma}.
    \item
        The second object is the charge conjugation matrix $\MAT{C}$, which is defined such that it satisfies the relation
            \begin{align}
                \bm{\Gamma}_{I}^\top &
                    =(-1)^{n}\,\MAT{C}\,\bm{\Gamma}_{I}\,\MAT{C}^{-1},
                \label{eq:C-defining-property}
            \end{align}
        i.e.~it is the basis change that transforms between gamma matrices and their transposes. The charge conjugation matrix is guaranteed to exist, since the set $\bm{\Gamma}_{I}^{\top}$ also satisfies the Clifford algebra relation in Eq.~\eqref{eq:definition-clifford}, but it is not unique, and its explicit definition expressed in terms of gamma matrices is basis dependent. 
        \par
        The chiral element transforms under transposition in the same way as individual gamma matrices:
        \begin{align}
            \bm{\chi}^{\top} &= (-1)^{n}\,\MAT{C} \bm{\chi} \MAT{C}^{-1}.
        \end{align}
\end{enumerate}
            

\subsection{An explicit construction of gamma matrices} \label{app:spinors-gamma}

In the previous subsection, we considered gamma matrices as purely abstract objects in a Clifford algebra. We now consider their $2^n$-dimensional irreducible representation, i.e.~their realization as $2^n\times 2^n$ matrices. We label the underlying vector space as $\mathcal{S}\equiv \mathbb{C}^{2^n}$ and refer to it as spinor space, so they map $\bm{\Gamma}_I: \mathcal{S}\to\mathcal{S}$ for every $I=1\ldots 2n$.
 
A very convenient explicit realization/representation of gamma matrices is due to Brauer and Weyl~\cite{Weyl-Brauer}, known in HEP literature from Wilczek and Zee \cite{Wilczek:1981iz}. The construction provides gamma matrices in a basis which already exhibits almost all the desired properties, but a slight basis modification is nevertheless necessary to implement the standard conventions for spinorial indices. We provide below all necessary details. To avoid notational confusion, we label all objects in their original basis with a prime, while the objects in the final basis will be unprimed. The transformation between the two is performed by an orthogonal transformation $\MAT{S}$, such that $\MAT{X}'\mapsto \MAT{X}=\MAT{S}\MAT{X}'\MAT{S}^{\top}$ for any Clifford object. The transformation $\MAT{S}$ is an as-of-yet unspecified \textit{signed permutation matrix}, i.e.~a composition of a permutation matrix and a sign redefinition of some basis elements.

The initial gamma matrices $\bm{\Gamma}'_{I}$ are constructed as $2^n\times 2^n$ complex matrices via (see \cite{Wilczek:1981iz})
    \begin{align}
        \bm{\Gamma}'_{2K-1} &:= 
                \mathbb{1}_{2}^{\otimes(K-1)}\;\otimes\;
                \bm{\tau}_{1}\;\otimes\;
                \bm{\tau}_{3}^{\otimes (n-K)},
                \label{eq:BW1}\\
        \bm{\Gamma}'_{2K} &:= 
                \mathbb{1}_{2}^{\otimes(K-1)}\;\otimes\;
                \bm{\tau}_{2}\;\otimes\;
                \bm{\tau}_{3}^{\otimes (n-K)},
                \label{eq:BW2}
    \end{align}
where $K=1,\ldots,n$. The symbol $\mathbb{1}_{2}$ denotes the $2\times 2$ identity matrix, while $\bm{\tau}_i$ denotes the Pauli matrices
    \begin{align}
        \bm{\tau}_{1}&:=
            \begin{pmatrix}
                0 & 1\\
                1 & 0\\
            \end{pmatrix},&
        \bm{\tau}_{2}&:=
            \begin{pmatrix}
                0 & -i\\
                i & 0\\
            \end{pmatrix},&
        \bm{\tau}_{3}&:=
            \begin{pmatrix}
                1 & 0\\
                0 & -1\\
            \end{pmatrix}. \label{eq:Pauli-matrices}
    \end{align}
The symbol $\otimes$ in this context denotes the Kronecker product of matrices, with the $m$-th Kronecker power of a matrix $\MAT{A}$ for $m\in\mathbb{N}$ interpreted as
    \begin{align}
        \MAT{A}^{\otimes m}
            &\equiv \underbrace{\MAT{A}\otimes\MAT{A}\otimes\cdots\otimes\MAT{A}}_{m}.
    \end{align}
The factorization in terms of Pauli matrices of this construction allows for easy computation, since the Kronecker product is multilinear, associative and satisfies the following convenient properties (written for square matrices of appropriate dimensions):
    \begin{align}
        (\MAT{A}_{1}\otimes\MAT{B}_{1})(\MAT{A}_{2}
        \otimes\MAT{B}_{2}) &
            = (\MAT{A}_{1}\MAT{A}_{2})\otimes(\MAT{B}_{1}\MAT{B}_{2}), 
            \label{eq:Kronecker-mixed}\\
        (\MAT{A}\otimes\MAT{B})^{\top} &
            = \MAT{A}^\top \otimes\MAT{B}^\top,
            \label{eq:Kronecker-transpose}\\
        (\MAT{A}\otimes\MAT{B})^{*} &
            = \MAT{A}^* \otimes\MAT{B}^*.
            \label{eq:Kronecker-conjugation}
    \end{align}
The Clifford algebra relation from Eq.~\eqref{eq:definition-clifford} can be easily checked by repeated use of the mixed-product property in Eq.~\eqref{eq:Kronecker-mixed} and the relation $\bm{\tau}_{i}\bm{\tau}_{j}=\delta_{ij}\mathbb{1}+i\epsilon_{ijk} \bm{\tau}_{k}$ for Pauli matrices. Complex conjugation and transposition properties from Eq.~\eqref{eq:Kronecker-transpose} and \eqref{eq:Kronecker-conjugation} imply that $\bm{\Gamma}'_{I}$ are all Hermitian (since Pauli matrices are Hermitian); in particular they are symmetric and real for $I=2K-1$, but antisymmetric and imaginary for $I=2K$. The Lie algebra generators $\MAT{T}'_{IJ}$ formed via Eq.~\eqref{eq:definition-gen-spinor} from Hermitian gamma matrices $\bm{\Gamma}'_{I}$ are also Hermitian, so the spinorial representation of the Lie group $\SO(2n)$ (or more rigorously $\mathrm{Spin}(2n)$) is unitary.

Explicitly computing the chiral element from Eq.~\eqref{eq:chi-definition} gives
    \begin{align}
        \bm{\chi}'&= (-\bm{\tau}_{3})^{\otimes n}, \label{eq:chiral-element-explicitly}
    \end{align}
so $\bm{\chi}'$ in the original basis is already a diagonal matrix with values $\pm 1$. Furthermore, the $n$ Cartan generators $\MAT{T}'_{2K-1,2K}$ for $K=1\ldots n$ based on the definition from Eq.~\eqref{eq:definition-gen-spinor} become explicitly
    \begin{align}
        \MAT{T}'_{2K-1,2K} = \mathbb{1}_{2}^{\otimes (K-1)}\otimes \left(-\frac{\bm{\tau}_{3}}{2}\right) \otimes \mathbb{1}_{2}^{\otimes (n-K)}. 
        \label{eq:cartan-explicitly}
    \end{align}
The Cartan generators are therefore already diagonal in the original basis, i.e.~the basis states already have well-defined quantum numbers. Every such state is uniquely characterized by $n$ quantum numbers with values $\pm 1/2$, which is consistent with having a total of $2^n$ states. These states are thus often denoted in the literature by a series of $n$ $\pm$ symbols specifying the signs of these quantum numbers, i.e.~by $|\pm\pm\cdots\pm \rangle$. In the $i^n$ convention of Eq.~\eqref{eq:chi-definition} for $\bm{\chi}$, a state $|s_{1}s_{2}\ldots s_{n} \rangle$ has chirality ($\bm{\chi}$-eigenvalue) $\prod_{i=1}^{n} s_{i}$, as can be inferred from the $-\bm{\tau}_{3}$ structure appearing in both Eqs.~\eqref{eq:chiral-element-explicitly} and \eqref{eq:cartan-explicitly}.

The charge conjugation matrix can be chosen to be proportional to the product of even-indexed gamma matrices, obtaining
    \begin{align}
        \MAT{C}'&
            =  i^{n}\,\bm{\Gamma}'_{2}\bm{\Gamma}'_{4}\cdots \bm{\Gamma}'_{2n} 
            \label{eq:charge-BW-result1}\\
        &= 
            \begin{cases}
                \left(i\,\bm{\tau}_{2}\otimes \,\bm{\tau}_{1}\right)^{\otimes k}
                &; n=2k\\
                \left(i\,\bm{\tau}_{2}\otimes \,\bm{\tau}_{1}\right)^{\otimes (k-1)}\otimes (i\,\bm{\tau}_{2})
                &; n=2k-1\\
            \end{cases}.
            \label{eq:charge-BW-result2}
    \end{align}
The form of Eq.~\eqref{eq:charge-BW-result2} indicates $\MAT{C}'$ is a signed permutation matrix, since it is a Kronecker product of signed permutation matrices (one non-zero entry $\pm 1$ in each column and row). This means it is orthogonal, i.e.~$\MAT{C}'{}^{-1}=\MAT{C}'{}^{\top}$. The form of Eq.~\eqref{eq:charge-BW-result1}, on the other hand, implies that
    \begin{align}
        \MAT{C}'^{\top} &
            = \MAT{C}'^{\dagger} 
            = (-1)^{n(n+1)/2}\;\MAT{C}',
            \label{eq:C-property-1}\\
        \{\MAT{T}'_{2K-1,2K},\MAT{C}'\}&
            = 0, 
            \label{eq:C-property-2}\\
        \bm{\chi}'\,\MAT{C}'&
            =(-1)^{n}\;\MAT{C}'\,\bm{\chi}'.
            \label{eq:C-property-3}
    \end{align}
Notice that the behavior of $\MAT{C}'$ is periodic with $n$ and depends only on $(n \bmod 4)$, in accordance with the well-known Bott periodicity of $(N\bmod 8)$ for $\SO(N)$ familiar from textbooks, see e.g.~\cite{Georgi:1999wka}. Charge conjugation anticommutes with Cartan generators and thus flips all quantum numbers (aka weight vector) of a basis state. It also commutes (anticommutes) with the chiral element for even (odd) $n$, and thus preserves (flips) the chirality in such a case. 

The listed properties of $\bm{\Gamma}'_{I}$, $\bm{\chi}'$ and $\MAT{C}'$ imply the existence of a basis change in spinor space $\mathcal{S}$ via a signed permutation matrix $\MAT{S}$, under which $\MAT{X}'\mapsto \MAT{X}=\MAT{S}\MAT{X}'\MAT{S}^\top$, such that all Clifford objects related to $\SO(2n)$ are simultaneously brought into their canonical forms:
    \begin{align}
        \bm{\Gamma}_{I}&=
            \begin{pmatrix}
                0 & \bm{\gamma}_{I}\\
                \overline{\bm{\gamma}}_{I} & 0\\
            \end{pmatrix}, &
        \bm{\chi} &=
            \begin{pmatrix}
                -\mathbb{1}_{2^{n-1}} & 0 \\
                0 & \mathbb{1}_{2^{n-1}} \\
            \end{pmatrix},
        \label{eq:canonical-gamma-chi}
    \end{align}
and for $\MAT{C}_{(n\bmod 4)}$ the form\footnote{
    A quick reference: $\SO(4)\mapsto \MAT{C}_{2}$, $\SO(6)\mapsto \MAT{C}_{3}$, $\SO(8)\mapsto \MAT{C}_0$, $\SO(10)\mapsto \MAT{C}_{1}$.
}
    \begin{align}
        \MAT{C}_{0} &=
            \left(\begin{smallmatrix}
                0 & \mathbb{1}_{2^{n-2}} & & \\
                \mathbb{1}_{2^{n-2}} & 0 & & \\
                && 0 & \mathbb{1}_{2^{n-2}}  \\
                && \mathbb{1}_{2^{n-2}} & 0   \\
            \end{smallmatrix}\right), &
        \MAT{C}_{1} &=
            \begin{pmatrix}
                0 & \mathbb{1}_{2^{n-1}} \\
                -\mathbb{1}_{2^{n-1}} & 0 \\
            \end{pmatrix}, 
            \label{eq:canonical-C-1}\\[3pt]
        \MAT{C}_{2} &=
            \left(\begin{smallmatrix}
                0 & \mathbb{1}_{2^{n-2}} & & \\
                -\mathbb{1}_{2^{n-2}} & 0 & & \\
                && 0 & \mathbb{1}_{2^{n-2}}  \\
                && -\mathbb{1}_{2^{n-2}} & 0   \\
            \end{smallmatrix}\right), &
        \MAT{C}_{3} &=
            \begin{pmatrix}
                0 & \mathbb{1}_{2^{n-1}} \\
                \mathbb{1}_{2^{n-1}} & 0 \\
            \end{pmatrix}.  
            \label{eq:canonical-C-2}
    \end{align}
Since a real orthogonal basis change retains properties such as hermiticity, reality, symmetry and antisymmetry of a matrix, the $\bm{\Gamma}_{I}$ remain Hermitian, and hence the $2^{n-1}\times 2^{n-1}$ off-diagonal blocks are related via $\overline{\bm{\gamma}}_{I}=\bm{\gamma}_{I}^{\dagger}$. 

The basis change $\MAT{S}$ can be performed in two steps, the details of which prove illuminating for understanding the underlying structures:
    \begin{enumerate}[leftmargin=0.3cm,itemsep=0pt,label=(\roman*)]
        \item \label{item:reordering1}  
            \textit{A reordering into chiral blocks:}\\
                As already noted earlier, the $\bm{\chi}'$ of Eq.~\eqref{eq:chiral-element-explicitly} is already diagonal and has an equal number of $+1$ and $-1$ eigenvalues, implying a split of spinor space $\mathcal{S}$ into two eigenspaces $\mathcal{S}_{\pm}$, i.e.~$\mathcal{S}=\mathcal{S}_{-}\oplus\mathcal{S}_{+}$. The subspaces can be constructed via the projection operators of Eq.~\eqref{eq:projection-operator} as $\mathcal{S}_{\pm}=\MAT{P}_{\pm}\mathcal{S}$.
                We sort the original basis vectors in accordance with their $\bm{\chi}$-eigenvalue, obtaining the canonical form $\bm{\chi}$ in Eq.~\eqref{eq:canonical-gamma-chi}. Since $\bm{\Gamma}_{I}$ anticommute with $\bm{\chi}$, cf.~Eq.~\eqref{eq:chi-property-1}, they flip chirality. They must thus be block off-diagonal in the new basis, i.e.~they are already in their canonical form of Eq.~\eqref{eq:canonical-gamma-chi}, with mappings $\bm{\gamma}_I: \mathcal{S}_{+}\to\mathcal{S}_{-}$ and $\overline{\bm{\gamma}}_I: \mathcal{S}_{-}\to\mathcal{S}_{+}$.
        \item \label{item:reordering2} 
            \textit{A signed reordering within chiral blocks:}\\
            We shall now perform a further signed reordering of the basis within $\mathcal{S}_{-}$ and $\mathcal{S}_{+}$, which will bring $\MAT{C}$ to its canonical form. Such a reordering always retains the already achieved block form of $\bm{\gamma}_I$ and exact form of $\bm{\chi}$ in Eq.~\eqref{eq:canonical-gamma-chi}.
            \par
            Signed reorderings retain the form of $\MAT{C}$ as a signed permutation matrix; also, they are orthogonal transformations, and thus the properties in Eqs.~\eqref{eq:C-property-1}--\eqref{eq:C-property-3} are retained. We investigate these now
            in more detail. The property of Eq.~\eqref{eq:C-property-2} 
            implies that $\MAT{C}$ flips between basis states of $\mathcal{S}$ with opposite quantum numbers (each one has a unique weight vector). The action of $\MAT{C}$ restricted to such a 2-dimensional subspace, being a signed permutation, thus implies one of four possible forms: symmetric 
                $\pm \bm{\tau}_{1}=\pm\left(\begin{smallmatrix}
                    0&1\\
                    1&0\\
                \end{smallmatrix}\right)$ or antisymmetric 
                $\pm\epsilon_{2}=\pm\left(\begin{smallmatrix}
                    0&1\\
                    -1&0\\
                \end{smallmatrix}\right)$, where Eq.~\eqref{eq:C-property-3}
            specifies whether $\MAT{C}$ is symmetric ($n\bmod 4=0,3$) or antisymmetric ($n\bmod 4=1,2$). We can always obtain the canonical $+$ versions of the $2\times 2$ blocks by a sign redefinition of one of the two states. Furthermore, the property of Eq.~\eqref{eq:C-property-3} implies $\MAT{C}$ retains the chirality
            of a basis state for even $n$ and flips it for odd $n$, implying 
            that after the reordering of item~\ref{item:reordering1} into the split $\mathcal{S}_{-}\oplus\mathcal{S}_{+}$, the matrix $\MAT{C}$ is already block diagonal for even $n$ and block off-diagonal
            for odd $n$. 
            
            The above properties straightforwardly lead to the canonical forms of $\MAT{C}_{n\bmod 4}$ in Eqs.~\eqref{eq:canonical-C-1} and \eqref{eq:canonical-C-2}. The final reorderings within the chiral blocks are as follows. For odd $n$, one can choose an arbitrary order of basis elements in $\mathcal{S}_{-}$, but then take a matching order in $\mathcal{S}_{+}$ with respect to flipping the quantum numbers. For even $n$, one can reorder within $\mathcal{S}_{-}$ and $\mathcal{S}_{+}$ independently; in each of these chiral blocks, one splits the basis states into two sets, such that each set receives one of a pair of states with opposite quantum numbers, and then ensures the two sets have matching order.
    \end{enumerate}

\noindent
We now conclude with a few final observations and remarks: 
\begin{itemize}[itemsep=-0.1cm,leftmargin=\DIMLEFTMARGIN]
    \item 
        Taking the form of $\bm{\Gamma}_I$ from Eq.~\eqref{eq:canonical-gamma-chi}, and inserting it into the definition of generators $\MAT{T}_{IJ}$ in Eq.~\eqref{eq:definition-gen-spinor}, the generators take the canonical form
        \begin{align}
            \MAT{T}_{IJ}&= \frac{i}{4}
                \begin{pmatrix}
                    \bm{\gamma}_{I}\bm{\gamma}^\dagger_{J}-\bm{\gamma}_{J}\bm{\gamma}_{I}^{\dagger}&0 \\
                    0& \bm{\gamma}_{I}^{\dagger} \bm{\gamma}_{J}-\bm{\gamma}_{J}^\dagger \bm{\gamma}_{I}
                \end{pmatrix}
            \equiv
                \begin{pmatrix}
                    \MAT{t}_{IJ}& 0\\
                    0&\MAT{\bar{t}}_{IJ}\\
            \end{pmatrix}. \label{eq:canonical-generator}
        \end{align}
        Since $\MAT{T}_{IJ}$ are block diagonal, the spinor space $\mathcal{S}=\mathcal{S}_{-}\oplus\mathcal{S}_{+}$ is a split of a reducible representation. 
    \item
        Inserting the forms of $\MAT{C}_{1}$ and $\MAT{C}_{3}$ into the implicit definition of $\MAT{C}$ in Eq.~\eqref{eq:C-defining-property}, it immediately follows that $\bm{\gamma}_{I}$ is symmetric for $n\bmod 4=1$ (e.g.~for $\SO(10)$) and antisymmetric for $n\bmod 4=3$ (e.g.~for $\SO(6)$). These symmetry properties lead in both cases $\MAT{\bar{t}}_{IJ}=-(\MAT{t}_{IJ})^\top$ in Eq.~\eqref{eq:canonical-generator}, implying that the spinor representation on $\mathcal{S}_{+}$ is conjugate to that on $\mathcal{S}_{-}$ for odd $n$. 
    \item  
        For even $n$ the $\bm{\gamma}_{I}$ have mixed symmetry. The structure of the representations $\mathcal{S}_{+}$ and $\mathcal{S}_{-}$, however, is revealed by considering the map $\rho(v)=\MAT{C}^{-1}\,v^\ast$ for $v\in\mathcal{S}$. It is block diagonal, so maps $\mathcal{S}_{+}\to\mathcal{S}_{+}$ and $\mathcal{S}_{-}\to\mathcal{S}_{-}$. Furthermore it is antilinear, with $\rho^{2}=+\mathrm{Id}$ (real structure, cf.~item~\ref{item:real-structure} in Section~\ref{subsec:group-theory}) for $\MAT{C}_{0}$ and $\rho^{2}=-\mathrm{Id}$ (pseudoreal structure) for $\MAT{C}_{2}$. Finally, it is $\SO(2n)$-equivariant: Eq.~\eqref{eq:C-defining-property} for even $n$ implies $\MAT{C}\MAT{T}_{IJ}\MAT{C}^{-1}=-\MAT{T}_{IJ}^{\top}$ and under the action of a group element $\MAT{X}=e^{\tfrac{i}{2}\alpha_{IJ} \MAT{T}_{IJ}}$ for real parameters $\alpha_{[IJ]}$ we have
        \begin{align}
            \rho(\MAT{X}v)
                =\MAT{C}^{-1}(\MAT{X}v)^*  
                = \MAT{C}^{-1} e^{-\tfrac{i}{2}\alpha_{IJ} \MAT{T}_{IJ}^\top} v^*
                = \MAT{C}^{-1} e^{\MAT{C} \alpha_{IJ}\MAT{T}_{IJ} \MAT{C}^{-1}} v^*
                = \MAT{X}\MAT{C}^{-1}v^*
                =\MAT{X}\,\rho(v).
        \end{align}
        \par 
        The above considerations show that representations $\mathcal{S}_{\pm}$ admit a $\SO(2n)$-equivariant real structure for $n\bmod 4=0$ and a pseudoreal structure for $n\bmod 4=2$, and are hence irreps of real and pseudoreal type, respectively.
    \item 
        So far we used the bold font to denote matrices in spinor space. We now consider how to translate these into index notation.

        Suppose we have an element of spinor space $v\in\mathcal{S}$, and denote its components by $v^X$, where $X=1\ldots 2^n$; we use upper capital indices starting with the label $X$. The spinor space is a complex vector space, so we use lower indices for the conjugate basis. This notation is borrowed from Table~\ref{tab:indices} and the case of $\SO(10)$, but note that the formalism applies completely generally. 
        
        Since the gamma matrices are linear maps $\mathcal{S}\to\mathcal{S}$, they transform components to components, implying an upper-lower index structure: $\bm{\Gamma}_{I}\to \Gamma_{I}{}^{X}{}_{Y}$ (we use an arrow to indicate the translation from matrix to index notation). Any object, which is structurally a product of gamma matrices (as per the abstract definitions of Appendix~\ref{app:spinors-Clifford}), retains the same upper-lower index structure. This applies to generators, the chiral element and the projection matrix: $\MAT{T}_{IJ}\to T_{IJ}{}^{X}{}_{Y}$, $\bm{\chi}\to \chi^{X}{}_{Y}$, and $\MAT{P}_{\pm}\to (P_{\pm})^{X}{}_{Y}$. The exception is the charge conjugation matrix, which is structurally defined via Eq.~\eqref{eq:C-defining-property}, implying the structure $\MAT{C}\to C_{XY}$ and $\MAT{C}^{-1} \to C^{XY}$, such that $C^{XY}C_{YZ}=\delta^{X}{}_{Z}$. 
        Note: although Eq.~\eqref{eq:charge-BW-result1} expresses $\MAT{C}$ as a product of even-indexed gamma matrices, it is valid only in a particular basis (the construction from Eqs.~\eqref{eq:BW1} and \eqref{eq:BW2}), and thus cannot be used to infer the index structure of $\MAT{C}$.
    \item 
    While we use the same construction as Ref.~\cite{Wilczek:1981iz} for $\bm{\Gamma}'_{I}$ in Eqs.~\eqref{eq:BW1}--\eqref{eq:BW2}, our convention for  the chiral element $\bm{\chi}$ differs already at the abstract level of Eq.~\eqref{eq:chi-definition} by a factor $(-1)^{n}$.

    The reason for our convention is that for a state $|s_{1}\ldots s_n\rangle$ in spinor space, the chirality eigenvalue becomes $\chi=\prod_{i=1}^{n} s_{i}$, as we already argued below Eq.~\eqref{eq:cartan-explicitly} --- this is the usual convention.
    Ref.~\cite{Wilczek:1981iz} claims the same expression, but in their conventions one in fact obtains $\chi=(-1)^{n}\prod_{i=1}^{n} s_{i}$, as can be easily derived from their equations (A16)--(A18). In~\cite{Aulakh:2002zr}, this issue is circumvented by using the convention for $\bm{\chi}$ from \cite{Wilczek:1981iz}, but also shifting the definition of generators by $-1$, cf.~our Eq.~\eqref{eq:definition-gen-spinor} and their Eq.~(69), thus flipping all $s_{i}$ and recovering the usual expression.

    As for $\MAT{C}$, our definition and result also matches that of~\cite{Aulakh:2002zr}, and we agree with their comment that the explicit form reported in (A19) of Ref.~\cite{Wilczek:1981iz} is not correct.
\end{itemize}

\noindent
This concludes our self-contained description of the general theory; the conclusions and constructions are utilized in the specific cases of $\SO(4)$, $\SO(6)$ and $\SO(10)$, cf.~Section~\ref{subsec:group-theory}.

\section{Technical details of the $\SO(10)$ calculation \label{app:basis}}

We provide here some of the technical information regarding our explicit $\SO(10)$ calculation of Section~\ref{subsec:result-SO10}.

Our goal is to explicitly compute the terms in Eq.~\eqref{eq:Yukawa-Lagrangian-SO10}. That calculation is best performed in the complex basis for the fundamental $\SO(10)$ indices, cf.~item~\ref{item:complex-basis} in Section~\ref{subsec:group-theory}, since every entry of the complex basis has well-defined quantum numbers. 

The main information of this section of the Appendix is how the tensors $\GIRREP{10}^{i}$, $\GIRREP{120}^{ijk}$ and $\GIRREP{126}^{ijklm}$ for the scalar irreps are populated in the complex basis. For this information to be meaningful, we first fix the order of fields in the complex basis: schematically we order them via $\SO(10)\to\SU(5)$ decomposition as $\GIRREP{10}=(\GIRREP{5},\GIRREP{\overline{5}})$, where the order within the $\mathbf{5}$ is specified by the $\SU(5)\to\SU(3)_C\times\SU(2)_L$ decomposition $\GIRREP{5}=(\GIRREP{3}_{\SU(3)},\GIRREP{2}_{\SU(2)})$, with analogous order within $\GIRREP{\overline{5}}$. This information is also reflected in listing the quantum numbers for each state in Table~\ref{tab:qn10}.

\begin{table}[htb]
\centering
\begin{tabular}{r@{$\qquad$}rrrrr}
    \toprule
    $i$& \LISTQ{t^C_{3}}{t^C_{8}}{t^L_{3}}{Y/2}{X}\\
    \midrule
    $1$ & \LISTQ{+\tfrac{1}{2}}{+\tfrac{1}{2\sqrt{3}}}{0}{-\tfrac{1}{3}}{+2}\\
    $2$ & \LISTQ{-\tfrac{1}{2}}{+\tfrac{1}{2\sqrt{3}}}{0}{-\tfrac{1}{3}}{+2}\\
    $3$ & \LISTQ{0}{-\tfrac{1}{\sqrt{3}}}{0}{-\tfrac{1}{3}}{+2}\\
    $4$ & \LISTQ{0}{0}{+\tfrac{1}{2}}{+\tfrac{1}{2}}{+2}\\
    $5$ & \LISTQ{0}{0}{-\tfrac{1}{2}}{+\tfrac{1}{2}}{+2}\\[6pt]
    $6$ & \LISTQ{-\tfrac{1}{2}}{-\tfrac{1}{2\sqrt{3}}}{0}{+\tfrac{1}{3}}{-2}\\
    $7$ & \LISTQ{+\tfrac{1}{2}}{-\tfrac{1}{2\sqrt{3}}}{0}{+\tfrac{1}{3}}{-2}\\
    $8$ & \LISTQ{0}{+\tfrac{1}{\sqrt{3}}}{0}{+\tfrac{1}{3}}{-2}\\
    $9$ & \LISTQ{0}{0}{-\tfrac{1}{2}}{-\tfrac{1}{2}}{-2}\\
    $10$ & \LISTQ{0}{0}{+\tfrac{1}{2}}{-\tfrac{1}{2}}{-2}\\
    \bottomrule
\end{tabular}
\caption{The quantum numbers for each component $i$ in $\GIRREP{10}^{i}$ in the complex basis. The group $\SO(10)$ has rank $5$, hence that many quantum numbers: $\{t_{3}^{C},t_{8}^{C}\}$ are diagonal generators of $\SU(3)_C$, $t_{3}^L$ is the diagonal generator of $\SU(2)_L$, $Y/2$ is the hypercharge, while $\mathrm{U}(1)_X$ is the remaining charge orthogonal to the other four, i.e.~the one in $\SU(5)\times\mathrm{U}(1)_{X}\subset\SO(10)$. \label{tab:qn10}}
\end{table}

Let us label the vector of all 
five quantum numbers for the state $i$ by $\vec{q}^{i}$. It is a basic result of group theory that the quantum numbers for a tensor entry is computed by adding the vectors of quantum numbers related to each upper index, and subtracting for each lower index. The vector of quantum numbers for the entry $\GIRREP{120}^{ijk}$ is thus computed as $\vec{q}^{i}+\vec{q}^{j}+\vec{q}^{k}$, and for $\GIRREP{\overline{126}}^{ijklm}$ as $\vec{q}^{i}+\vec{q}^{j}+\vec{q}^{k}+\vec{q}^{l}+\vec{q}^{m}$.

If we know the vector $\vec{q}$ for a given field in a scalar irrep, we can thus immediately obtain candidates for tensor entries that match its quantum numbers. The number of candidates is usually larger than one, since other degrees of freedom in the irrep may have the same quantum numbers (but they transform differently under the generators), and some tensor entry combinations may not be part of the irrep at all. We remove the unwanted degrees of freedom
by imposing any constraints the irrep tensor may satisfy (such as antisymmetry, tracelessness, or self-duality), and impose a sufficient number of relations involving group generators to isolate the desired degree of freedom.

We apply this procedure to finding the neutral component of Higgs doublets $\SMIRREP{1}{2}{\pm\tfrac{1}{2}}$,\footnote{
    The $u$-type VEVs are located in the SM irrep $\SMIRREP{1}{2}{+\tfrac{1}{2}}$, while the $d$-type VEVs are in $\SMIRREP{1}{2}{-\tfrac{1}{2}}$.
} so that we can insert into those entries the EW VEV. The quantum numbers of this entry will be of $\vec{q}=(0,0,\pm\tfrac{1}{2},\mp\tfrac{1}{2},x)$ for an arbitrary $x$ (that ultimately depends on the irrep; knowing it can help in tracking down relevant entries faster). Since $Q=t_{3}^L+Y/2$, the VEV with EM charge $Q=0$ must have a vanishing sum of the 3rd and 4th entry. The irrep constraint on $\GIRREP{120}^{ijk}$ is its antisymmetry in all indices, while $\overline{126}^{ijklm}$ must be completely antisymmetric and anti-selfdual (complex conjugate of the relation in Eq.~\eqref{eq:126-selfduality}).
Furthermore, one ensures that state is part of a color singlet by demanding that the action of every $\SU(3)_C$ generator destroys the state, while it being a  doublet of $\SU(2)$ can be determined by repeatedly applying the raising/lowering generators $t^L_{\pm}$. These SM generators have obvious explicit forms inferred from the placement of color (anti)triplets and weak doublets in the basis given by Table~\ref{tab:qn10}; more generally, all $\SO(10)$ generators in the irrep $\GIRREP{10}$ can be constructed with the help of methods described e.g.~in Appendix~A of \cite{Antusch:2019avd}.

In the $\GIRREP{120}$ of $\SO(10)$, there are two $u$- and two $d$-type VEVs. In accordance with our PS-based calculation, we choose a basis of states in each 2D space which have well-defined transformation properties under the PS group $4_C\,2_L\,2_R$: one state is part of $\PSIRREP{1}{2}{2}$ and one of the $\PSIRREP{15}{2}{2}$ irrep. We can determine which is which by applying generators from $\SU(4)_C$ and noting that their action will always destroy the $\PSIRREP{1}{2}{2}$ VEV, while the $\PSIRREP{15}{2}{2}$ VEV is the $B-L$ direction within $\SU(4)_C$ and must be orthogonal in the sense that there are no mixed terms in Eq.~\eqref{eq:normalization-120}.

Going through the procedure described above for all the scalar irreps, we obtain the following results:
\begin{itemize}
    \item The VEV entries in a complex $\GIRREP{10}^{[i]}$ are
        \begin{align}
            [5]&=\tfrac{1}{\sqrt{2}}\;\VEVEW{10}{u}{1}, &
            [10]&=\tfrac{1}{\sqrt{2}}\;\VEVEW{10}{d}{1}, \label{eq:EWVEVs-10}
        \end{align}
        Imposing the reality condition of Eq.~\eqref{eq:reality-condition-10-c} gives the relation
        \begin{align}
            \VEVEW{10}{u}{1}{}^*&=\VEVEW{10}{d}{1},
        \end{align}
        and the VEV normalization is consistent with Eq.~\eqref{eq:normalization-10} once the reality condition is imposed.
    \item
        The VEV entries $[i,j,k]$ in a complex irrep $\GIRREP{120}^{[ijk]}$, populated in accordance with antisymmetry under index exchange, are 
        \begin{align}
                [4,5,9]&= \tfrac{1}{\sqrt{2}}\;\VEVEW{120}{u}{1}, \label{eq:EWVEVs-120-1u}\\
                [9,10,4]&= \tfrac{1}{\sqrt{2}}\;\VEVEW{120}{d}{1}, \label{eq:EWVEVs-120-1d}\\
                [1,5,6]=[2,5,7]=[3,5,8]&=-\tfrac{1}{\sqrt{6}}\;\;\VEVEW{120}{u}{15}, \label{eq:EWVEVs-120-15u}\\
                [6,10,1]=[7,10,2]=[8,10,3]&=-\tfrac{1}{\sqrt{6}}\;\;\VEVEW{120}{d}{15}.\label{eq:EWVEVs-120-15d}
        \end{align}
        The above sign conventions lead to the simplest/standard form of $\MAT{Y}_{U}$ in Eq.~\eqref{eq:result-MU}. Imposing the reality condition of Eq.~\eqref{eq:reality-condition-120-rc} gives the VEV relations\footnote{
            A remark to avoid confusion: Eq.~\eqref{eq:VEV-reality-condition-so10} represents the reality conditions for the VEVs in the $\GIRREP{120}$ of $\SO(10)$, using their labels defined by Eq.~\eqref{eq:EWVEVs-120-1u}--\eqref{eq:EWVEVs-120-15d}. No minus sign appears in Eq.~\eqref{eq:VEV-reality-condition-so10}, but instead arises when explicitly computing $\SO(10)$ invariants, see Section~\ref{subsec:result-SO10}. Whenever we refer to a non-trivial minus sign in reality conditions, e.g.~in the Introduction and in Sections~\ref{subsec:result-PS} and \ref{subsec:result-SO10-PS},  we instead mean conditions obtained in pure Pati-Salam calculations and the associated VEV-labeling therein.
        }
        \begin{align}
            \VEVEW{120}{u}{1}{}^*&=\VEVEW{120}{d}{1}, &
            \VEVEW{120}{u}{15}{}^*&=\VEVEW{120}{d}{15}. \label{eq:VEV-reality-condition-so10}
        \end{align}
These conditions can be confirmed by noting that Eq.~\eqref{eq:EWVEVs-120-1u} becomes \eqref{eq:EWVEVs-120-1d} by change of label $u\mapsto d$ and change of every index value (retaining its relative placement in the $[i,j,k]$ label) by $l\mapsto (l+5)\mod 10$, where the identification $0\equiv 10$ is assumed. The same transformations turn Eq.~\eqref{eq:EWVEVs-120-15u} into \eqref{eq:EWVEVs-120-15d}. The normalization of all VEVs is consistent with Eq.~\eqref{eq:normalization-120} after imposing the reality condition.
        \item
            The VEV entries $[i,j,k,l,m]$ in a complex irrep $\overline{126}^{[ijklm]}$ are as follows:
            \begin{align}
                [1,2,5,6,7]=
                    [1,3,5,6,8]=
                    [1,4,5,6,9]&=  \\\nonumber
                [2,3,5,7,8]= 
                    [2,4,5,7,9]=
                    [3,4,5,8,9] &=  \tfrac{1}{\sqrt{6}}\;\VEVEW{\overline{126}}{u}{15},\\[3pt]
                [1,2,6,7,10]=
                    [1,3,6,8,10] =
                    [1,4,6,10,9] &= \\\nonumber
                [2,3,7,8,10]= 
                    [2,4,7,10,9] =
                    [3,4,8,10,9] &=  \tfrac{1}{\sqrt{6}}\;\VEVEW{\overline{126}}{d}{15},\\[3pt]
                [1,2,3,4,5]&=-\tfrac{1}{\sqrt{2}}\;\VEV{\sigma}
            \end{align}
        We included in this list the $(B-L)$-breaking VEV $\VEV{\sigma}$. Note that the irrep $\GIRREP{\overline{126}}$ is complex, so no reality condition is (or can be) imposed. The irrep must, however, be anti-selfdual. A remnant of this property can be seen by comparing the entries associated to its EW VEVs $\VEVEW{\overline{126}}{u}{15}$ and $\VEVEW{\overline{126}}{d}{15}$: comparing the populated entries under $u\mapsto d$, we again see that they satisfy the rule of transforming every index via the $l\mapsto (l+5)\mod 10$ rule, but the index order can now be permuted. 
        After restoring the chosen index ordering, some entries acquire an additional minus sign, as required by anti-self-duality.
        Finally, the VEVs have been normalized in accordance with Eq.~\eqref{eq:normalization-126}.
\end{itemize}

Once a basis for the irrep $\GIRREP{10}$ and the VEV normalization for each scalar irrep is fixed, the above results for the entries become unique up to a sign convention for the VEV, i.e.~one can reparametrize any VEV $v$ by taking $v=-v'$. This sign convention is irrelevant from the point of view of the final result (since it just reparametrizes it), but it will provide crucial information for an alternative third derivation of the reality condition on the $\GIRREP{120}$ that is explored in Appendix~\ref{subapp:result-alternative-su5xu1}.

\section{More alternative derivations of the reality conditions/Yukawa sector} \label{app:alternatives}

Our main claim of the paper is the appearance of a relative minus sign between the Yukawa contributions of $\PSIRREP{1}{2}{2}$ and $\PSIRREP{15}{2}{2}$ in the down-sector when originating from a real irrep $\GIRREP{120}$ of $\SO(10)$. We have provided two independent derivations of this claim in Sections~\ref{subsec:result-SO10} and \ref{subsec:result-SO10-PS}. Since both involve some technical tedium, presenting alternative and independent derivations can prove insightful; this is the purpose of Appendix~\ref{app:alternatives}. 

In Appendix~\ref{subapp:result-alternative-su5xu1}, we revisit the calculation of the $\GIRREP{16}\cdot\GIRREP{16}\cdot\GIRREP{120}_{\mathbb{R}}$ invariant using the language of the $\SU(5)$-subgroup rather than PS and implement the oscillator method. Conversely, Appendix~\ref{app:rtransport} introduces a novel algorithm for determining how an embedding transports the real structure from parent to daughter, which we apply to our $4_C\,2_L\,2_R\hookrightarrow\SO(10)$ case.

\subsection{An $\SU(5)\times\mathrm{U}(1)$ calculation \label{subapp:result-alternative-su5xu1}}

In the main derivations of Section~\ref{sec:reality-conditions} we used a description of VEVs in terms of the Pati-Salam maximal subgroup of $\SO(10)$. We now cross-check the result using a description based on the $\SU(5)\times\mathrm{U}(1)$ subgroup of $\SO(10)$. This alternative choice turns out to manifest surprisingly different issues in the derivation. 

Before proceeding to the calculation, we illuminate the conceptual overview of what a derivation involves by 
considering the diagram of mappings in Figure~\ref{fig:map-diagram}. The map $i_{422}$ is the embedding of the PS group $4_C\,2_L\,2_R$ into $\SO(10)$, $i_{51}$ is the embedding map of the group $\SU(5)\times\mathrm{U}(1)_X \equiv 5\,1_X$ into $\SO(10)$, while $\mathrm{Id}$ denotes the identity map on $\SO(10)$ representing a basis-change from states adapted to the PS subgroup to those adapted to $5\,1_X$ (colloquially we can refer to this as a change from PS to $\SU(5)$ language). All these maps are Lie group homomorphisms. The embedding maps can be defined by specifying explicitly how the defining (fundamental) irrep $\GIRREP{10}$ of $\SO(10)$ decomposes into subgroup irreps. The embedding map $i_{422}$ of PS, for example, has been effectively specified in Section~\ref{subsec:result-SO10-PS} by identifying the $\PSIRREP{1}{2}{2}$ and $\PSIRREP{\bar{6}}{1}{1}$ of PS in the $\GIRREP{10}$ of $\SO(10)$.

\begin{figure}[htb]
\begin{center}
\[
\begin{tikzcd}
    \makebox[3.5cm][r]{$\SU(4)_C\times\SU(2)_L\times\SU(2)_R$} 
    \arrow[r, hook, "i_{422}"] & \makebox[1.5cm][c]{$\SO(10)$} \arrow[d, "\mathrm{Id}"] \\ 
    \makebox[3.5cm][r]{$\SU(5)\times\mathrm{U}(1)_{X}$}
    \arrow[r, hook, "i_{51}"] & \makebox[1.5cm][c]{$\SO(10)$}
\end{tikzcd}
\]
\caption{A diagram with embedding maps of the Pati-Salam and $5\,1_X$ subgroups into $\SO(10)$; it is intended as visual support for conceptual discussions in the main text. \label{fig:map-diagram}}
\end{center}
\end{figure}

The main derivations vis-a-vis the diagram manifest themselves as follows:
\begin{enumerate}[itemsep=-0.1cm,leftmargin=\DIMLEFTMARGIN,label=$\lfloor$\arabic*$\rfloor$ ]
    \item \label{item:strategy1} Our first derivation, see Section~\ref{subsec:result-SO10}, performs an explicit $\SO(10)$ calculation using PS language, i.e.~it is performed in the codomain of the map $i_{422}$. 
    \item \label{item:strategy2} Our second derivation performs a simpler calculation in the Pati-Salam theory, see Section~\ref{subsec:result-PS},
    then ports the missing $\SO(10)$ information (in that case the reality condition) via the embedding $i_{422}$ in Section~\ref{subsec:result-SO10-PS}. In other words, the computation is performed in the domain of $i_{422}$, with the missing $\SO(10)$ information then supplemented via the map $i_{422}$. 
\end{enumerate}

In the case of the maximal subgroup $5\,1_X$, applying the analog of strategy~\ref{item:strategy1} would simply repeat the $\SO(10)$ calculation of Section~\ref{subsec:result-SO10} in a different basis (codomain of $i_{51}$). This alone, however, would not constitute an independent cross-check: a possible coding mistake in the implementation of $\SO(10)$ gamma matrices, for example, would remain. Therefore the analog of strategy~\ref{item:strategy2} is more promising as a cross-check: perform the calculation assuming $5\,1_X$ symmetry, supply the missing $\SO(10)$ information from the map $i_{51}$ to effectively obtain the result in the codomain of $i_{51}$, then translate the result to PS language using $\mathrm{Id}^{-1}$ for comparison with the main sections.

We start by writing the relevant irrep decompositions under $\SO(10)\to\SU(5)\times\mathrm{U}(1)_{X}$, cf.~\cite{Slansky:1981yr}:
\begin{align}
    \GIRREP{16} &= 
        \SUUIRREP{1}{-5} \oplus 
        \SUUIRREP{\overline{5}}{+3} \oplus
        \SUUIRREP{10}{-1},\\
    \GIRREP{120} &= 
        \SUUIRREP{5}{+2} \oplus 
        \SUUIRREP{\overline{5}}{-2} \oplus
        \SUUIRREP{10}{-6} \oplus
        \SUUIRREP{\overline{10}}{+6} \oplus
        \SUUIRREP{45}{+2} \oplus 
        \SUUIRREP{\overline{45}}{-2}.
\end{align}
The $u$-type doublets are located in the irreps $\SUUIRREP{5}{+2}$ and $\SUUIRREP{45}{+2}$, their VEVs respectively labeled by $\VEVEW{120}{u}{5}$ and $\VEVEW{120}{u}{45}$, while the $d$-type doublets are in the irreps $\SUUIRREP{\overline{5}}{-2}$ and $\SUUIRREP{\overline{45}}{-2}$, with VEVs  labeled as $\VEVEW{120}{d}{\overline{5}}$ and $\VEVEW{120}{d}{\overline{45}}$.

Using $\SU(5)$ index notation with labels $\{x,y,\ldots\}$, cf.~Table~\ref{tab:indices}, and indices $\{a,b\}$ as generation indices, we can write the relevant irreps as follows.
The fermions are written\footnote{
    Regarding notation: $a$-th generation fermion irreps are labeled by $M_{a}$ and scalar irreps by $h$; $\SU(5)$ index placement distinguishes the irrep type. Notation has been adopted from \cite{Nath:2001uw}, whose results will be relevant later in this section.
}  
as
\begin{align}
    \SUUIRREP{1}{-5}&\sim M_{a0} = (\nu^{c})_a, &
    \SUUIRREP{\overline{5}}{+3}&\sim M_{ax} =
        \begin{pmatrix}
            d^c_1\\d^c_2\\d^c_3\\e\\ -\nu
        \end{pmatrix}_a, &
    \SUUIRREP{10}{-1} & \sim M_{a}{}^{xy} =
        \begin{pmatrix}
            0&u^c_3&-u^c_2&u_1&d_1\\
            -u^c_3&0&u^c_1&u_2&d_2\\
            u^c_2&-u^c_1&0&u_3&d_3\\
            -u_1&-u_2&-u_3&0&e^c\\
            -d_1&-d_2&-d_3&-e^c&0
        \end{pmatrix}_a, \label{eq:su5-irreps-fermions}
\end{align}
while the scalar Higgses with their VEVs are written as 
\begin{align}
    \SUUIRREP{5}{+2} &\sim \langle h^{x} \rangle = (0,0,0,0,\VEVEW{120}{u}{5}), \label{eq:su5-scalar-1}\\
    \SUUIRREP{\overline{5}}{-2} &\sim \langle h_{x} \rangle = (0,0,0,0,\VEVEW{120}{d}{\overline{5}}), \\
    \SUUIRREP{45}{+2}&\sim \langle h^{xy}{}_{z}\rangle = -\tfrac{1}{2\sqrt{3}}\;\VEVEW{120}{u}{45}\,
        (\delta^{x5}\,\mathrm{diag}[1,1,1,-3,0]^{y}{}_{z}-\delta^{y5}\,\mathrm{diag}[1,1,1,-3,0]^{x}{}_{z}),\\
    \SUUIRREP{\overline{45}}{-2}&\sim \langle h^{z}{}_{xy}\rangle = -\tfrac{1}{2\sqrt{3}}\, \VEVEW{120}{d}{\overline{45}}\,
        (\delta_{x5}\,\mathrm{diag}[1,1,1,-3,0]_{y}{}^{z}-\delta_{y5}\,\mathrm{diag}[1,1,1,-3,0]_{x}{}^{z}),
        \label{eq:su5-scalar-4}
\end{align}
where $\delta$ is the Kronecker delta symbol.
Note that the $\GIRREP{45}$ of $\SU(5)$ is antisymmetric in its upper indices $h^{[xy]}{}_{z}$, as well as subject to the tracelessness condition $h^{xy}{}_{y}=0$. Analogous conditions hold for $\GIRREP{\overline{45}}\sim h^{z}{}_{xy}{}$. Note that Eq.~\eqref{eq:su5-irreps-fermions}--\eqref{eq:su5-scalar-4} fix both the sign and normalization convention of the fields/VEVs, where the latter has been deliberately chosen for convenient use later in the $\SO(10)$ oscillator method. 

The $5\,1_X$-invariant Lagrangian terms involving EW VEVs arising from the $\SO(10)$ invariant $\GIRREP{16}_F\cdot \GIRREP{16}_F\cdot \GIRREP{120}$ are then the following:
\begin{align}
    \mathcal{L}_{Y}^{\SU(5)} &=
        Y^{1}_{ab}\; M_{a0}\,M_{bx}\,h^{x}
        +Y^{5}_{ab}\; M_{a}{}^{xy}\,M_{bx}\,h_{y}
        + Z^{5}_{ab}\; M_{ax}\,M_{b}{}^{yz}\,h^{x}{}_{yz}
        + Z^{10}_{ab}\; M_{a}{}^{xy}\,M_{b}{}^{uv}\,h^{zt}{}_{v}\,\epsilon_{xyztu} + h.c.,
        \label{eq:su5-Lagrangian}
\end{align}
where $\epsilon_{xyztu}$ is the $\SU(5)$-invariant Levi-Civita tensor.
The Yukawa matrices $Y^{1,5}_{ab}$ and $Z^{5,10}_{ab}$ are antisymmetric under the $a\leftrightarrow b$ exchange and related to one another by Clebsch coefficients arising from the larger $\SO(10)$ symmetry and a single antisymmetric Yukawa coupling $\MAT{Y}_{120}$. We omitted the $\SU(5)$-invariant term $M_{a}{}^{xy}\,M_{b}{}^{zt}\,h^{u}\,\epsilon_{xyztu}$, since it is symmetric under the exchange of $M$'s, and thus necessarily symmetric in generation indices. For all terms in this calculation, invariance under $\mathrm{U}(1)_X$ provides no additional constraints on their existence beyond those related to $\SU(5)$ invariance, so we henceforth omit mentioning the $\mathrm{U}(1)$ factor. 

A comparison of pure PS terms from Eq.~\eqref{eq:Yukawa-PS} and pure $\SU(5)$ terms from Eq.~\eqref{eq:su5-Lagrangian} shows that albeit equal in number, their structure yields a crucial conceptual difference between the languages of the two subgroups: 
\begin{itemize}[itemsep=0.1cm,leftmargin=\DIMLEFTMARGIN ] 
    \item For PS Yukawa terms, the scalars containing the Higgs VEVs are always in self-conjugate representations, and hence invariants come in structurally-related pairs: one member of the pair involves a scalar irrep, and the other its conjugate. Once reality conditions are applied, the four linearly independent terms collapse down to two, as can be seen by inserting their explicit form from Eqs.~\eqref{eq:reality-phi} and \eqref{eq:reality-varphi} into the terms of Eq.~\eqref{eq:Yukawa-PS} and using elementary $\epsilon$-relations for $\SU(2)$ indices. Consequently, the resulting fermion mass matrices in Eqs.~\eqref{eq:Yukawa-PS-real-U}--\eqref{eq:Yukawa-PS-real-N} contain only $2$ independent terms, so that two $\SO(10)\to 4_C\,2_L\,2_R$ Clebsch coefficients can be absorbed by a different normalization of (the VEVs in) scalar irreps $\phi$ and $\varphi$. This feature is a double-edged sword: 
    not needing Clebsch factors makes the computation with the PS subgroup very attractive, but information from $\SO(10)$ is still needed for properly implementing reality conditions. The omission of the latter led to erroneous relations in the original PS-based calculation in the literature~\cite{Babu:2016bmy}.  
    \item For $\SU(5)$ Yukawa terms, the Higgs VEVs reside in complex representations. The reality conditions thus relate VEVs in two different irreps rather than one (as was the case in PS). Indeed, a $\GIRREP{120}$ with real entries in $\SO(10)$ leads to simple conditions $\GIRREP{\overline{5}}=\GIRREP{5}^*$ and $\GIRREP{\overline{45}}=\GIRREP{45}^*$, written in the notation of Eqs.~\eqref{eq:su5-scalar-1}--\eqref{eq:su5-scalar-4} as
    \begin{align}
        (h^{x})^{*}&=h_{x}, &
        (h^{xy}{}_{z})^* &= h_{xy}{}^{z}, \label{eq:su5-reality-condition-general} 
    \end{align}
    leading to the explicit VEV conditions
    \begin{align}
        (\VEVEW{120}{u}{5})^* &= \VEVEW{120}{d}{\overline{5}}, &
        (\VEVEW{120}{u}{45})^* &= \VEVEW{120}{d}{\overline{45}}. \label{eq:su5-reality-condition} 
    \end{align}
    Furthermore, all four Yukawa terms are structurally different, since they involve different pairs of $\SU(5)$ fermion irreps. Four relevant $\SO(10)\to 5\,1_X$ Clebsch coefficients cannot be absorbed by the two independent VEVs (counted after imposing reality conditions), hence Clebsch factors are now required for obtaining the result. In summary, the change to $\SU(5)$ language shifted where the non-trivial part of the calculation occurs: it simplifies the reality conditions, but makes knowing the Clebsches essential.  
\end{itemize}

Given the above discussion, the embedding map $i_{51}$ is now required to provide Clebsch factors rather than port reality conditions like $i_{422}$, which is a more difficult task to perform independently of the $\SO(10)$-based explicit calculation of Section~\ref{subsec:result-SO10}. 

Desiring an independent derivation/confirmation, we look to the literature and find that an $\SO(10)$ calculation for the invariant $\GIRREP{16}\cdot\GIRREP{16}\cdot\GIRREP{120}$ using $\SU(5)$ language has been performed in~\cite{Nath:2001uw}, whereby the ``oscillator method'' was used, see e.g.~\cite{Mohapatra:1979nn}. Unfortunately, recalculating the result in \cite{Nath:2001uw} ourselves by using their methods, we find a sign discrepancy in one of their terms, which we identify later in the text and is not related to the sign discrepancy in the PS calculation of \cite{Babu:2016bmy}. Since we believe modifying a cited result without proper commentary is not credible, we proceed as follows: we lay out the main ideas of the oscillator method and illuminate its connections to other parts of our paper, we provide the relevant relations for performing the calculation (taken from \cite{Nath:2001uw}, following their notation closely), and we state our modified result. 
To visually separate this digression from the main line of thought, we formulate the oscillator method calculation as a sequence of steps:
\begin{enumerate}[itemsep=0.0cm,leftmargin=\DIMLEFTMARGIN,label=(\alph*)]
    \item 
        The oscillator method defines new objects out of $\Gamma$ matrices. For $x,y=1\ldots 5$, the gamma matrices are written as
            \begin{align}
                \bm{\Gamma}_{2x}&= \mathbf{b}_x + \mathbf{b}_{x}^\dagger, &
                \bm{\Gamma}_{2x-1} &= -i(\mathbf{b}_{x}-\mathbf{b}_{x}^\dagger),
            \end{align}
        implicitly defining the operators $\mathbf{b}_x$ and $\mathbf{b}_x^\dagger$.
        The Clifford algebra relation of Eq.~\eqref{eq:definition-clifford} for $\SO(10)$ ($n=5$) then yields the following anti-commutation relations for the new objects:
            \begin{align}
                \{\mathbf{b}_{x},\mathbf{b}_{y}\}=
                \{\mathbf{b}_{x}^\dagger,\mathbf{b}_{y}^\dagger\}&=0, &
                \{\mathbf{b}_{x},\mathbf{b}_{y}^\dagger\}&=\delta_{xy}. \label{eq:oscillator-anticom-relations}
            \end{align}
    \item
        The oscillator method is ultimately just a rewriting of $\SO(10)$ expressions using operators $\mathbf{b}_{x}$ and $\mathbf{b}_{x}^\dagger$ instead of gamma matrices. As such, it is conceptually not truly independent from the methods used in the explicit $\SO(10)$ computation of Section~\ref{subsec:result-SO10}. To mitigate this limitation, we refrain from reusing the gamma matrix construction of Appendix~\ref{app:spinors-gamma} for constructing $\mathbf{b}$-operators.
        \par
        Luckily, one advantage of the oscillator formalism is that computation is tractable through symbolic manipulation by the use of relations in Eq.~\eqref{eq:oscillator-anticom-relations}, so no matrix construction is necessary. 
        Alternatively, for an explicit matrix realization, one can recognize that the relations of Eq.~\eqref{eq:oscillator-anticom-relations} are mathematically equivalent to those of creation operators $\mathbf{b}_x^\dagger$ and annihilation operators $\mathbf{b}_x$ of a system of five fermionic modes $x=1\ldots 5$. One can thus use, for example, the Jordan-Wigner construction~\cite{Jordan1928} 
            \begin{align}
                \mathbf{b}_{x}&= (\bm{\tau}_{3})^{\otimes (x-1)}\otimes \MAT{T}_{-}\otimes (\mathbb{1}_{2})^{\otimes (5-x)}, &
                \mathbf{b}_{x}^\dagger&= (\bm{\tau}_{3})^{\otimes (x-1)}\otimes \MAT{T}_{+}\otimes (\mathbb{1}_{2})^{\otimes (5-x)} \label{eq:oscillator-JordanWigner}
            \end{align}
        with $\MAT{T}_{\pm}:=\tfrac{1}{2}(\bm{\tau}_1\pm i\bm{\tau}_{2})$ interpreted as one-particle raising (creation) and lowering (annihilation) operators, where the Pauli matrices $\bm{\tau}_{1,2,3}$ have been defined in Eq.~\eqref{eq:Pauli-matrices}. Using this construction, we achieve independence in implementation compared to Appendix~\ref{app:spinors-gamma} and Section~\ref{subsec:result-SO10}. 
    \item The oscillator method is particularly well-suited for using the $\SU(5)$-subgroup language in $\SO(10)$. The $\phi\sim \GIRREP{10}$ of $\SO(10)$ can be written (see Eq.~(6) of \cite{Nath:2001uw}) as
        \begin{align}
            \bm{\Gamma}_{p} \phi_{p} &= \bm{\Gamma}_{i} \phi^{i} 
                = \MAT{b}_{x} \phi_{\bar{c}_{x}} + \MAT{b}_{x}^\dagger \phi_{c_{x}}, \label{eq:oscillator-10}
        \end{align}
    where indices $p$ and $i$ refer to the real and complex basis of the $\GIRREP{10}$, respectively, while $x$ is an $\SU(5)$ index, cf.~Table~\ref{tab:indices}. Crucially, we embed $\GIRREP{5}^{x}\equiv \tfrac{1}{\sqrt{2}}\phi_{c_{x}}$ and $\GIRREP{\overline{5}}_{x}\equiv \tfrac{1}{\sqrt{2}}\phi_{\bar{c}_{x}}$ of $\SU(5)$ into the $\GIRREP{10}\equiv \phi^{i}$ of $\SO(10)$ via $\GIRREP{10}^{i}=(\GIRREP{5}^{x},\GIRREP{\overline{5}}_{x})$, defining the embedding map $i_{51}$ from Figure~\ref{fig:map-diagram}. The explicit factor $\tfrac{1}{\sqrt{2}}$ is part of the conversion from $c/\bar{c}$ indices to our indices given the normalization used in the oscillator method, see paragraph below Eq.~(6) in \cite{Nath:2001uw}, so the $\SU(5)\to \SO(10)$ index conversion is $\phi^{c_{x}}=\sqrt{2}\phi^x$ and $\phi^{\bar{c}_{x}}=\sqrt{2}\phi^{x+5}$, where the right-hand sides involve components in our normalization of the complex basis defined in Appendix~\ref{app:basis}.
    Also, the operators $\MAT{b}^\dagger_{x}$ and $\MAT{b}_{x}$ in Eq.~\eqref{eq:oscillator-10} act as the basis for the components $\GIRREP{5}^{x}$ and $\GIRREP{\overline{5}}_{x}$, respectively, hence they themselves transform conjugately as $\GIRREP{\overline{5}}$ and $\GIRREP{5}$, respectively. Although we retain the index placement conventions from \cite{Nath:2001uw}, an alternative notation $(\MAT{b}^\dagger)^{x}=(\MAT{b}_{x})^\dagger$ for creation operators would manifestly capture their $\SU(5)$ transformation properties. 
    \item We now move on to writing the ansatz for relevant $\SO(10)$ irreps in terms of their $\SU(5)$ constituents. The chiral spinor denoted as $\GIRREP{16}$ of $\SO(10)$ in \cite{Nath:2001uw} is written in braket notation according to their Eq.~(7) as 
        \begin{align}
            \KET{\Psi^{+}_{a}} &= 
                M_{a0}\,\KET{0} +
                \tfrac{1}{2!}\;\mathbf{b}_x^\dagger\mathbf{b}_{y}^\dagger\KET{0}\,M_{a}{}^{xy} + 
                \tfrac{1}{4!}\;\epsilon^{xyzuv}\,
                    \mathbf{b}_{x}^\dagger
                    \mathbf{b}_{y}^\dagger
                    \mathbf{b}_{z}^\dagger
                    \mathbf{b}_{u}^\dagger
                    \KET{0}\,M_{av}, \label{eq:oscillator-16}
        \end{align}
        where $a$ labels the generation and the $M$-coefficients are the spinorial fields of Eq.~\eqref{eq:su5-irreps-fermions}. The ket $\KET{0}$ is the vacuum state defined by the property that all $\MAT{b}$-operators annihilate it: 
            \begin{align}
                \mathbf{b}_{x}\KET{0}=0
            \end{align}
        for all $x=1\ldots 5$.
        \par
        The coefficients for terms in Eq.~\eqref{eq:oscillator-16} reflect the number of antisymmetric indices summed over. The normalizations of fermions in Eq.~\eqref{eq:su5-irreps-fermions}, once inserted into Eq.~\eqref{eq:oscillator-16}, yield every individual Weyl fermion field with the same normalization coefficient $1$. This consistent normalization of fermions ensures that derivatives with respect to these fields yield proper fermion mass matrices.  
        \par 
        The superscript $+$ in Eq.~\eqref{eq:oscillator-16} 
        follows the convention of Ref.~\cite{Nath:2001uw} and denotes the chiral spinor constructed from an even number of creation operators $\MAT{b}_x^\dagger$. 
        We now use the notation introduced in the discussion below Eq.~\eqref{eq:cartan-explicitly}: states are labeled by five signs $\KET{\pm\pm\pm\pm\pm}$, where the $x$-th symbol signifies the sign of the $\pm1/2$ eigenvalue under the diagonal generator $\MAT{D}_{x}\equiv\MAT{T}_{2x-1,2x}=(i/4)[\bm{\Gamma}_{2x-1},\bm{\Gamma}_{2x}]$. The chirality eigenvalue of a state equals the product of its signs.
        The operator $\mathbf{b}^\dagger_x$ turns the $x$-th sign in the label from a $+$ into a $-$ (and annihilates a state carrying a $-$), and $\mathbf{b}_{x}$ does the opposite. 
        The vacuum state is thus identified as $\KET{0}=\KET{+++++}$, the states with an even number of $\MAT{b}^{\dagger}$-operators have positive chirality, and thus $\KET{\Psi^{+}}$ belongs to the space $\mathcal{S}_{+}$ as defined in Appendix~\ref{app:spinors-gamma}. For 
        the Jordan-Wigner construction of Eq.~\eqref{eq:oscillator-JordanWigner}, the implied one-particle basis is $\binom{-}{+}$, and hence the vacuum $\KET{0}$ in that basis is written in terms of components as $\binom{0}{1}^{\otimes 5}$.  
        \par
        Note that the $\GIRREP{16}$ of $\SO(10)$ (as opposed to the $\GIRREP{\overline{16}}$) is defined as the chiral spinor decomposing as  $\PSIRREP{4}{2}{1}\oplus\PSIRREP{\overline{4}}{1}{2}$ under the PS subgroup. 
        With the conventions of Appendix~\ref{app:spinors-gamma} and our embedding of PS into spinor space, we have the identification $\GIRREP{16}=\mathcal{S}_{-}$. 
        The conventions of Ref.~\cite{Nath:2001uw}, on the other hand, label the even-occupation spinor, which corresponds to $\mathcal{S}_{+}$ in our chirality convention, as $\GIRREP{16}$.
        This difference is purely conventional and 
        can be viewed as arising from a different choice of subgroup embedding, e.g.~a conjugate embedding of the $\SU(4)_C$ factor; it does not affect the oscillator calculation once the component identifications are translated consistently.
    \item The gamma-matrix structure involving the $\GIRREP{120}\equiv \phi$ irrep of $\SO(10)$ can be rewritten using $\MAT{b}$-operators, see Eq.~(11) of \cite{Nath:2001uw}:
        \begin{align}
            \bm{\Gamma}_{p}\bm{\Gamma}_{q}\bm{\Gamma}_{r}\,\phi_{pqr} = \bm{\Gamma}_{i}\bm{\Gamma}_{j}\bm{\Gamma}_{k}\,\phi^{ijk} &= 
                \mathbf{b}_{x}\mathbf{b}_{y}\mathbf{b}_{z}\;\phi_{\bar{c}_{x}\bar{c}_{y}\bar{c}_{z}}
                +\mathbf{b}^\dagger_{x}\mathbf{b}^\dagger_{y}\mathbf{b}^\dagger_{z}\;\phi_{c_{x}c_{y}c_{z}}
                \nonumber \\
            &\quad + 3 
                (\mathbf{b}^\dagger_{x}\mathbf{b}_{y}\mathbf{b}_{z}\;\phi_{c_{x}\bar{c}_{y}\bar{c}_{z}}
                +\mathbf{b}^\dagger_{x}\mathbf{b}^\dagger_{y}\mathbf{b}_{z}\;\phi_{c_{x}c_{y}\bar{c}_{z}})
                +3(\mathbf{b}_x\;\phi_{\bar{c}_{z}c_{z}\bar{c}_{x}}+\mathbf{b}^\dagger_{x}\;\phi_{\bar{c}_{z}c_{z}c_{x}}).
            \label{eq:oscillator-120-expansion}
        \end{align}
        The various entries of the rank 3 tensor $\phi$ are referred to with un-barred and barred indices ($c_{x}$ and $\bar{c}_x$), as was the case for the $\GIRREP{10}$ in Eq.~\eqref{eq:oscillator-10}. The entries are populated with the $\GIRREP{5}$, $\GIRREP{10}$ and $\GIRREP{45}$ of $\SU(5)$ (and their conjugates) according to 
        Eq.~(21) in \cite{Nath:2001uw}:\footnote{We list only the independent entries of $\GIRREP{120}=\phi$ up to antisymmetry, in a format suitable for insertion into Eq.~\eqref{eq:oscillator-120-expansion}.}
        \begin{align}
            \phi_{c_{x}c_{y}\bar{c}_{z}} &= f^{xy}{}_{z} + \tfrac{1}{4} (\delta^{x}{}_{z}\,f^{y}-\delta^{y}{}_{z}\,f^{x}), &
            \phi_{c_{x}c_{y}c_{z}}&= \epsilon^{xyzuv} f_{uv}, &
            \phi_{\bar{c}_{z}c_{z}c_{x}}&= f^{x}, \label{eq:oscillator-120-phi-1}\\
            \phi_{c_{x}\bar{c}_{y}\bar{c}_{z}} &= f^{x}{}_{yz} + \tfrac{1}{4} (\delta^{x}{}_{y}\,f_{z}-\delta^{x}{}_{z}\,f_{y}), &
            \phi_{\bar{c}_{x}\bar{c}_{y}\bar{c}_{z}}&= \epsilon_{xyzuv} f^{uv}, &
            \phi_{\bar{c}_{z}c_{z}\bar{c}_{x}}&= f_{x}, \label{eq:oscillator-120-phi-2}
        \end{align}
        where the $f$-quantities are improperly normalized Higgs fields from Eq.~\eqref{eq:su5-scalar-1}--\eqref{eq:su5-scalar-4}, and are defined in Eq.~(22) of \cite{Nath:2001uw}:
            \begin{align}
                f^{x} &= \tfrac{4}{\sqrt{3}}\,h^{x}, &
                f^{xy} &= \tfrac{1}{\sqrt{3}}\,h^{xy}, &
                f^{xy}{}_{z} &= \tfrac{2}{\sqrt{3}}\, h^{xy}{}_{z}, \label{eq:oscillator-120-f-1}\\
                f_{x} &= \tfrac{4}{\sqrt{3}}\,h_{x}, &
                f_{xy} &= \tfrac{1}{\sqrt{3}}\,h_{xy}, &
                f^{x}{}_{yz} &= \tfrac{2}{\sqrt{3}}\, h^{x}{}_{yz}. \label{eq:oscillator-120-f-2}
            \end{align}
        \par
        The above construction is such that the $h$-fields have the kinetic term given by Eq.~(23) in \cite{Nath:2001uw}, where $\mu$ is the Lorentz index (where we use the $+---$ signature for the metric tensor):
            \begin{align}
                \mathcal{L}_{kin}^{(120)}&= \partial_{\mu}\phi_{pqr}\,\partial^{\mu}\phi_{pqr}^\dagger 
                = \partial_{\mu}\phi^{ijk}\,\partial^{\mu}(\phi^\dagger)_{ijk}
                \label{eq:oscillator-kinetic120-1}\\
                &=
                \tfrac{1}{2}\partial_{\mu}h^{xy}\,\partial^{\mu}(h^{xy})^{\dagger} +
                \tfrac{1}{2}\partial_{\mu}h_{xy}\,\partial^{\mu}(h_{xy})^{\dagger} +
                \tfrac{1}{2}\partial_{\mu}h^{xy}{}_{z}\,\partial^{\mu}(h^{xy}{}_{z})^\dagger +
                \tfrac{1}{2}\partial_{\mu}h^{x}{}_{yz}\,\partial^{\mu}(h^{x}{}_{yz})^{\dagger} \nonumber \\
                &\quad + \partial_{\mu} h^{x}\,\partial^{\mu} (h^{x})^\dagger +
                \partial_{\mu} h_{x}\,\partial^{\mu} (h_{x})^\dagger, \label{eq:oscillator-kinetic120-2}
            \end{align}
        where terms with $h^{xy}\sim \GIRREP{10}$ and $h_{xy}\sim\GIRREP{\overline{10}}$ of $\SU(5)$ have also been included. The VEVs defined in Eqs.~\eqref{eq:su5-scalar-1}--\eqref{eq:su5-scalar-4} are canonically normalized with respect to this kinetic term (before applying reality conditions), meaning that modulus-squared terms of individual VEVs all come with the numeric coefficient $1$. 
        \par
        Finally, notice that given the reality conditions of Eq.~\eqref{eq:su5-reality-condition-general}, the expressions of Eq.~\eqref{eq:oscillator-120-f-2} are simply complex conjugates of those in Eq.~\eqref{eq:oscillator-120-f-1}, and Eq.~\eqref{eq:oscillator-120-phi-2} is conjugate to Eq.~\eqref{eq:oscillator-120-phi-1}.
    \item We are interested in the invariant $\GIRREP{16}_F\cdot\GIRREP{16}_F\cdot\GIRREP{120}$ given by Eq.~(10) in \cite{Nath:2001uw}:
            \begin{align}
                W^{(120)} &= \tfrac{1}{3!}\,f_{ab} {\psi}_{a}^T \MAT{B} \bm{\Gamma}_{p}\bm{\Gamma}_{q}\bm{\Gamma}_{r}\psi_{b}\,\phi_{pqr}, \label{eq:oscillator-invariant-ansatz}
            \end{align}
        where $f_{ab}$ labels the Yukawa matrix. The matrix $\MAT{B}$ plays the role of the charge conjugation matrix, defined in \cite{Nath:2001uw} as
            \begin{align}
                \mathbf{B} &:=\bm{\Gamma}_{1}\bm{\Gamma}_{3}\bm{\Gamma}_{5}\bm{\Gamma}_{7}\bm{\Gamma}_{9} \\
                    &= -i\,(\MAT{b}_1-\MAT{b}_1^\dagger)(\MAT{b}_2-\MAT{b}_2^\dagger)(\MAT{b}_3-\MAT{b}_3^\dagger)
                    (\MAT{b}_4-\MAT{b}_4^\dagger)(\MAT{b}_5-\MAT{b}_5^\dagger). \label{eq:oscillator-Boperator-2}
            \end{align}
        Note that this construction differs from $\mathbf{C}'$ in Eq.~\eqref{eq:charge-BW-result1} of Appendix~\ref{app:spinors-gamma}, which is the charge conjugation used in this paper outside this appendix section, but nevertheless satisfies the required property of Eq.~\eqref{eq:C-defining-property}.
        \par 
        We can now compute the invariant of Eq.~\eqref{eq:oscillator-invariant-ansatz} by inserting all objects expressed with $\MAT{b}$-operators, namely $\Psi_a$ and $\Psi_b$ (columns in spinor space) based on Eq.~\eqref{eq:oscillator-16}, $\bm{\Gamma}_p\bm{\Gamma}_q\bm{\Gamma}_r\,\phi_{pqr}$ from Eq.~\eqref{eq:oscillator-120-expansion} supplemented by Eqs.~\eqref{eq:oscillator-120-phi-1}--\eqref{eq:oscillator-120-f-2}, and $\MAT{B}$ from Eq.~\eqref{eq:oscillator-Boperator-2}. We obtain the following result:
            \begin{align}
                W^{(120)} & = i\frac{2}{\sqrt{3}}\,f_{ab} \Bigg(
                    \,2 M_{a0}\,M_{bx}\,h^{x}
                    + M_{a}{}^{xy}\,M_{b0}\,h_{xy}
                    + M_{ax}\,M_{by}\,h^{xy}
                    \nonumber\\
                &+ M_{a}{}^{xy}\,M_{bx}\,h_{y}
                    + M_{ax}\,M_{b}{}^{yz}\,h^{x}{}_{yz}
                    -\frac{1}{4} \epsilon_{xyztu}\,M_{a}{}^{xy}\,M_{b}{}^{uv}\,h^{zt}{}_{v}
                    \Bigg). \label{eq:oscillator-invariant-result}
            \end{align}
        Comparing with the pure $\SU(5)$ setup of Eq.~\eqref{eq:su5-Lagrangian}, we see that our result gave the following Clebsch factors:
        \begin{align}
            Y^{1}_{ab}/\tilde{f}_{ab} &= 2, &
            Y^{5}_{ab}/\tilde{f}_{ab} &= 1, &
            Z^{5}_{ab}/\tilde{f}_{ab} &= 1, &
            Z^{10}_{ab}/\tilde{f}_{ab} &= -1/4, 
        \end{align}
        where we used the shorthand $\tilde{f}_{ab}:= i\tfrac{2}{\sqrt{3}} f_{ab}$.
        \par
        The result of Eq.~\eqref{eq:oscillator-invariant-result} matches the one given in Eq.~(12) of \cite{Nath:2001uw}, except for the aforementioned minus sign in the first term of our second line, i.e.~in the coefficient $Y^{5}_{ab}$. We believe this is a typo in \cite{Nath:2001uw}, given that we later show consistency of our oscillator result for fermion mass matrices with those obtained by other methods in Section~\ref{sec:reality-conditions}. Crucially, the minus discrepancy is in the context of the $\SU(5)$ basis, 
        so it translates into a non-trivial change in the combination of PS-based VEVs, and thus has nothing to do with the derived relative minus sign in reality conditions for the PS basis.    
\end{enumerate}

We now return to our $\SU(5)$-based fermion mass calculation. The oscillator method provided the Clebsch coefficients, yielding the result of Eq.~\eqref{eq:oscillator-invariant-result}. Inserting the fermion fields from Eq.~\eqref{eq:su5-irreps-fermions} and scalar VEVs from Eq.~\eqref{eq:su5-scalar-1}--\eqref{eq:su5-scalar-4}, and accounting for the reality conditions via Eq.~\eqref{eq:su5-reality-condition}, we obtain the following contributions to the fermion mass matrices from the $\GIRREP{16}_F\cdot\GIRREP{16}_F\cdot\GIRREP{120}_{\mathbb{R}}$ term:
\begin{align}
    \MAT{M}_{U}&\supset \tfrac{i}{\sqrt{3}}\,f_{ab}\;\big(-\tfrac{2}{\sqrt{3}}\,\VEVEW{120}{u}{45}\big), \label{eq:su5-MU}\\
    \MAT{M}_{D}&\supset \tfrac{i}{\sqrt{3}}\,f_{ab}\;\big(-\tfrac{1}{\sqrt{3}}\,\VEVEW{120}{u}{45}^{*}+\VEVEW{120}{u}{5}^{*}\big),\\
    \MAT{M}_{E}&\supset \tfrac{i}{\sqrt{3}}\,f_{ab}\;\big(-\sqrt{3}\,\VEVEW{120}{u}{45}^{*}-\VEVEW{120}{u}{5}^{*}\big),\\
    \MAT{M}_{\nu_{D}}&\supset \tfrac{i}{\sqrt{3}}\,f_{ab}\;\big(2\,\VEVEW{120}{u}{5}\big). \label{eq:su5-MnuD}
\end{align}

We want to compare these contributions to the results in Section~\ref{sec:reality-conditions}, so we also need to derive the relation between the VEVs in the PS and $\SU(5)$ basis. The explicit components of the $\GIRREP{120}^{ijk}$ with PS-basis VEVs have been given in Eqs.~\eqref{eq:EWVEVs-120-1u}--\eqref{eq:EWVEVs-120-15d}. Analogously, the components for the $\SU(5)$-based VEVs employed in the oscillator method can be extracted by inserting
the VEV ansatz of Eqs.~\eqref{eq:su5-scalar-1}--\eqref{eq:su5-scalar-4} first into Eqs.~\eqref{eq:oscillator-120-f-1}--\eqref{eq:oscillator-120-f-2}, and then into Eqs.~\eqref{eq:oscillator-120-phi-1}--\eqref{eq:oscillator-120-phi-2}. The components of $\GIRREP{120}^{c_{x}c_{y}c_{z}}$ are matched to our complex basis via $\GIRREP{120}^{c_{x}c_{y}c_{z}}=\sqrt{8}\cdot\GIRREP{120}^{xyz}$ (and analogous equations for $\bar{c}$-components using the $\bar{c}_{u}\mapsto u+5$ rule), where the factor $\sqrt{8}=\sqrt{2}^3$ arises due to how each index corresponding to the $\GIRREP{10}$ is normalized, cf.~comment below Eq.~\eqref{eq:oscillator-10}. For $\GIRREP{120}^{ijk}$ we obtain 
\begin{align}
    [4, 5, 9] & = \tfrac{1}{\sqrt{8}}\,
        (\tfrac{1}{\sqrt{3}}\,\VEVEW{120}{u}{5}
        -\VEVEW{120}{u}{45}), 
        \label{eq:EWVEVs-120-su5-1}\\
    [9, 10, 4] & = \tfrac{1}{\sqrt{8}}\,
        (\tfrac{1}{\sqrt{3}}\,\VEVEW{120}{d}{\overline{5}}
        -\VEVEW{120}{d}{\overline{45}}),\\
    [1, 5, 6]=[2, 5, 7]=[3, 5, 8] &= \tfrac{1}{\sqrt{8}}\,
        (\tfrac{1}{\sqrt{3}}\,\VEVEW{120}{u}{5}
        +\tfrac{1}{3}\,\VEVEW{120}{u}{45}),\\
    [6, 10, 1]=[7, 10, 2]=[8, 10, 3] &= \tfrac{1}{\sqrt{8}}\,
        (\tfrac{1}{\sqrt{3}}\,\VEVEW{120}{d}{\overline{5}}
        +\tfrac{1}{3}\,\VEVEW{120}{d}{\overline{45}}), 
        \label{eq:EWVEVs-120-su5-4}
\end{align}
written before the application of reality conditions. Note that the $d$-type assignments should not be independent from the $u$-type assignments: taking the reality condition of Eq.~\eqref{eq:su5-reality-condition} and using the $l\mapsto (l+5)\mod 10$ conjugation rule for each index, the $u$-VEV assignments should become the $d$-VEV assignments. This is indeed the case, which serves as a consistency check.

Now we compare the entries in the PS VEV-basis of Eqs.~\eqref{eq:EWVEVs-120-1u}--\eqref{eq:EWVEVs-120-15d} and $\SU(5)$ VEV-basis of Eqs.~\eqref{eq:EWVEVs-120-su5-1}--\eqref{eq:EWVEVs-120-su5-4}, and obtain
\begin{align}
    \VEVEW{120}{u}{1} &= \alpha
        \big(\tfrac{1}{2}\,\VEVEW{120}{u}{5} 
        -\tfrac{\sqrt{3}}{2}\,\VEVEW{120}{u}{45}\big), \label{eq:VEV-PS-SU5-u1}\\
    \VEVEW{120}{u}{15} &= \alpha
        \big(-\tfrac{\sqrt{3}}{2}\,\VEVEW{120}{u}{5}
        -\tfrac{1}{2}\,\VEVEW{120}{u}{45}\big), \label{eq:VEV-PS-SU5-u2}\\
    \VEVEW{120}{d}{1} &= \alpha
        \big(\tfrac{1}{2}\,\VEVEW{120}{d}{\overline{5}} 
        -\tfrac{\sqrt{3}}{2}\,\VEVEW{120}{d}{\overline{45}}\big), \label{eq:VEV-PS-SU5-d1}\\
    \VEVEW{120}{d}{15} &= \alpha
        \big(-\tfrac{\sqrt{3}}{2}\,\VEVEW{120}{d}{\overline{5}}
        -\tfrac{1}{2}\,\VEVEW{120}{d}{\overline{45}}\big), \label{eq:VEV-PS-SU5-d2}
\end{align}
where $\alpha=\sqrt{1/3}$. We make the following observations: 
\begin{itemize}[itemsep=-0.1cm,leftmargin=\DIMLEFTMARGIN]
    \item 
        The normalization coefficient $\alpha\neq 1$ arises due to comparing our $\SO(10)$ computation with the oscillator method, where kinetic terms 
        are normalized differently, i.e.~Eq.~\eqref{eq:normalization-120} uses a factor $1/3!$ after imposing reality conditions, while Eqs.~\eqref{eq:oscillator-kinetic120-1}--\eqref{eq:oscillator-kinetic120-2} use a factor $1$ before imposing reality conditions. The overall ratio of kinetic terms after imposing reality conditions is thus $1/3$, yielding a field/VEV normalization factor $\alpha=1/\sqrt{3}$. 
    \item 
        Notice that Eq.~\eqref{eq:VEV-PS-SU5-d1}--\eqref{eq:VEV-PS-SU5-d2} turn into complex conjugates of Eq.~\eqref{eq:VEV-PS-SU5-u1}--\eqref{eq:VEV-PS-SU5-u2} once the reality conditions of Eqs.~\eqref{eq:VEV-reality-condition-so10} and \eqref{eq:su5-reality-condition} are taken into account, which again serves as a consistency check of the calculation. 
    \item
        Assuming the kinetic term is normalized the same in both methods, the relations yield $\alpha = 1$ and represent an orthogonal transformation. This transformation is unique up to a sign convention for each of the VEV labels. Comparing e.g.~with Eq.~(2.20) and (2.21) in \cite{Babu:2025wop}, we see the relations to be consistent with a single change of sign convention, namely for the VEV $\VEVEW{120}{u}{15}$. Obtaining the relations on our part required careful handling of sign conventions for VEVs in each method. This is why writing the defining relations for the VEV labels in Eqs.~\eqref{eq:VEV-PS-SU5-u1}--\eqref{eq:VEV-PS-SU5-d2} and the ansatz of Eqs.~\eqref{eq:su5-scalar-1}--\eqref{eq:su5-scalar-4} is necessary for our result to be reproducible.
\end{itemize}

Solving Eqs.~\eqref{eq:VEV-PS-SU5-u1} and \eqref{eq:VEV-PS-SU5-u2} for the $\SU(5)$-basis VEVs, and inserting into the mass matrices of Eq.~\eqref{eq:su5-MU}--\eqref{eq:su5-MnuD}, we indeed recover the $\MAT{Y}_{120}$ terms from our main result in Eqs.~\eqref{eq:result-MU}--\eqref{eq:result-MND} by matching the overall Yukawa coefficients via
    \begin{align}
        f_{ab} &= -i\,(\MAT{Y}_{120})_{ab}.
    \end{align}
Given this matching, we consistently reproduce the fermion mass relations, including the crucial minus sign in the coefficient for $\VEVEW{120}{u}{15}{}^{*}$ in $\MAT{M}_D$ arising from $\SO(10)$-consistency of reality conditions on VEVs in the PS-basis. The factor $-i$ in the matching condition arises due to the oscillator method implementing the charge conjugation matrix as $\MAT{B}$ of Eq.~\eqref{eq:oscillator-Boperator-2} rather than $\MAT{C}'$ of Eq.~\eqref{eq:charge-BW-result1}.  

Given the above results, the $\SU(5)$-subgroup based approach with the oscillator method thus represents a third consistent derivation of the fermion mass matrices.

\subsection{Transporting the reality structure \label{app:rtransport}}

We present here another alternative derivation: a new general method for determining how the reality structure is transported from parent to daughter irreps, which avoids specifying all the details of the embedding map. This method is the most direct way to obtain the specific answer to inheritance of real structures.

We present this material in the following sequence: we motivate the method conceptually, we state the explicit algorithm, we apply it to our case, and finally we provide a general proof why it works and discuss its robustness.

First, let us conceptually motivate the procedure. 
In Section~\ref{subsec:result-SO10-PS} we determined the real-structure inheritance by deriving reality conditions from the full specification of the embedding $i_{422}$ of Figure~\ref{fig:map-diagram}. While full knowledge of the embedding $i_{422}$ enables the investigation of essentially any group-theoretic question, it is overkill for our specific task. Compare now the $\SO(10)$ reality conditions of e.g.~Eq.~\eqref{eq:VEV-reality-condition-so10} and those in pure Pati-Salam in Eq.~\eqref{eq:reality-constraint-options-Pati-Salam}, which link the conjugate of the $u$-type VEV to the $d$-type in each of the PS irreps. Let's take the PS irrep $\PSIRREP{1}{2}{2}$ as an example. One could hope to somehow compute the unknown inherited sign $s'_{1}$ in PS if one knew how to relate the VEV labels $\VEVEW{120}{u}{1}$ and $\VEVEW{120}{d}{1}$ of the $\SO(10)$ calculation to VEVs $\VEVEWPS{1}{u}$ and $\VEVEWPS{1}{d}$ in Pati-Salam. If the determination of the states underlying the VEVs has been performed correctly in both $\SO(10)$ and PS, then the embedding map $i_{422}$ should send $\VEVEWPS{1}{u}$ into $\VEVEW{120}{u}{1}$ up to some normalization factor (in the same way in all relevant tensor entries), and analogously for the $d$-VEVs. Since we can normalize the $u$- and $d$-VEVs the same within every irrep, the main ambiguity in reality conditions lies in sign conventions for all the VEV labels. The relative sign between $u$'s could be absorbed into a redefinition of the irrep space (on the PS side), making only the relative sign between the $u$- and $d$-type VEV on both sides of $i_{422}$ the relevant information. It turns out that anchoring such a relative sign convention between the $u$- and $d$-type VEV with a PS-equivariant condition is indeed all that is necessary.

Stated concisely, it turns out the following general procedure can be used:
\begin{theorem}
    \label{the:procedure}
    Let $\mathcal{V}^{H}_{r}$ be the underlying vector space of the irrep $r$ for the 
    Lie group $H$, and $\mathcal{V}^{G}_{R}$ for irrep $R$ of Lie group $G$, where $H$ is a subgroup of $G$. Let $i_{r}:\mathcal{V}^H_{r}\hookrightarrow\mathcal{V}^{G}_{R}$
    be an $H$-equivariant embedding map of $r$ into $R$, with the image labeled as $\mathcal{V}^{G}_{r}:=i_{r}(\mathcal{V}^H_{r})$. Let $\rho^{H}_{r}$ and $\rho^{G}_{R}$ be $H$- and $G$-equivariant real structures on $\mathcal{V}^{H}_{r}$ and $\mathcal{V}^{G}_{R}$, respectively, and assume the embedding $i_r$ is real, i.e.~$i_r\circ\rho^{H}_{r}=\rho^{G}_{R}\circ i_{r}$.\footnote{The embedding $i_{r}$ being real implies that $\rho^{H}_{r}$ is the inherited real structure from the parent $\rho^{G}_{R}$.} Suppose we find non-trivial vectors $\MAT{e}_{u},\MAT{e}_{d}\in \mathcal{V}^{H}_{r}$ and $\MAT{e}_{U},\MAT{e}_{D}\in \mathcal{V}^{G}_{r}$ as well as $\hat{\MAT{T}}\in \mathcal{U}(\mathfrak{h})$, such that the following hold simultaneously:
        \begin{align}
            i_{r}(\MAT{e}_{u})&=a_{r}\, \MAT{e}_{U}, &
            i_{r}(\MAT{e}_{d})&\propto \MAT{e}_{D}, 
            \label{eq:theorem-0a}\\
            \rho^{H}_{r}(\MAT{e}_{u})&=\eta^{H}_{r}\,\MAT{e}_{d}, &
            \MAT{T}^{H}_{r}\MAT{e}_{u}&=\alpha^{H}_{r}\,\MAT{e}_{d},
            \label{eq:theorem-0b}\\
            \rho^{G}_{R}(\MAT{e}_{U})&=\eta^{G}_{r,R}\,\MAT{e}_{D}, &
            \MAT{T}^{G}_{R}\MAT{e}_{U}&=\alpha^{G}_{r,R}\,\MAT{e}_{D}, 
            \label{eq:theorem-0c}
        \end{align}
    where $a_r,\eta^{H}_{r},\eta^{G}_{r,R},\alpha^{H}_{r},\alpha^{G}_{r,R}\in  \mathbb{C}\setminus\{0\}$ are coefficients, $\mathcal{U}(\mathfrak{h})$ is the universal enveloping algebra of the Lie algebra of $H$, and $\MAT{T}^{H}_{r}=r(\hat{\MAT{T}})$ and $\MAT{T}^{G}_{R}=R(\hat{\MAT{T}})$ are the irrep realizations of its element $\hat{\MAT{T}}$. Then the following holds:
    \begin{enumerate}[nosep,leftmargin=\DIMLEFTMARGIN,label=(\alph*)]
        \item If the embedding coefficient $a_{r}$ is real, the $\eta/\alpha$ ratio is invariant under the $i_{r}$ embedding, implying
            \begin{align}
                \frac{\eta^{H}_{r}}{\alpha^{H}_{r}} &= \frac{\eta^{G}_{r,R}}{\alpha^{G}_{r,R}}. \label{eq:theorem-1a}
            \end{align}
        \item Suppose $\rho^{H}_{r}$ is unknown, but we instead take a predefined $H$-equivariant real structure $\tilde{\rho}^{H}_{r}$ on $\mathcal{V}^{H}_{r}$, for which 
            \begin{align}
                \tilde{\rho}^{H}_{r} (\MAT{e}_{u}) &= \tilde{\eta}^{H}_{r}\,\MAT{e}_{d}, \label{eq:theorem-2a}
            \end{align}
            with $\tilde{\eta}^{H}_{r}\in\mathbb{C}\setminus\{0\}$. Then the inherited real structure must be $\rho^{H}_{r}=s_{r,R}\,\tilde{\rho}^{H}_{r}$, where the coefficient $s_{r,R}\in\mathrm{U}(1)$ is computed by
            \begin{align}
                s_{r,R} &= \frac{a_{r}^*}{a_{r}}\;\frac{\alpha^{H}_{r}}{\alpha^{G}_{r,R}}\;\frac{\eta^{G}_{r,R}}{\tilde{\eta}^{H}_{r}}, \label{eq:theorem-2bp}
            \end{align}
            reducing in the case $a_{r}\in \mathbb{R}$ to
            \begin{align}
                s_{r,R} &= \frac{\alpha^{H}_{r}}{\alpha^{G}_{r,R}}\;\frac{\eta^{G}_{r,R}}{\tilde{\eta}^{H}_{r}}.\label{eq:theorem-2b}
            \end{align}
    \item If we multiply the vectors with field/VEV coefficients $U,D,u,d\in\mathbb{C}$ by denoting
        \begin{align}
            \MAT{U}&=U\,\MAT{e}_{U}, &
            \MAT{D}&=D\,\MAT{e}_{D}, &
            \MAT{u}&=u\,\MAT{e}_{u}, &
            \MAT{d}&=d\,\MAT{e}_{d}, \label{eq:theorem-3a}
        \end{align}
        the reality conditions become 
        \begin{align}
            \eta^{G}_{r,R}\,U^* &= D, & s_{r,R}\,\tilde{\eta}^{H}_{r}\,u^{*} &= d,
            \label{eq:theorem-3b}
        \end{align}
        where the right-side expression is the inherited reality condition of the left-side one.
    \end{enumerate}
\end{theorem}

Let us first make some clarifications for intuition. The idea is to choose a pair of conjugate states $\MAT{e}_{u},\MAT{e}_{d}$ in the irrep $r$, and corresponding states $\MAT{e}_{U},\MAT{e}_{D}$ up to some normalization in $R$.
We use small (capital) letters in subscripts related to the $H$-side ($G$-side) of the calculation. Applying the real structure map (the ``complex conjugation on the irrep'') turns the $u$-type state to a multiple of the $d$-type state. Since the states are part of the same $H$-irrep, an identical operation up to an overall factor should be accessible via a product of ladder operators of $H$, denoted by $\hat{\MAT{T}}$ (products/polynomials of generators live in $\mathcal{U}(\mathfrak{h})$ rather than $\mathfrak{h}$). The parent and reference daughter real structures are predefined, therefore we are testing which phase-transformed version of the predefined real structure on $H$ is the inherited one through the computation of the coefficient $s_{r,R}$ in Eq.~\eqref{eq:theorem-2bp}. 
To perform the calculation, we need to find a suitable $\hat{\MAT{T}}$, and compute two coefficients on the $G$-side and two coefficients on the $H$-side of the computation, as well as know the phase of the embedding coefficient $a_{r}$ (unless it is real).

Now we go through the procedure explicitly in great detail for the three cases of our paper, which should serve as a good pedagogical demonstration of its use.
Note that the procedure itself is conceptually independent from all previous methods, but the calculation we perform below does reuse the explicit form of generators for the $\GIRREP{10}$ of $\SO(10)$ implemented in  Section~\ref{subsec:result-SO10} and the state identifications within scalar irreps (but crucially does not involve the spinorial irrep and gamma matrix construction). 
Clearly the Lie groups in the algorithm are set to \hbox{$G=\SO(10)$} and \hbox{$H=4_C\,2_L\,2_R$}, and we are to compute the $s_{r,R}$ coefficient for three valid pairs of irreps $r\subset R$, where $r$ is $\PSIRREP{1}{2}{2}$ or $\PSIRREP{15}{2}{2}$, while $R$ is $\GIRREP{10}$ or $\GIRREP{120}$. Given that throughout this paper we populated these tensors with pairs of EW VEVs related by complex conjugation, it is most convenient to choose the underlying directions of these VEVs as our state pairs needed in the procedure. The correspondence of Eq.~\eqref{eq:theorem-0a} is guaranteed by correctly identifying the EW VEVs states on each side of the calculation based on quantum numbers and transformation properties under PS-generators --- these considerations already turned out to pick unique states in each tensor up to an overall factor. 
Moreover, no complex phases are involved in our embedding (discussed in greater detail after the proof), so $a_{r}\in\mathbb{R}$ and we can use the simpler expression of Eq.~\eqref{eq:theorem-2b}. 
For the series of ladder operators $\hat{\MAT{T}}$, we pick the same one for all three cases under consideration: the $u$-VEV directions become the $d$-VEV directions under
\begin{align}
    \hat{\MAT{T}} := \hat{\MAT{T}}^{R}_{-}\,\hat{\MAT{T}}^{L}_{+},
\end{align}
where
\begin{align}
    \hat{\MAT{T}}^{L}_{\pm} &:= \hat{\MAT{T}}^{L}_{1}\pm i\hat{\MAT{T}}^{L}_{2},&
    \hat{\MAT{T}}^{R}_{\pm} &:= \hat{\MAT{T}}^{R}_{1}\pm i\hat{\MAT{T}}^{R}_{2}
\end{align}
are the raising/lowering operators of the groups $\SU(2)_L$ and $\SU(2)_R$, respectively. Such a $\hat{\MAT{T}}$ is clearly a product of generators of the PS algebra. For concreteness, we specify the explicit forms these generators take in the fundamental representation of $\SU(2)_{L,R}$ and the $\GIRREP{10}$ of $\SO(10)$, with index conventions consistent with Table~\ref{tab:indices}:\footnote{
    Since ladder operators are not Hermitian, one would have to be careful how lower indices are transformed if this appeared in the calculation. The basis used for $\SO(10)$ is the complex basis of Table~\ref{tab:qn10}.
}
\begin{align}
    (\MAT{T}^{L}_{+})^{\alpha}{}_{\beta}&= 
        \left(\begin{smallmatrix}
            0&1\\
            0&0\\
        \end{smallmatrix}\right),&
    (\MAT{T}^{R}_{-})^{\dot{\alpha}}{}_{\dot{\beta}}&= 
        \left(\begin{smallmatrix}
            0&0\\
            1&0\\
        \end{smallmatrix}\right), 
        \label{eq:Telement-PS}\\
    (\MAT{T}^{L}_{+})^{i}{}_{j} &= 
        \delta^{i4} \delta_{j5} - 
        \delta^{i,10} \delta_{j9}, &
    (\MAT{T}^{R}_{-})^{i}{}_{j} &= 
        \delta^{i9} \delta_{j5} - 
        \delta^{i,10} \delta_{j4}, \label{eq:Telement-SO10}
\end{align}
Note that the minus signs in Eq.~\eqref{eq:Telement-SO10} appear when the state $i=10$ is involved, since it is part of the $\SU(2)_L$ and $\SU(2)_R$ doublets that are embedded as a conjugate $\GIRREP{\overline{2}}$ into $\GIRREP{10}$; in other words, that state is the last entry of the $\SU(5)$-irrep $\GIRREP{\overline{5}}=(d^{c}_{1},d^{c}_{2},d^{c}_3,e,-\nu)$ in the $\GIRREP{10}$, where we borrowed the notation from fermion states. 
Finally, for a tensor $\phi$ populated with field/VEV variables, we denote by $\phi|_x$ the direction of the $x$-variable, i.e.~we take $x=1$ and zero for all other variables. We now proceed to each of the three individual cases $r\subset R$: 
\begin{itemize}[itemsep=0.0cm,leftmargin=\DIMLEFTMARGIN]
    \item \underline{The case $\PSIRREP{1}{2}{2}\subset \GIRREP{10}$:}\\[3pt]
        On the Pati-Salam side we choose the VEV ansatz $\phi$ in Eq.~\eqref{eq:ansatz-PS-scalars}, so that the pair of states is explicitly 
        \begin{align}
            \MAT{e}_{u}&=\phi|_{\VEVEWPS{1}{u}}= \MATS{0&0\\ 1&0\\}, &
            \MAT{e}_{d}&=\phi|_{\VEVEWPS{1}{d}}= \MATS{0&-1\\ 0&0\\}.
        \end{align}    
        The predefined real structure in PS is taken with the $s'_{1}=1$ coefficient from Eq.~\eqref{eq:reality-phi}, which is equivalent to
        \begin{align}
            \tilde{\rho}^{\text{PS}}_{1}(\phi^{\alpha\dot{\alpha}}):= \left[\epsilon_{\alpha\beta} \,\epsilon_{\dot{\alpha}\dot{\beta}}\, \phi^{\beta\dot{\beta}}\right]^{*}. \label{eq:C2procedure-112-rho}
        \end{align}
        The two alternate ways of relating the $\MAT{e}_{u}$ and $\MAT{e}_{d}$ via Eq.~\eqref{eq:theorem-0b} take the form
        \begin{align}
            \big(
                \epsilon_{\alpha\beta}\, 
                \epsilon_{\dot{\alpha}\dot{\beta}}\,
                (\MAT{e}_{u})^{\beta\dot{\beta}}\big)^{*} 
            &= \tilde{\eta}^{\text{PS}}_{1}\,(\MAT{e}_{d})^{\alpha\dot{\alpha}}, &
            (\MAT{T}^{L}_{+})^{\alpha}{}_{\beta}\,
                (\MAT{T}^{R}_{-})^{\dot{\alpha}}{}_{\dot{\beta}}\,
                (\MAT{e}_{u})^{\beta\dot{\beta}} &= 
                    \alpha^{\text{PS}}_{1}\,
                    (\MAT{e}_{d})^{\alpha\dot{\alpha}}.
        \end{align}
        Explicit computation gives $\tilde{\eta}^{\text{PS}}_{1}=+1$ and $\alpha^{\text{PS}}_{1}=-1$.
        \par
        On the $\SO(10)$ side of the calculation, we take the state $\GIRREP{10}^{i}\equiv \Phi^{i}$ defined by Eq.~\eqref{eq:EWVEVs-10}, so that 
        \begin{align}
            (\MAT{e}_{U})^{i} &=(\Phi|_{\VEVEW{10}{u}{1}})^{i} = \tfrac{1}{\sqrt{2}}\,\delta^{i5}, &
            (\MAT{e}_{D})^{i} &=(\Phi|_{\VEVEW{10}{d}{1}})^{i} = \tfrac{1}{\sqrt{2}}\,\delta^{i,10}, &
        \end{align}
        and take the parent real structure from Eq.~\eqref{eq:reality-condition-10-c}, defining
        \begin{align}
            \rho^{\SO(10)}_{10} (\Phi^i)&= (P_{ij}\,\Phi^{j})^*  
        \end{align}
        where $P_{ij}$ is defined in \eqref{eq:Pmatrix}. The two relations between $\MAT{e}_{U}$ and $\MAT{e}_{D}$ from Eq.~\eqref{eq:theorem-0c}
        then explicitly become
        \begin{align}
            \big(P_{ij}\,(\MAT{e}_{U})^{j}\big)^{*} &
            = \eta^{\SO(10)}_{1,10}\, (\MAT{e}_D)^{i}, &
            (\MAT{T}^{R}_{-})^{i}{}_{j}\,
                (\MAT{T}^{L}_{+})^{j}{}_{k}\,
                (\MAT{e}_{U})^{k} &= 
                    \alpha^{\SO(10)}_{1,10}\,
                (\MAT{e}_D)^{i}.
        \end{align}
        Explicit computation and comparison of tensor components give $ \eta^{\SO(10)}_{1,10}=+1$ and $\alpha^{\SO(10)}_{1,10}=-1$.
    \par
    Using Eq.~\eqref{eq:theorem-2b}, we get
    \begin{align}
        s_{1,10}&= \frac{\alpha^{\text{PS}}_{1}}{ \alpha^{\SO(10)}_{1,10}} \frac{\eta^{\SO(10)}_{1,10}}{\tilde{\eta}^{\text{PS}}_{1}} =\frac{-1}{-1}\,\frac{+1}{+1}=+1.
    \end{align}
    This implies the inherited reality condition is equal to the predefined (standard) one, consistent with other derivations. The VEVs in PS should be related according to Eq.~\eqref{eq:theorem-3b} via 
    \begin{align}
        \VEVEWPS{1}{u}^{*}=s_{1,10}\cdot\tilde{\eta}^{\text{PS}}_{1}\cdot\VEVEWPS{1}{d} = +\VEVEWPS{1}{d},
    \end{align}
    when they are coming from $\PSIRREP{1}{2}{2}$ embedded in the $\GIRREP{10}$.
    \item \underline{The case $\PSIRREP{1}{2}{2}\subset\GIRREP{120}$:}\\[3pt]
        The PS-side of the calculation is not influenced by the irrep $\PSIRREP{1}{2}{2}$ embedded in $\GIRREP{120}$ instead of $\GIRREP{10}$:
        we take the same VEV ansatz and same predefined real structure to obtain $\tilde{\eta}^{\text{PS}}_{1}=+1$ and $\alpha^{\text{PS}}_{1}=-1$.
        \par
        On the $\SO(10)$ side of the calculation, we take $\GIRREP{120}^{ijk}=\Phi^{ijk}$ to be populated by the $\VEVEW{120}{u}{1}$ and $\VEVEW{120}{d}{1}$ VEVs of Eqs.~\eqref{eq:EWVEVs-120-1u}--\eqref{eq:EWVEVs-120-1d}, so that the states are defined as
        $(\MAT{e}_{U})^{ijk}=\Phi^{ijk}|_{\VEVEW{120}{u}{1}}$ and
        $(\MAT{e}_{D})^{ijk}=\Phi^{ijk}|_{\VEVEW{120}{d}{1}}$.
        The real structure is specified by Eq.~\eqref{eq:reality-condition-120-rc}, or equivalently
        \begin{align}
            \rho^{\SO(10)}_{120} (\Phi^{ijk})&= (P_{il}\,P_{jm}\,P_{kn}\,\Phi^{lmn})^* .
        \end{align}
        The $UD$-relations of Eqs.~\eqref{eq:theorem-0c} take the form
        \begin{align}
            \big(P_{il}\,P_{jm}\,P_{kn}\,(\MAT{e}_{U})^{ijk}\big)^{*} &=
                \eta^{\SO(10)}_{1,120}\,(\MAT{e}_{D})^{ijk}, &
            (\hat{\MAT{T}}^{R}_{-}\,\hat{\MAT{T}}^{L}_{+}\,\MAT{e}_{U})^{ijk} &= \alpha^{\SO(10)}_{1,120}\,(\MAT{e}_{D})^{ijk},
            \label{eq:procedure-122-120-cond}
        \end{align}
        where the action of the generators must be applied tensorially one after the other, so that for every action each index is transformed, e.g.
        \begin{align}
            (\hat{\MAT{T}}^{L}_{+}\MAT{e}_{U})^{ijk} &= 
                (\MAT{T}^{L}_{+})^{i}{}_{l} (\MAT{e}_{U})^{ljk} 
                +  (\MAT{T}^{L}_{+})^{j}{}_{l} (\MAT{e}_{U})^{ilk}
                + (\MAT{T}^{L}_{+})^{k}{}_{l} (\MAT{e}_{U})^{ijl}.
        \end{align}
        Explicit computation and comparison of tensor components in Eq.~\eqref{eq:procedure-122-120-cond} gives $\eta^{\SO(10)}_{1,120}=+1$ and $\alpha^{\SO(10)}_{1,120}=-1$, yielding the $s_{1,120}$ sign from the expression in Eq.~\eqref{eq:theorem-2b} to be
        \begin{align}
            s_{1,120}&= \frac{\alpha^{\text{PS}}_{1}}{ \alpha^{\SO(10)}_{1,120}} \frac{\eta^{\SO(10)}_{1,120}}{\tilde{\eta}^{\text{PS}}_{1}}
                = \frac{-1}{-1}\,\frac{+1}{+1} = +1,
        \end{align}
        so the standard predefined real structure is the inherited one, i.e.~$s'_{1}=+1$ in Eq.~\eqref{eq:reality-phi}.
    \item \underline{The case $\PSIRREP{15}{2}{2}\subset\GIRREP{120}$}\\[3pt]
        On the PS-side, we take for $\PSIRREP{15}{2}{2}$ the VEV ansatz $\varphi^{A}{}_{B}{}^{\alpha\dot{\alpha}}$ from Eq.~\eqref{eq:ansatz-PS-scalars}, thus obtaining the states 
        \begin{align}
            (\MAT{e}_{u})^{A}{}_{B}{}^{\alpha\dot{\alpha}}&=
                (\varphi|_{\VEVEWPS{15}{u}})^{A}{}_{B}{}^{\alpha\dot{\alpha}} 
                =\tfrac{1}{2\sqrt{3}}\,\mathrm{diag}(1,1,1,-3)\otimes 
                    \MATS{0&0\\1&0\\},\\
            (\MAT{e}_{d})^{A}{}_{B}{}^{\alpha\dot{\alpha}}&=
                (\varphi|_{\VEVEWPS{15}{d}})^{A}{}_{B}{}^{\alpha\dot{\alpha}} 
                =\tfrac{1}{2\sqrt{3}}\,\mathrm{diag}(1,1,1,-3)\otimes 
                    \MATS{0&-1\\ 0&0\\},
        \end{align}
        while the predefined real structure $\tilde{\rho}^{\text{PS}}_{15}$ is taken according to Eq.~\eqref{eq:reality-varphi} with $s_{15}'=+1$ as
        \begin{align}
            \tilde{\rho}^{\text{PS}}_{15}(\varphi^{A}{}_{B}{}^{\alpha\dot{\alpha}}) &= \big(
                \epsilon_{\alpha\beta} \,\epsilon_{\dot{\alpha}\dot{\beta}} \,\varphi^{B}{}_{A}{}^{\beta\dot{\beta}}
                \big)^{*}.
        \end{align}
        The $ud$-relations of Eq.~\eqref{eq:theorem-0b} then take the form
        \begin{align}
             \big(
                \epsilon_{\alpha\beta} \,\epsilon_{\dot{\alpha}\dot{\beta}} \,(\MAT{e}_{u})^{B}{}_{A}{}^{\beta\dot{\beta}}
                \big)^{*} &
                    = \tilde{\eta}^{\text{PS}}_{15}\,
                        (\MAT{e}_{d})^{A}{}_{B}{}^{\alpha\dot{\alpha}}, &
                (\MAT{T}^{L}_{+})^{\alpha}{}_{\beta}\,
                (\MAT{T}^{R}_{-})^{\dot{\alpha}}{}_{\dot{\beta}}\,
                (\MAT{e}_{u})^{A}{}_{B}{}^{\beta\dot{\beta}} &= 
                    \alpha^{\text{PS}}_{15}\,
                    (\MAT{e}_{d})^{A}{}_{B}{}^{\alpha\dot{\alpha}}.
        \end{align}
        Explicit computation and comparison of tensor components yields $\tilde{\eta}^{\text{PS}}_{15}=+1$ and $\alpha^{\text{PS}}_{15}=-1$.
        \par
        Now for the $\SO(10)$ side. The $UD$-relations must take exactly the same expressions as in Eq.~\eqref{eq:procedure-122-120-cond} of the previously performed $\SO(10)$ calculation, since the real structure and generator action depend only on the irrep $\GIRREP{120}$, except that the pair of states $\MAT{e}_{U,D}$ and unknown coefficients are now different. We populate $\Phi^{ijk}$ by Eqs.~\eqref{eq:EWVEVs-120-15u} and \eqref{eq:EWVEVs-120-15d} and define $(\MAT{e}_U)^{ijk}=(\Phi|_{\VEVEW{120}{u}{15}})^{ijk}$ and $(\MAT{e}_D)^{ijk}=(\Phi|_{\VEVEW{120}{d}{15}})^{ijk}$, and replace the coefficient labels $\eta^{\SO(10)}_{1,120}\mapsto \eta^{\SO(10)}_{15,120}$ and $\alpha^{\SO(10)}_{1,120}\mapsto \alpha^{\SO(10)}_{15,120}$. Explicit computation yields $\eta^{\SO(10)}_{15,120}=\alpha^{\SO(10)}_{15,120}=+1$, and hence according to Eq.~\eqref{eq:theorem-2b} of our algorithm, the sign associated to the inherited real structure $s_{15,120}$ is computed as
        \begin{align}
            s_{15,120}&= \frac{\alpha^{\text{PS}}_{15}}{ \alpha^{\SO(10)}_{15,120}} \frac{\eta^{\SO(10)}_{15,120}}{\tilde{\eta}^{\text{PS}}_{15}}
                = \frac{-1}{+1}\,\frac{+1}{+1} = -1.
        \end{align}
        The inherited real structure in the irrep $\PSIRREP{15}{2}{2}$ is thus opposite to the (standard) predefined one, consistent with taking $s'_{15}=-1$ in Eq.~\eqref{eq:reality-varphi} for the PS calculation, which is indeed our main claim of Section~\ref{sec:reality-conditions}. 
\end{itemize}

\noindent
\textbf{Proof of Theorem~\ref{the:procedure}.}
Having demonstrated how our method works, we now present the proof of our theorem. It does not involve much beyond the application of basic definitions from representation theory and some simple algebra. For a review of definitions involving real structures, the reader is invited to consult item~\ref{item:reality-conditions} in Section~\ref{subsec:group-theory}. 

The assumptions of matching states in irreps $r$ and $R$ up to normalization in Eq.~\eqref{eq:theorem-0a} can be explicitly written out as
\begin{align}
    i_r(\MAT{e}_{u})&=a_{r}\,\MAT{e}_U, &
    i_r(\MAT{e}_{d})&=b_{r}\,\MAT{e}_D, \label{eq:proof-ir}
\end{align}
introducing $b_{r}\in\mathbb{C}\setminus\{0\}$. We deliberately avoided this label in the formulation of the theorem, since it drops out in the final result. We now perform a series of algebraic steps after applying $\MAT{T}^{G}_{R}$ to the left-side expression of Eq.~\eqref{eq:proof-ir}:
\begin{align}
    \MAT{T}^{G}_{R}\circ i_{r}(\MAT{e}_{u}) &=
        i_{r}\circ\MAT{T}^{H}_{r}(\MAT{e}_{u})=
        i_{r}(\alpha^{H}_{r}\,\MAT{e}_{d})=
        \alpha^{H}_{r} i_{r}(\MAT{e}_{d}) =
        \alpha^{H}_{r}\,b_{r}\,\MAT{e}_{D}, \label{eq:proof-t1}\\
    \MAT{T}^{G}_{R}\circ i_{r}(\MAT{e}_{u}) &= \MAT{T}^{G}_{R} (a_{r}\,\MAT{e}_{U}) =
            a_{r}\,\MAT{T}^{G}_{R}(\MAT{e}_{U}) =
            a_{r}\,\alpha^{G}_{r,R}\,\MAT{e}_{D}. \label{eq:proof-t2}
\end{align}
In order from left to right, the equalities of Eq.~\eqref{eq:proof-t1} reflect the $H$-equivariance of the embedding map $i_{r}$ (which applies also to elements of the universal enveloping algebra $\mathcal{U}(\mathfrak{h})$), the right-side assumption of Eq.~\eqref{eq:theorem-0b}, the linearity of the embedding map $i_{r}$, and the right-side embedding of Eq.~\eqref{eq:proof-ir}.
Similarly, the equalities of Eq.~\eqref{eq:proof-t2} arise from the left-side embedding of Eq.~\eqref{eq:proof-ir}, the linearity of the action of $\MAT{T}^{G}_{R}$, and the right-side assumption in Eq.~\eqref{eq:theorem-0c}. 
An analogous relation can be derived by applying $\rho^{G}_{R}$ to the 
left-side expression of Eq.~\eqref{eq:proof-ir}:
\begin{align}
    \rho^{G}_{R}\circ i_{r}(\MAT{e}_{u}) &=
        i_{r}\circ\rho^{H}_{r}(\MAT{e}_{u})=
        i_{r}(\eta^{H}_{r}\,\MAT{e}_{d})=
        \eta^{H}_{r}\,i_{r}(\MAT{e}_{d}) =
        \eta^{H}_{r}\,b_{r}\,\MAT{e}_{D}, \label{eq:proof-c1}\\
    \rho^{G}_{R}\circ i_{r}(\MAT{e}_{u}) &= \rho^{G}_{R} (a_{r}\,\MAT{e}_{U}) =
            a_{r}^{*}\,\rho^{G}_{R}(\MAT{e}_{U}) =
            a_{r}^{*}\,\eta^{G}_{r,R}\,\MAT{e}_{D}. \label{eq:proof-c2}
\end{align}
In order from left to right, the equalities of Eq.~\eqref{eq:proof-c1} follow from $i_{r}$ being a real embedding (commutes with real structure), the left-side assumption of Eq.~\eqref{eq:theorem-0b}, the linearity of the embedding map $i_{r}$, and the right-side embedding relation in Eq.~\eqref{eq:proof-ir}.
Similarly, the equalities of Eq.~\eqref{eq:proof-c2} take into account the left-side embedding in Eq.~\eqref{eq:proof-ir}, the anti-linearity of the real structure $\rho^{G}_{R}$, and the left-side assumption of Eq.~\eqref{eq:theorem-0c}.
\par
Eqs.~\eqref{eq:proof-t1} and \eqref{eq:proof-t2} describe the same object, so we can relate the coefficients of their final expressions, and analogously for Eqs.~\eqref{eq:proof-c1} and \eqref{eq:proof-c2}. We obtain
\begin{align}
    \alpha^{H}_{r}\,b_{r}&= \alpha^{G}_{r,R}\,a_{r}, &
    \eta^{H}_{r}\,b_{r}&= \eta^{G}_{r,R}\,a_{r}^*.
\end{align}
Dividing the right-side by the left-side relation eliminates $b_{r}$, yielding
\begin{align}
    \frac{\eta^{H}_{r}}{\alpha^{H}_{r}} &= \frac{a_{r}^*}{a_{r}}\;\frac{\eta^{G}_{r,R}}{\alpha^{G}_{r,R}}. \label{eq:proof-ratio1}
\end{align}
We now have everything prepared to derive the statements of the theorem:
\begin{enumerate}[itemsep=0cm,leftmargin=\DIMLEFTMARGIN,label=(\alph*)]
    \item Suppose $a_{r}\in\mathbb{R}$. Then $a_{r}^*=a_{r}$, so it drops out of Eq.~\eqref{eq:proof-ratio1}, yielding the desired $\eta/\alpha$-invariance 
    of Eq.~\eqref{eq:theorem-1a}.
    \item Suppose we have a real structure $\tilde{\rho}^{H}_{r}$ on $\mathcal{V}^{H}_{r}$ that is $H$-equivariant. Since such real structures are unique up to a phase, cf.~item~\ref{item:real-structure} in Section~\ref{subsec:group-theory}, it must be related to the inherited real structure $\rho^{H}_{r}$ via $\rho^{H}_{r}=s_{r,R}\,\tilde{\rho}^{H}_{r}$ for some coefficient $s_{r,R}\in\mathrm{U}(1)$. We can relate
    \begin{align}
        \eta^{H}_{r}\,\MAT{e}_{d}=\rho^{H}_{r}(\MAT{e}_{u})=s_{r,R}\cdot \tilde{\rho}^{H}_{r}(\MAT{e}_{u}) = s_{r,R}\,\tilde{\eta}^{H}_{r}\,\MAT{e}_{d}, \label{eq:proof-ratio2}
    \end{align}
    where in order from left to right the equalities follow from the left-side assumption of Eq.~\eqref{eq:theorem-0b}, the relation between the inherited and predefined real structure on the vector space $\mathcal{V}^{H}_{r}$, and the assumption of Eq.~\eqref{eq:theorem-2a}. Relating the coefficients $\eta^{H}_{r}=s_{r,R}\tilde{\eta}^{H}_{r}$ in Eq.~\eqref{eq:proof-ratio2}, and inserting $\eta^{H}_{r}$ expressed from Eq.~\eqref{eq:proof-ratio1}, we arrive at the desired claim of Eq.~\eqref{eq:theorem-2bp}. 
    \item Assume the expansion of Eq.~\eqref{eq:theorem-3a}. Define linear combinations 
    \begin{align}
        \MAT{X}&:=\MAT{U}+\MAT{D}=U\,\MAT{e}_{U}+D\,\MAT{e}_D, &
        \MAT{x}&:=\MAT{u}+\MAT{d}=u\,\MAT{e}_{u}+d\,\MAT{e}_d.
    \end{align}
    Let us first focus on $\MAT{X}$. The combination $\MAT{X}$ will be by definition part of the real subspace $(\mathcal{V}^{G}_{R})_{\mathbb{R}}$ if it is a stationary point of the reality condition, i.e.~$\rho^{G}_{R}(\MAT{X})=\MAT{X}$. We study this relation more closely:
    \begin{align}
        U\,\MAT{e}_{U}+D\,\MAT{e}_{D}&
            =\rho^{G}_{R}(U\,\MAT{e}_{U}+D\,\MAT{e}_{D})
            =U^*\,\rho^{G}_{R}(\MAT{e}_{U})+D^{*}\,\rho^{G}_{R}(\MAT{e}_{D})
            = U^*\,\eta^{G}_{r,R}\,\MAT{e}_{D} + 
            D^*/(\eta^{G}_{r,R})^*\,\MAT{e}_{U}. \label{eq:proof-linear1}
    \end{align}
    From left to right, the equalities follow from the stationary property of $\MAT{X}$ under the real structure, the anti-linearity of $\rho^{G}_{R}$, and from the left-side assumption of Eq.~\eqref{eq:theorem-0c}. Note that $\rho^{G}_{R}(\MAT{e}_D)$ can be extracted by applying another $\rho^{G}_{R}$ to the left-side relation in Eq.~\eqref{eq:theorem-0c}: 
    \begin{align}
        \MAT{e}_{U}&=\rho^{G}_{R}\circ\rho^{G}_{R}(\MAT{e}_U) = \rho^{G}_{R}(\eta^{G}_{r,R}\,\MAT{e}_{D})=(\eta^{G}_{r,R})^*\cdot \rho^{G}_{R}(\MAT{e}_D),
    \end{align}
    where we took into account in the first relation that real structures are by definition an involution, i.e.~$\rho^{G}_{R}\circ\rho^{G}_{R}=\mathrm{Id}$.
    Matching the coefficients in Eq.~\eqref{eq:proof-linear1}, we obtain the left-side relation $\eta^{G}_{r,R}\,U^*=D$ in Eq.~\eqref{eq:theorem-3b}.
    By a completely analogous sequence of steps, we derive the reality condition $\eta^{H}_{r} u^*=d$ for the inherited real structure $\rho^{H}_{r}$, and matching the coefficients of Eq.~\eqref{eq:proof-ratio2} yields the right-side relation of Eq.~\eqref{eq:theorem-3b}.
\end{enumerate}
\textbf{End of proof.}

\vskip 0.5cm

We finish this appendix section with a discussion of our theorem/algorithm. The inherited real structure $\rho^{H}_{r}$ is uniquely determined by the embedding $i_{r}$ and the parent real structure $\rho^{G}_{R}$. The coefficient $s_{r,R}$ computed from Eq.~\eqref{eq:theorem-2bp} then tells us how to compare the predefined $\tilde{\rho}^{H}_{r}$ with the inherited $\rho^{H}_{r}$. The method is robust against various changes of conventions or setup:
\begin{itemize}[itemsep=0cm,leftmargin=\DIMLEFTMARGIN]
    \item One advantage of the algorithm is that not much is needed to be known explicitly about the embedding map $i_{r}$, only the phase of the embedding coefficient $a_{r}$ for the state $\MAT{e}_{u}$. Provided the matching states on the $H$- and $G$-side of the computation have been identified correctly, the remainder of the algorithm involves computation purely on the $H$- and purely on the $G$-side.  
    \item Suppose we normalize the states $\MAT{e}_{u,d,U,D}$ differently, which is tantamount to individually reparametrizing the VEV variables with a factor. This changes the coefficients $a_{r}$, $b_{r}$, $\alpha$ and $\eta$ in such a way that $s_{r,R}$ remains unchanged, as is shown in  
    Table~\ref{tab:theorem-rephasing}. The expression for $s_{r,R}$ is thus manifestly invariant under changing the (complex) normalization of states,  or equivalently under VEV reparametrizations, as it should be. The simpler expression of Eq.~\eqref{eq:theorem-2b} is unchanged provided $a_{r}\in\mathbb{R}$, i.e.~when no complex phases were introduced in reparametrizing the $u$-type states/VEVs.
    \begin{table}[htb]
        \centering
        \begin{tabular}{l@{$\qquad$}l@{$\qquad$}l}
            \toprule
                state change&coefficient change&change in $s_{r,R}$\\
            \midrule
            $\MAT{e}_{u}\mapsto c\cdot\MAT{e}_{u}$ &
                $a_{r}\mapsto c\cdot a_{r},\ 
                \alpha^{H}_{r}\mapsto c\cdot\alpha^{H}_{r},\ 
                \tilde{\eta}^{H}_{r}\mapsto c^*\cdot\tilde{\eta}^{H}_{r}$&
                $s_{r,R}\mapsto s_{r,R}$\\
            $\MAT{e}_{U}\mapsto c\cdot \MAT{e}_{U}$ &
                $a_{r}\mapsto c^{-1}\cdot a_{r},\ 
                \alpha^{G}_{r,R}\mapsto c\cdot\alpha^{G}_{r,R},\ 
                \eta^{G}_{r,R}\mapsto c^*\cdot\eta^{G}_{r,R}
                $&
                $s_{r,R}\mapsto s_{r,R}$\\
            $\MAT{e}_{d}\mapsto c\cdot \MAT{e}_{d}$&
                $b_{r}\mapsto c\cdot b_{r},\ 
                \alpha^{H}_{r}\mapsto c^{-1}\cdot\alpha^{H}_{r},\ 
                \tilde{\eta}^{H}_{r}\mapsto c^{-1}\cdot\tilde{\eta}^{H}_{r}
                $ &
                $s_{r,R}\mapsto s_{r,R}$\\
            $\MAT{e}_{D}\mapsto c\cdot \MAT{e}_{D}$&
                $b_{r}\mapsto c^{-1}\cdot b_{r},\ 
                \alpha^{G}_{r,R}\mapsto c^{-1}\cdot\alpha^{G}_{r,R},\ 
                \eta^{G}_{r,R}\mapsto c^{-1}\cdot\eta^{G}_{r,R}
                $&
                $s_{r,R}\mapsto s_{r,R}$\\
            \bottomrule
        \end{tabular}
        \caption{Multiplying any state with a complex number  $c\neq 0$ modifies the coefficients in Eqs.~\eqref{eq:proof-ir}, \eqref{eq:theorem-0b} and \eqref{eq:theorem-0c}, but the result $s_{r,R}$ in Eq.~\eqref{eq:theorem-2bp} remains invariant. To track coefficient changes, we take into account that $i_{r}$ and $\hat{\MAT{T}}$ are linear maps, while $\tilde{\rho}^{H}_{r}$ and $\rho^{G}_{R}$ are antilinear. \label{tab:theorem-rephasing}}
    \end{table}
    \item In many practical applications, the calculation is set up so that $a_{r}$ can be inferred to be real (without necessarily knowing its exact value), and so the simpler expression of Eq.~\eqref{eq:theorem-2b} can be used. One helpful criterion in practice is if the defining/fundamental representation of $G$ has all states from the $H$-language embedded into it with real coefficients, as happens explicitly in the conventions used here. Then all tensor constructions from that embedding and subsequent real-linear restrictions (such as symmetry conditions or tracelessness) also have exclusively real entries in their embedding matrix $i_{r}$. Note that embedding matrices are basis dependent, i.e.~they depend on the basis chosen for both $\mathcal{V}^{H}_{r}$ and $\mathcal{V}^{G}_{r}$, and so $a_{r}\in\mathbb{R}$ is a basis dependent condition. For the above tensorial argument to hold, one must retain the original basis of both sides in its construction (or at least not introduce complex phases if there is a basis change).
    \par
    In our $\SO(10)$ case, for example, we use the embedding where PS fields are simply inserted into the fundamental $\GIRREP{10}$ up to a sign: the embedding 
        \begin{align}
            \PSIRREP{1}{2}{2}\oplus\PSIRREP{6}{1}{1} \hookrightarrow \GIRREP{10}
        \end{align}
        is realized as
        \begin{align}
            \phi^{\alpha\dot{\alpha}}=\MATS{\bar{e}&\nu\\ \bar{\nu}&e\\}, \quad
            \phi^{[AB]}= \MATS{
                0&d^{c}_{3}&-d^{c}_{2}&\bar{d}^{c}_{1}\\
                -d^{c}_{3}&0&d^{c}_{1}&\bar{d}^{c}_{2}\\
                d^{c}_{2}&-d^{c}_{1}&0&\bar{d}^{c}_{3}\\
                -\bar{d}^{c}_{1}&-\bar{d}^{c}_{2}&-\bar{d}^{c}_{3}&0\\
                } 
            &\quad\hookrightarrow\quad
            \Phi^{i}=(\bar{d}^{c}_{1},\bar{d}^{c}_{2},\bar{d}^{c}_{3},\bar{e},\bar{\nu},d^{c}_{1},d^{c}_{2},d^{c}_{3},e,-\nu),
        \end{align}
        where a fermion-like notation for the states is used, labeling the indices according to Table~\ref{tab:indices} and using the $\SO(10)$ complex basis of Table~\ref{tab:qn10}. Using the tensor argument, this guarantees that our descriptions for the embeddings into $\GIRREP{120}$ also have only real embedding coefficients. 
    \item Suppose we set up our calculation so that $i_{r}$ has real matrix entries in a certain basis for the $r$- and $R$-irreps, e.g.~as described in the previous point. If each of the two real structures can be written as a product $\rho=AK$, where $A$ is a real linear map and $K$ is complex conjugation of entries, then the states $\MAT{e}_{u,d,U,D}$ can be chosen with real tensor entries. If we normalize the states according to
    $\|\MAT{e}_{u}\|=\|\MAT{e}_{d}\|$ and $\|\MAT{e}_{U}\|=\|\MAT{e}_{D}\|$, which requires a suitable normalization of the operator $\hat{\MAT{T}}$ as well, and the linear part of the real structure $A$ is norm-preserving, then all $\alpha$- and $\eta$-coefficients of Eq.~\eqref{eq:theorem-2b} have values $\pm 1$, giving $s_{r,R}=\pm 1$. Introducing no complex phases in such calculations thus implies that the ambiguity for the inherited real structure was only in its sign. Our calculations for various irrep cases of $4_C\,2_L\,2_R\subset \SO(10)$ were performed exactly under the circumstances described above. 
    \item The computation of $\alpha$- and $\eta$-coefficients has a certain robustness in practice. For tensor constructions, Eqs.~\eqref{eq:theorem-0b} and \eqref{eq:theorem-0c}    
    become tensor equations, for which all entries must match. In more complicated constructions, the states typically have many non-vanishing entries, and thus the $\alpha$- and $\eta$-coefficients must satisfy a large number of conditions; the existence of a solution then serves as a consistency check of the calculation. An inconsistency typically indicates a misidentification or erroneous implementation of the states or raising/lowering operators. Furthermore, an independent cross-check of $s_{r,R}$ can be performed if the calculation is repeated on an alternative set of states $\MAT{e}'_{u,d,U,D}$ (e.g.~with different quantum numbers) and a different element $\hat{\MAT{T}}'$.
    \item Changing the sign/phase of the real structure $\rho^{G}_{R}$ or $\tilde{\rho}^{H}_{r}$ changes the sign/phase of $\eta^{G}_{r,R}$ and $\tilde{\eta}^{H}_{r}$, respectively, and thus also the sign/phase of $s_{r,R}$. This is fully consistent with expectations, since the former changes the parent real structure and hence must change the inherited real structure, while the latter changes the predefined real structure relative to which the inherited one is being compared to via $s_{r,R}$. 
    \par
    Since all phases $\theta$ for the parent $e^{i\theta}\rho^{G}_{R}$ give equivalent physics, cf.~item~(\ref{item:real-structure}) in Section~\ref{subsec:group-theory}, the computation of $s_{r,R}$ for a single irrep pair $r\hookrightarrow R$ is merely bookkeeping. A product $s_{r,R}s_{r',R}^{*}$ for any two irreps $r,r'\hookrightarrow R$, however, is invariant under $\theta$, providing physical information on how the inherited real structures of all daughter irreps of $R$ are related.
    \item The method assumes the same abstract sequence of ladder operators $\hat{\MAT{T}}$ is taken both on the $H$- and $G$-side of the calculation.  
    The method is not robust under making a sign mistake in $\hat{\MAT{T}}$ only on one side, since either $\alpha^{H}_{r}$ or $\alpha^{G}_{r,R}$ changes sign, and thus $s_{r,R}\mapsto - s_{r,R}$. In our $\SO(10)$ calculation, however, the product $s_{1,120}s_{15,120}=-1$ has a robustness against this kind of mistake, since we used the same ladder element $\hat{\MAT{T}}$
    for computing $s_{1,120}$ and $s_{15,120}$, and even the same $2D$ representations of this element from Eq.~\eqref{eq:Telement-PS} in the tensor formalism on the PS-side of the computation (the $\SO(10)$ side is automatically the same). Therefore, any sign mistake in $\hat{\MAT{T}}$ either on the PS-side or $\SO(10)$-side changes both $s'_{1}$ and $s'_{15}$ simultaneously, leaving the product invariant.
\end{itemize}

\section{Proton Decay} \label{app:PD}
In this Appendix, we summarize the relevant equations for computing the gauge-mediated proton decay widths, following the analyses of Refs.~\cite{FileviezPerez:2004hn,Nath:2006ut}. Within the Standard Model context, the relevant dimension-six ($d=6$) operators contributing to proton decay are given by 
\begin{align}
    \mathcal{O}^{B-L}_I&
        =k_1^2 \; \epsilon_{ijk}\; \epsilon_{\alpha\beta}\; 
        \big(\overline{u^C_{ia}}_L\, \gamma^\mu\, Q_{j\alpha aL}\big)\;
        \big(\overline{e^C_{b}}_L \,\gamma_\mu \, Q_{k \beta bL}\big),\\
    \mathcal{O}^{B-L}_{II}&
        =k_1^2 \; \epsilon_{ijk}\; \epsilon_{\alpha\beta}\; 
        \big(\overline{u^C_{ia}}_L\, \gamma^\mu \, Q_{j\alpha aL}\big)\; \big(\overline{d^C_{kb}}_L \;\gamma_\mu \;L_{\beta bL}\big),\\
    \mathcal{O}^{B-L}_{III}&
        =k_2^2 \; \epsilon_{ijk}\; \epsilon_{\alpha\beta}\; 
        \big(\overline{d^C_{ia}}_L\, \gamma^\mu\, Q_{j\beta aL}\big)\; 
        \big(\overline{u^C_{kb}}_L \,\gamma_\mu \, L_{\alpha bL}\big), \\
    \mathcal{O}^{B-L}_{IV}&
        =k_2^2 \; \epsilon_{ijk}\; \epsilon_{\alpha\beta}\; 
        \big(\overline{d^C_{ia}}_L\, \gamma^\mu\, Q_{j\beta aL}\big)\; 
        \big(\overline{\nu^C_{b}}_L \,\gamma_\mu \,Q_{k \alpha bL}\big).
\end{align}
\noindent 
Here, the SM fermion doublets are defined as 
$ Q_L = (u_L, d_L) $ and $ L_L = (\nu_L, e_L) $.  The indices $i,j,k$ are color indices, $a,b$ are family indices and, $\alpha,\beta$ are $SU(2)_L$ indices, and Dirac bilinears are enclosed in parentheses. 
In the above equation, the factors $k_{1,2}$ encode information 
about the superheavy gauge boson masses as well as the unified 
coupling $g_{\mathrm{GUT}} $. They are given by
\begin{align}
    k_1 &= \frac{g_\mathrm{GUT}}{\sqrt{2}\, M_{(X,Y)}}, &
    k_2 &= \frac{g_\mathrm{GUT}}{\sqrt{2}\, M_{(X',Y')}},
\end{align}
where \( M_{(X,Y)} \) and \( M_{(X',Y')} \) denote the masses of the 
\( (X,Y) \) and \( (X',Y') \) gauge bosons, respectively.    Proton decay rates must be computed in the physical basis, in which the above operators take the following form:
\begingroup
\allowdisplaybreaks
\begin{align}
    \mathcal{O}(e_\alpha^C,d_\beta) &
        = c(e_\alpha^C,d_\beta)\; \epsilon_{ijk}\; 
        \big(\overline{u^C_i}_L\, \gamma^\mu \, u_{jL} \big)\;
        \big(\overline{e^C_\alpha}_L\,\gamma_\mu \, d_{k\beta L}\big),  \\
    \mathcal{O}(e_\alpha,d_\beta^C) &
        = c(e_\alpha,d_\beta^C)\; \epsilon_{ijk}\; 
        \big(\overline{u^C_i}_L\, \gamma^\mu \, u_{jL}\big)\; 
        \big(\overline{d^C_{k \beta}}_L\, \gamma_\mu \, e_{\alpha L}\big),  \\
    \mathcal{O}(\nu_l, d_\alpha,d_\beta^C)&
        = c(\nu_l, d_\alpha,d_\beta^C)\; \epsilon_{ijk}\; 
        \big(\overline{u^C_i}_L\, \gamma^\mu\, d_{j\alpha L}\big)\; 
        \big(\overline{d^C_{k \beta}}_L\, \gamma_\mu\, \nu_{lL}\big), \\
    \mathcal{O}(\nu_l^C, d_\alpha,d_\beta^C) &
        = c(\nu_l^C, d_\alpha,d_\beta^C)\; \epsilon_{ijk}\; 
        \big(\overline{d^C_{i \beta}}_L \, \gamma^\mu\, u_{j L}\big)\; 
        \big(\overline{\nu_l^C}_L \,\gamma_\mu \, d_{k \alpha L}\big).
\end{align}
\endgroup
In the above equations, the quantities \( c \) capture the information about the fermion mixing parameters. By defining   $c= k^2_1\,\hat{c}$ and $K^2=k^2_2/k^2_1=M^2_X/M^2_{X'}$, the detailed flavor structure is contained in \( \hat{c} \):
\begingroup
\allowdisplaybreaks
\begin{align}
    \hat c(e^C_\alpha,d_\beta) &
        =  \left[ V_1^{11} V_2^{\alpha \beta} +\left( V_1 V_{UD}\right)^{1\beta}\left(V_2 V^\dagger_{UD}\right)^{\alpha 1} \right], \label{CIA} \\
    \hat c(e_\alpha,d^C_\beta) &
        =  V_1^{11} V_3^{\beta \alpha}+K^2 \left(V_4V^\dagger_{UD}\right)^{\beta 1} \left( V_1 V_{UD} V_4^\dagger V_3\right)^{1 \alpha}, \label{CIA2} \\
    \hat c(\nu_l,d_\alpha,d_\beta^C)&
        = \left( V_1 V_{UD}\right)^{1\alpha}\left( V_3 V_{EN}\right)^{\beta l} +K^2 V_4^{\beta \alpha} \left( V_1 V_{UD} V_4^\dagger V_3 V_{EN}\right)^{1l},\\
    \hat c(\nu_l^C,d_\alpha,d_\beta^C)&
        = K^2 \left[\left(V_4 V_{UD}^\dagger \right)^{\beta 1} \left(U_{EN}^\dagger V_2\right)^{l \alpha} + V_4^{\beta \alpha} \left(U_{EN}^\dagger V_2 V_{UD}^\dagger\right)^{l 1}\right],\quad \alpha=\beta \neq 2. \label{CIB}
\end{align}
\endgroup

\noindent
The mixing matrices, $V_i$, are defined such  that
\begin{align}
    V_1&= U_R^\top U_L, &
    V_2&=E_R^\top D_L, &
    V_3&=D_R^\top E_L, &
    V_4&=D_R^\top  D_L, \\
    V_{UD}&=U_L^{\dagger}D_L, &
    V_{EN}&=E_L^{\dagger}N_L, &
    U_{EN}&= E^\top_R N_L,   
\end{align}
where $U,D,E$ define the  diagonalizing matrices given by
\vspace{-2pt}
\begin{align}
    U^{\dagger}_R \; \MAT{M}_U \; U_L &= M_U^{diag}, &
    D^{\dagger}_R \; \MAT{M}_D \; D_L &= M_D^{diag}, &
    E^{\dagger}_R \; \MAT{M}_E \; E_L &= M_E^{diag}, &
    N^\top_L \; \MAT{M}_N \ N_L &= M_N^{diag}.
\end{align}
Therefore, the CKM (Cabibbo–Kobayashi–Maskawa) and PMNS (Pontecorvo–Maki–Nakagawa–Sakata) matrices are defined by $U_\mathrm{CKM}=U^\dagger_L D_L$ and 
$U_\mathrm{PMNS}=E^\dagger_L N_L$, respectively.

Then the partial decay width of the decay $N\to P+ \overline{l}$, where $N=(p, n)$, $P= (\pi, K, \eta)$ and $\overline{l}$ is an anti-lepton, is given by~\cite{Aoki:2017puj}:
\begin{align}
\Gamma(N\to P+ \overline{l})= \frac{m_N}{32 \pi} \left[1-\left(\frac{m_P}{m_N}\right)^2\right]^2 \left| \sum_I C^I W^I_0(N\to P) \right|^2 .
\end{align}
Writing explicitly for all gauge-mediated proton decay modes, we obtain:
\begingroup
\allowdisplaybreaks
\begin{align}
&\Gamma(p^{+}\to\pi^{0}e^{+}_\beta)= \frac{\left(m^2_p-m^2_{\pi^0}\right)^2}{32\pi m^3_p} \frac{g^4 A^2_L}{M^4_X} 
\nonumber\\& \hspace{60pt} \times
\bigg\{ 
 A^2_{SR} \left|  \langle  \pi^0|(ud)_Ru_L|p \rangle  \right|^2  \left|  \hat c(e_\beta,d^C)  \right|^2
+
 A^2_{SL} \left|  \langle   \pi^0|(ud)_Lu_R|p \rangle \right|^2 \left|  \hat c(e_\beta^C,d)  \right|^2
\bigg\},\label{eq:decay-pi-e}
\\
&\Gamma(p^{+}\to K^{0}e^{+}_\beta)= \frac{\left(m^2_p-m^2_{K^0}\right)^2}{32\pi m^3_p} \frac{g^4 A^2_L}{M^4_X} 
\nonumber\\& \hspace{60pt} \times
\bigg\{ 
 A^2_{SR} \left|  \langle  K^0|(us)_Ru_L|p \rangle  \right|^2  \left|  \hat c(e_\beta,s^C)  \right|^2
+
 A^2_{SL} \left|  \langle   K^0|(us)_Lu_R|p \rangle \right|^2 \left|  \hat c(e_\beta^C,s)  \right|^2
\bigg\},
\\
&\Gamma(p^{+}\to\eta e^{+}_\beta)= \frac{\left(m^2_p-m^2_{\eta}\right)^2}{32\pi m^3_p} \frac{g^4 A^2_L}{M^4_X} 
\nonumber\\&  \hspace{60pt} \times
\bigg\{ 
 A^2_{SR} \left|  \langle  \eta|(ud)_Ru_L|p \rangle  \right|^2  \left|  \hat c(e_\beta,d^C)  \right|^2 
+
 A^2_{SL} \left|  \langle   \eta|(ud)_Lu_R|p \rangle \right|^2  \left|  \hat c(e_\beta^C,d)  \right|^2
\bigg\}, \label{eq:decay-eta-e}
\\
&\Gamma(p^{+}\to\pi^{+}\overline{\nu})= \frac{\left(m^2_p-m^2_{\pi^+}\right)^2}{32\pi m^3_p} \frac{g^4 A^2_L}{M^4_X} A^2_{SR}  
  \left|  \langle  \pi^+|(ud)_Rd_L|p \rangle  \right|^2  \sum_{l=1}^{3}
\left|  \hat c(\nu_l,d,d^C)  \right|^2,
\\
&\Gamma(p^{+}\to K^{+}\overline{\nu})= \frac{\left(m^2_p-m^2_{K^+}\right)^2}{32\pi m^3_p} \frac{g^4 A^2_L}{M^4_X} A^2_{SR} 
\nonumber\\&  \hspace{60pt} \times
\sum_{l=1}^{3}
\bigg\{   
\left|  \langle  K^+|(us)_Rd_L|p \rangle   \hat c(\nu_l,d,s^C)  
+  
\langle  K^+|(ud)_Rs_L|p \rangle   \hat c(\nu_l,s,d^C)  \right|^2
\bigg\}.
\end{align}
\endgroup
The relevant proton decay matrix elements have been computed in Ref.~\cite{Aoki:2017puj} and are listed below (the rest can be obtained by the interchange $RL \leftrightarrow LR$): 
\begin{align}
    \langle \pi^0| (ud)_R u_L |p \rangle &= -0.131 \,\mathrm{GeV}^2, &
    \langle K^0| (us)_R u_L |p \rangle &= 0.103 \,\mathrm{GeV}^2, \\
    \langle \eta | (ud)_R u_L |p \rangle &= 0.006 \,\mathrm{GeV}^2, &
    \langle \pi^+| (ud)_R d_L |p \rangle &= -0.186 \,\mathrm{GeV}^2, \\
    \langle K^+| (us)_R d_L |p \rangle &= -0.049 \,\mathrm{GeV}^2, &
    \langle K^+| (ud)_R s_L |p \rangle &= -0.134 \,\mathrm{GeV}^2.
\end{align}

The long-distance enhancement factor $A_L$~\cite{Nihei:1994tx} is approximately $A_L \approx 1.2$, while the short-range renormalization factor can be taken as $A_S \approx 2$. Note that these two factors are not needed for computing the branching ratios. The relevant meson and baryon masses are listed below~\cite{ParticleDataGroup:2024cfk}:
\begin{align}
    m_p &= 938.27\,\mathrm{MeV}, &
    m_{\pi^0} &= 134.97\,\mathrm{MeV}, &
    m_{\pi^+} &= 139.57\,\mathrm{MeV},\\
    m_{K^0} &= 497.61\,\mathrm{MeV}, &
    m_{K^+} &= 493.67\,\mathrm{MeV}, &
    m_{\eta} &= 547.86\,\mathrm{MeV}.
\end{align}

A final remark: the structural similarity of Eq.~\eqref{eq:decay-pi-e} and \eqref{eq:decay-eta-e} implies the ratios of rates for certain channels to be model independent, i.e.~predicted solely based on baryon/meson masses and matrix elements:
\begin{align}
    \frac{\Gamma(p\to \eta e^+)}{\Gamma(p\to \pi^0 e^+)}
    = \frac{\Gamma(p\to \eta \mu^+)}{\Gamma(p\to \pi^0 \mu^+)}
    &= \frac{(m_{p}^{2}-m^{2}_{\eta})^2}{(m_{p}^{2}-m^{2}_{\pi^0})^2}\;\left|\frac{\langle  \eta|(ud)_Ru_L|p \rangle}{\langle  \pi^0|(ud)_Ru_L|p \rangle}\right|^{2} = 9.50\cdot 10^{-4}. \label{eq:decay-relation}
\end{align}

\bibliographystyle{style}
\bibliography{reference}
\end{document}